\documentclass[12pt,subeqn,a4paper]{article}

\usepackage{amssymb,amsmath,amsfonts,amsthm, amscd, mathrsfs,helvet}
\usepackage{bm}
\usepackage{graphicx,verbatim}
\usepackage{psfrag}
\usepackage[all]{xy}

\theoremstyle{plain}

\numberwithin{equation}{section}

\begin{document}

\begin{titlepage}
\begin{flushright}
{\bf January 2006} \\

hep-th/0601194 \\
\end{flushright}
\begin{centering}
\vspace{.2in}
 {\large {\bf
On the Dynamics of Finite-Gap Solutions \\ in Classical String Theory}}\\

\vspace{.3in}

Nick Dorey and Beno\^{\i}t Vicedo\\
\vspace{.1 in}
DAMTP, Centre for Mathematical Sciences \\
University of Cambridge, Wilberforce Road \\
Cambridge CB3 0WA, UK \\
\vspace{.2in}
%
%
\vspace{.4in}
{\bf Abstract} \\

We study the dynamics of finite-gap solutions in classical string
theory on $\mathbb{R}\times S^{3}$. Each solution is
characterised by a spectral curve, $\Sigma$, of genus $g$ and a
divisor, $\gamma$, of degree $g$ on the curve.
We present a complete reconstruction of the general solution and identify
the corresponding moduli-space, $\mathcal{M}^{(2g)}_{\mathbb{R}}$, as a
real symplectic manifold of dimension $2g$.
The dynamics of the general solution is shown to be equivalent to
a specific Hamiltonian integrable system with phase-space
$\mathcal{M}^{(2g)}_{\mathbb{R}}$.
The resulting description resembles the free motion of a rigid
string on the Jacobian torus $J(\Sigma)$. Interestingly, the
canonically-normalised action variables of the integrable system
are identified with certain filling fractions which play an important role
in the context of the AdS/CFT correspondence.
\end{centering}

\end{titlepage}

\setcounter{page}{0}

\tableofcontents

\input{epsf}

\setcounter{section}{-1}

\section{Introduction}

The AdS/CFT correspondence continues to inspire progress both in
gauge theory and in string theory. A particularly exciting recent
development is the emergence of integrability in classical string
theory on $AdS_{5}\times S^{5}$ \cite{Bena:2003wd}. Specifically,
the classical equations of motion of the string can be reformulated
as the zero-curvature condition for a Lax connection which
immediately leads to the existence of an infinite tower of conserved
quantities. A similar Lax representation exists for classical string
theory on a large family of backgrounds including symmetric spaces,
group manifolds and various supersymmetric
generalisations\footnote{Of course these spaces yield consistent
backgrounds for first-quantised string theory only in very special
cases such as that of $AdS_{5}\times S^{5}$}.

In concrete terms, the Lax representation allows one to obtain a
very large class of explicit solutions of the classical string
equations known as {\em finite-gap solutions}. Each of these
solutions is characterised by an auxiliary Riemann surface $\Sigma$
known as the spectral curve. Solutions are naturally classified by
the genus $g$ of $\Sigma$ or, equivalently, by the number $K=g+1$ of
gaps or forbidden zones in the spectrum of the Lax operator. We will
refer to configurations of fixed $K$ as $K$-gap solutions. As we
discuss below, it is plausible that arbitrary classical motions of
the string can be obtained from an appropriate $K\rightarrow \infty$
limit of the $K$-gap solutions. Moreover, these solutions yield a
description of the phase-space of the string analogous to that
provided by the mode expansion in flat-space string theory
\cite{Kazakov:2004qf,Beisert:2005bm}.

In this paper we will present a detailed study of the dynamics of
finite-gap solutions in integrable classical string theory.
Following \cite{Kazakov:2004qf} (see also \cite{Marsh}), 
we will focus on the case of
bosonic string theory on $\mathbb{R}\times S^{3}$. At the classical
level, this is a consistent truncation of the full string theory on
$AdS_{5}\times S^{5}$. In particular the solutions we study
correspond to strings located at the centre of $AdS_{5}$ moving on
an $S^{3}$ submanifold of $S^{5}$ (see \cite{Tseytlin} and references
therein). The methods we use should apply
equally to other cases where the string equations have a Lax
representation including the full Metsaev-Tseytlin action
\cite{Metsaev:1998it} for $AdS_{5}\times S^{5}$.

Our first result is an explicit reconstruction of the most general
$K$-gap solution\footnote{Similar solutions were obtained by a
  different method in \cite{Kazakov:2004qf, Krichever AdS}.}.
In particular, we identify the independent degrees of freedom of
the solution and construct the corresponding moduli-space ${\cal
M}^{(2g)}_{\mathbb{R}}$ as a real symplectic manifold of dimension
$2g=2K-2$. We find that the time evolution of the string admits a
Hamiltonian description as an integrable system with this moduli
space as its phase space. In fact the resulting integrable system
is one of a large class of models constructed by Krichever and
Phong \cite{Krichever+Phong}. As in other examples, the
finite-dimensional phase-space takes the form of a Jacobian
fibration over the moduli-space of the spectral curve $\Sigma$. At
generic points, the resulting system corresponds to the free
motion of an infinitely rigid, wound string on a flat torus of
dimension $K$. A similar picture emerged for the case $K\leq 3$ in
\cite{Gleb}. This construction naturally provides a set of
action-angle variables for the string. By definition the angle
variables are the flat coordinates on the Liouville torus
normalised to have period $2\pi$. Interestingly, the conjugate
action variables turn out to be exactly equal to certain {\em
filling fractions} \cite{Kazakov:2004qf,Beisert:2004ag}
which also play a key role in the context of AdS/CFT duality.
Action-angle variables for pulsating string solutions were also
discussed in \cite{KT}. 
In the rest of this introductory section we provide a brief
overview of our main results. Full details of the
calculations described are presented in the body of the paper.


\subsection{The model}

We consider a classical bosonic string moving on $\mathbb{R}\times
S^{3}$. The string carries conserved Noether charges, $Q_{L}$ and
$Q_{R}$ corresponding to the $SO(4)\simeq SU(2)_{L}\times
SU(2)_{R}$ isometry group of $S^{3}$. In a conformal, static gauge
the string is effectively described by a single worldsheet field
$g(\sigma,\tau)\in SU(2)\simeq S^{3}$ obeying the closed string
boundary condition $g(\sigma+2\pi,\tau)=g(\sigma,\tau)$.
Equivalently, we can describe the dynamics in terms of the
$SU(2)_{R}$ current $j=-g^{-1}dg$ which obeys the equations,
\begin{equation} \label{f&c}
dj - j \wedge j = 0, \quad d \ast j = 0.
\end{equation}
Physically-allowed classical motions of the string must also obey
the Virasoro constraint,
\begin{equation} \label{V}
\frac{1}{2} \text{tr} j_{\pm}^2 = -\kappa^{2}
\end{equation}
where $j_{\pm}$ are the components of the current $j$ in
worldsheet lightcone coordinates and $\kappa$ is a constant
related to the spacetime energy of the string. It will sometimes
be convenient to work with a complexified version of the model,
defined by the same equations, where the components of the current
$j$ take values in the Lie algebra $\mathfrak{sl}(2,\mathbb{C})$.
Solutions to the original problem are then obtained by applying
appropriate reality conditions. In the following, we will refer to
classical solutions in these two models as real solutions and
complex solutions respectively.

A starting point for demonstrating the integrability of the model
is the fact that the equations of motion \eqref{f&c} are
equivalent to the flatness condition for a one-parameter family of
conserved $\mathfrak{su}(2)$ currents,
\begin{equation} \label{LC}
J(x) = \frac{1}{1 - x^2} (j - x \ast j).
\end{equation}
Here $x\in \mathbb{C}$ is a complex spectral parameter. Using
these flat currents, we can construct a monodromy matrix,
\begin{equation} \label{MD}
\Omega(x,\sigma,\tau) = P \overleftarrow{\exp}
\int_{[\gamma(\sigma,\tau)]} J(x)\,\,\in\,\, SU(2)
\end{equation}
where $\gamma(\sigma,\tau)$ is a non-contractible loop on the
string worldsheet based at the point $(\sigma,\tau)$. An immediate
consequence of the flatness of the currents \eqref{LC}, is that
$\Omega(x)$ undergoes isospectral evolution in the world-sheet
coordinates. In other words the eigenvalues of the monodromy
matrix are independent of $\sigma$ and $\tau$. As $\Omega$ takes
values in $SU(2)$, it is convenient to parametrise the eigenvalues
as,
\begin{equation*}
\lambda_{\pm}=\exp\left(\pm ip(x)\right).
\end{equation*}
Here $p(x)$ is a (multi-valued) function of the spectral parameter
which is known as the {\em quasi-momentum} which yields a
one-parameter family of conserved quantities on the worldsheet.

It is a remarkable fact, familiar from the study of other
integrable PDEs \cite{Babelon, Belokolos}, that solutions to the
string equations of motion \eqref{f&c} are uniquely characterised
by the analytic behaviour of their monodromy matrix and its
eigenvalues in the spectral parameter. In particular the two
eigenvalues, $\lambda_{\pm}$, naturally correspond to the two
branches of an analytic function defined on a double-cover of the
complex $x$-plane. In general, the two sheets are joined by an
arbitrary number of branch cuts $\mathcal{C}_I$. As the
quasi-momentum appears in the exponent, it can have
discontinuities across the branch cuts of the form,
\begin{equation}\label{mode}
p(x+\epsilon) + p(x-\epsilon)=2\pi n_I, \quad x\in \mathcal{C}_I, \quad
n_I \in \mathbb{Z}, \quad I = 1,\ldots,K.
\end{equation}
More abstractly, the resulting double-cover of the $x$-plane
defines a Riemann surface $\Sigma$, known as the spectral curve,
on which $dp$ is a meromorphic differential. The integers $n_{I}$
then correspond to certain periods of the differential $dp$.

Solutions with a finite number of cuts are known as finite-gap
solutions. More specifically we will consider $K$-gap solutions
with $K$ cuts where the corresponding spectral curve, $\Sigma$,
has genus $g = K-1$. Roughly speaking, the size of each cut
provides a single continuous parameter of the solution. An
important special case occurs when all $K$ cuts shrink to zero
size. The resulting solution describes a massless point-like
string orbiting a great circle on $S^3$ at the speed of light.
Re-introducing cuts of infinitesimal size corresponds to studying
linearised oscillations around this point-like solution
\cite{Kazakov:2004qf}. More precisely, the resulting solution is a
linear superposition of $K$ eigenmodes of the small fluctuation
operator. The amplitude of each mode is set by the size of the
corresponding cut $\mathcal{C}_I$ while the corresponding mode
number is the integer $n_I$ appearing in \eqref{mode}. In fact,
allowing arbitrary $K$, the finite-gap solutions reproduce the
full space of linear fluctuations around the point-like string
\cite{WIP}. For cuts of finite size, one obtains instead a family
of fully non-linear solutions with the same number of parameters.
This suggests that the finite-gap solutions are essentially a
complete set of solutions of the equations of motion.

\subsection{The moduli space of solutions}

As mentioned above, finite-gap solutions are determined by the
analytic properties of the corresponding monodromy matrix in the
complex parameter $x$. In particular, the spectral curve
determines the spectrum of conserved quantities for a given
solution. On general grounds, for each conserved quantity, we also
expect to find a canonically conjugate variable which evolves
linearly in time. The extra information needed to specify a
particular solution are the initial values of these variables.

In the following we will identify the holomorphic data which
uniquely characterise the general finite-gap solution of
\eqref{f&c} and \eqref{V}. To simplify the problem we will focus
on `highest-weight' solutions where the Noether charges $Q_L$ and
$Q_R$ are chosen to point in a fixed direction in the internal
space,
\begin{equation*}
Q_R = \frac{1}{2i}\, R\sigma_3, \; Q_L =\frac{1}{2i}\, L\sigma_3
\end{equation*}
where $\sigma_3 = \text{diag}(1,-1)$ it the third Pauli matrix.
This choice is preserved by the corresponding Cartan subgroup of
the isometry group denoted $U(1)_{L}\times U(1)_{R}$. The
remaining generators of the isometry group rotate $Q_{L}$ and
$Q_{R}$ and sweep out the full set of solutions.

For fixed values of the constants $L$ and $R$, the full set of
data which characterise a highest-weight solution is,
\begin{equation}\label{data}
\left\{ \Sigma,\, dp,\,\gamma, \bar{\theta} \right\}.
\end{equation}
Starting from an arbitrary hyperelliptic Riemann surface $\Sigma$
and meromorphic differential $dp$, a set of necessary and
sufficient conditions for the pair $(\Sigma,dp)$ to correspond to
a finite-gap solution was given in \cite{Kazakov:2004qf}. The new
ingredients are a divisor $\gamma$ of degree $g$ on the spectral
curve $\Sigma$ and a complex ``angle'' $\bar{\theta}$ with
$W=\exp(i\bar{\theta})\in \mathbb{C}^{*}$. We find that any
finite-gap solution gives rise to a unique data set. Conversely,
given the data \eqref{data}, we can reconstruct a unique
$\mathfrak{sl}(2,\mathbb{C})$ current $j(\sigma,\tau)$ which obeys
\eqref{f&c} and \eqref{V} and satisfies closed-string boundary
conditions\footnote{More precisely the solutions we construct
satisfy all of the Virasoro constraints {\em except} the single
condition that the total worldsheet momentum should vanish. At
this stage there is also a residual redundancy  corresponding to
rigid translations of the spatial world-sheet coordinate $\sigma$.
As discussed below, the momentum constraint is imposed at the end
of the calculation.}. Explicit formulae for the resulting solution
are given in \eqref{explicit reconstruction of j}. To obtain real
solutions we must impose appropriate reality conditions on the
data which are described in sections \ref{section: reality of
curve} and \ref{section: reality of divisor}. We also discuss the
reconstruction of the original world-sheet field
$g(\sigma,\tau)\in SU(2)$.

To make the correspondence between data and solutions described
above precise, we define a {\em moduli space}
$\mathcal{M}_{\mathbb{C}}$ of holomorphic data
where each point corresponds to a distinct complex finite-gap
solution. We will show
that the reconstructed solution described above defines an injective
map,
\begin{equation} \label{GM}
\mathcal{G} : \mathcal{M}_{\mathbb{C}}
\rightarrow \mathcal{S}_{\mathbb{C}}
\end{equation}
where the target, $\mathcal{S}_{\mathbb{C}}$, is the space of
complex solutions.

The moduli space $\mathcal{M}_{\mathbb{C}}$ takes the form of a
product $\mathcal{M}^{(2g)}_{\mathbb{C}}\times \mathbb{C}^{*}$
where the $\mathbb{C}^{*}$ factor is parametrised by
$\bar{\theta}$ introduced above. The remaining factor
$\mathcal{M}^{(2g)}_{\mathbb{C}}$, which we call the {\em reduced
moduli space} is a complex manifold of dimension $2g$. We will
obtain an explicit description of the reduced moduli space as a
special case of a very general construction due to Krichever and
Phong \cite{Krichever+Phong}. The resulting space is defined as a
fibration,
\begin{equation} \label{EB}
J(\Sigma) \rightarrow \mathcal{M}_{\mathbb{C}}^{(2g)} \rightarrow
\mathcal{L}.
\end{equation}
The base ${\cal L}$ of the fibration is the moduli space of
admissible pairs $(\Sigma,dp)$ with fixed Noether charges $L$ and
$R$ and the fibre is the Jacobian torus,
\begin{equation*}
J(\Sigma) = \mathbb{C}^g / (2\pi \mathbb{Z}^g + 2\pi\Pi
\mathbb{Z}^g),
\end{equation*}
where $\Pi_{ij}$ is the period matrix of $\Sigma$. Following
\cite{Krichever+Phong}, the base ${\cal L}$ can be defined very
precisely as a
leaf in a certain smooth $g$-dimensional foliation of the universal moduli
space of all Riemann surfaces of genus $g=K-1$.

The construction of Krichever and Phong also provides a good set of
holomorphic coordinates on $\mathcal{M}_{\mathbb{C}}^{(2g)}$. To
define these coordinates, we introduce the standard basis of
one-cycles on $\Sigma$ with canonical intersections,
\begin{equation*}
a_i \cap a_j = b_i \cap b_j = 0, \; a_i \cap b_j = \delta_{ij}.
\end{equation*}
for $i=1, \ldots, g$. We also define the dual basis of holomorphic
differentials $\omega_i$, normalised as,
\begin{equation*}
\int_{a_i} \omega_j =\delta_{ij}, \quad \int_{b_i}
\omega_j=\Pi_{ij}.
\end{equation*}
With a convenient choice of normalisation, holomorphic coordinates on
the leaf $\mathcal{L}$ are given as,
\begin{equation}\label{FF}
S_i = \frac{1}{2\pi i}\, \frac{\sqrt{\lambda}}{4\pi}\,
\int_{a_i}\, \left(x+\frac{1}{x}\right)\, dp
\end{equation}
for $i=1, \ldots, g$. As we discuss in Section \ref{section:
moduli}, these integrals are essentially the filling fractions
defined in \cite{Kazakov:2004qf,Beisert:2004ag}. The divisor
$\gamma = \gamma_1 \ldots \gamma_g$ provides coordinates
$\bm{\theta} = (\theta_1, \ldots ,\theta_g)$ on the fibre
$J(\Sigma)$ via the Abel map,
\begin{equation}\label{am2}
\bm{\theta} = 2\pi \sum_{j=1}^g \int_{\infty^+}^{\gamma_j}\,
\bm{\omega} + \bm{\theta}_0
\end{equation}
where $\bm{\theta}_0 \in \mathbb{C}^g$ is a constant
vector\footnote{The precise value of $\bm{\theta}_0$
is fixed in section \ref{section: reality of
divisor} (see eqn \eqref{reality of theta})
to ensure that the coordinates $\theta_i$ real when the approriate reality
conditions are imposed on the corresponding divisor $\gamma$.}.
Together $\{S_i, \theta_i\}$ define a set of holomorphic
coordinates on the total space of the fibration \eqref{EB} and
hence $\{S_{i},\theta_{i},\bar{\theta}\}$ are holomorphic
coordinates on the moduli space $\mathcal{M}_{\mathbb{C}}$.

The reconstruction of explicit solutions reveals a remarkable
feature which is a recurrent theme in the theory of integrable
systems. The evolution of the solutions in the world-sheet
coordinates can be described very simply by promoting the
holomorphic data (i.e. the coordinates on
$\mathcal{M}_{\mathbb{C}}$) to dynamical variables depending on
$\sigma$ and $\tau$. The resulting evolution of the data
corresponds to linear motion on the fibre $J(\Sigma)$ (and on
$\mathbb{C}^{\ast}$) over a fixed point on the base $\mathcal{L}$.
In terms of the angular coordinates $\theta_i$ introduced above,
the evolution is,
\begin{equation}\label{lin}
\theta_i(\sigma,\tau) = \theta_i(0,0) - k_i \sigma - w_i \tau
\end{equation}
where the constant angular velocities are given as,
\begin{equation*}
k_i = \frac{1}{2\pi} \int_{b_i} \, dp, \quad w_i = \frac{1}{2\pi}
\int_{b_i} \, dq.
\end{equation*}
Here $dp$ is the differential of the quasi-momentum introduced
above. Correspondingly, $dq$ is the differential of the {\em
quasi-energy} which is defined in section \ref{section: spectral
curve} below. The $b$-periods $k_i$ of $dp$ are related to the
constant mode numbers $n_I$ appearing in \eqref{mode} as $k_i =
n_i - n_K$ for $i=1,\ldots, g$. In contrast, the $b$-periods $w_i$
of $dq$ are non-trivial functions of the moduli $S_i$.

The solution also involves the following linear motion of the
additional coordinate $\bar{\theta}$,
\begin{equation}\label{linbar}
\bar{\theta} = \bar{\theta}_0 - k_{\bar{\theta}} \sigma -
w_{\bar{\theta}} \tau
\end{equation}
As for the other coordinates, the
constant angular velocities $k_{\bar{\theta}}$ and
$\omega_{\bar{\theta}}$ are expressed as periods of the
differentials $dp$ and $dq$ respectively,
\begin{eqnarray}
k_{\bar{\theta}} = -\frac{1}{2 \pi} \int_{\mathcal{B}_{g+1}} dp,
& \qquad{} & w_{\bar{\theta}} = -\frac{1}{2\pi} \int_{\mathcal{B}_{g+1}}
dq \nonumber
\end{eqnarray}
where the one-cycle $\mathcal{B}_{g+1}$ is defined in the text and
$k_{\bar{\theta}}$ is equal to the mode number $n_K$.

With an appropriate choice for the constant $\bm{\theta}_0$ appearing in
\eqref{am2}, we find that
the condition for real solutions simply corresponds to restricting
the coordinates $\{S_{i},\theta_{i},\bar{\theta}\}$ on
$\mathcal{M}_{\mathbb{C}}$ to real values.
We denote the corresponding real
slice of the moduli space as $\mathcal{M}_{\mathbb{R}}=
\mathcal{M}_{\mathbb{R}}^{(2g)}\times S^1$. In particular, the
resulting real coordinates $\theta_i$ and $\bar{\theta}$ are
angular variables with period $2\pi$. The angle
$\bar{\theta}$ is a coordinate describing the orientation of the
string in the unbroken global symmetry group $U(1)_R$, while the
remaining angles correspond to internal degrees of
freedom of the string. Together, the variables $\theta_{i}$ and
$\bar{\theta}$ parametrise a real torus
$T^{g+1}\subset J(\Sigma)\times \mathbb{C}^{\ast}$ and the
linear evolution \eqref{lin} \eqref{linbar}
can be understood as the free motion of a rigid string on this torus.
The string is wound around a cycle on $T^{g+1}$ which is
determined by the mode numbers $n_I$ and moves on the torus with
constant angular velocity.



\subsection{The integrable system}

So far we have discussed the reduced moduli space
$\mathcal{M}^{(2g)}_{\mathbb{C}}$ as a complex manifold without
additional structure. However, the construction of Krichever and
Phong also provides a globally-defined holomorphic, non-degenerate
closed two-form with which $\mathcal{M}^{(2g)}_{\mathbb{C}}$ becomes
a complex symplectic manifold. Similarly the real slice
$\mathcal{M}^{(2g)}_{\mathbb{R}}$ becomes a real, symplectic
manifold. In the coordinates $\{S_{i}, \theta_{i}\}$ introduced
above, the real symplectic form is,
\begin{equation}\label{symp}
\omega_{2g} = \,\sum_{i=1}^g \, \delta S_i \wedge \delta \theta_i
\end{equation}
Here $\delta$ denotes the exterior derivative on the moduli space.
Together $\mathcal{M}^{(2g)}_{\mathbb{R}}$ and $\omega_{2g}$
define a Hamiltonian integrable system having $g$
Poisson-commuting conserved quantities $\{S_i\}$. The conjugate
angle variables $\{\theta_i\}$ evolve linearly under the flows
generated by these Hamiltonians. We will now show that this
integrable system provides an effective Hamiltonian description of
the dynamics of the string.
\paragraph{}
The gauge-fixed string action implies a non-trivial set of Poisson
brackets for the current components $j_{0}(\sigma,\tau)$ and
$j_{1}(\sigma,\tau)$ (see equations \eqref{PB} below). Using these
brackets, translations of the world-sheet coordinates $\sigma$ and
$\tau$ are generated by the momentum and energy of the 2d
principal chiral model defined as
\begin{equation*}
\begin{split}
\mathcal{P} & = -\frac{\sqrt{\lambda}}{4\pi}\, \int_{0}^{2\pi}\,
d\sigma\, {\rm tr}\left[ j_{0}j_{1}\right] \\
\mathcal{E} & = -\frac{\sqrt{\lambda}}{4\pi}\, \int_{0}^{2\pi}\,
d\sigma\, \frac{1}{2}{\rm tr}\left[ j_{0}^{2}+j_{1}^{2}\right]
\end{split}
\end{equation*}
respectively\footnote{Ultimately, the Virasoro constraints
\eqref{V} imply that $\mathcal{P}=0$. As mentioned above, we will
delay imposing this final constraint until the end of the
calculation.}.
As we have already accounted for all the moduli of the solution,
we should think of $\mathcal{P}$ and $\mathcal{E}$ as functions on
the space of solutions. Further, as these quantities are
manifestly conserved, they correspond to well-defined functions on
the base ${\cal L}$ of the fibration \eqref{EB}. We now define
corresponding Hamiltonian functions,
\begin{equation}\label{Ham}
\begin{split}
H_{\sigma} & = \mathcal{P}[S_1,\ldots, S_g]\\
H_{\tau} & = \mathcal{E}[S_1,\ldots, S_g]
\end{split}
\end{equation}
for the integrable system defined by the manifold
$\mathcal{M}^{(2g)}_{\mathbb{R}}$ equipped with the symplectic
form $\omega_{2g}$ given in \eqref{symp} above. Our main result is
that the linear flow on $\mathcal{M}^{(2g)}_{\mathbb{R}}$
generated by $H_{\sigma}$ and $H_{\tau}$, with parameters $\sigma$
and $\tau$ is precisely the same as the linear evolution
\eqref{lin} defined by the finite-gap solution.

It is striking that the Hamiltonians \eqref{Ham} reproduce the
correct evolution precisely when they act on the holomorphic data
with the Poisson brackets defined by the natural symplectic form
$\omega_{2g}$ defined in \cite{Krichever+Phong}. Of course,
imposing the correct evolution under two Hamiltonians does not
uniquely fix the symplectic form and it would be preferable to
derive the symplectic structure on the moduli space of finite-gap
solutions directly from the string action. In a forthcoming paper
\cite{BVII} we will show that the symplectic form $\omega_{2g}$ on
$\mathcal{M}^{(2g)}_{\mathbb{R}}$ is indeed a consequence of the
Poisson brackets implied by the string action. The latter define a
symplectic form $\omega_{\infty}$ on the full infinite-dimensional
phase space of the string. The form $\omega_{2g}$ then arises
naturally when $\omega_{\infty}$ is restricted to the space of
$K$-gap solutions. Similar results exist for several other
integrable non-linear PDEs including the equations of the KP and
KdV hierarchies and the Toda equations \cite{Babelon,
Krichever:1997sq, Krichever+Phong}. The present example is more
complicated as the infinite-dimensional symplectic structure is
non-ultra local, leading to ambiguities in the brackets of the
monodromy matrices which require regularisation. This problem was
addressed in a series of papers by Maillet \cite{Maillet}. In
\cite{BVII} we will show that the symmetric regularisation
prescription advocated in these papers, when applied to the
present model, leads directly to the symplectic form defined in
\eqref{symp} above.

Another striking feature is that the
correctly-normalised\footnote{By this we mean that the canonically
conjugate angle variables are normalised to have period $2\pi$.}
action variables of the integrable system are the filling
fractions defined in \eqref{FF} above. In a leading-order
semiclassical quantisation of the integrable system, the
Bohr-Sommerfeld condition dictates that these action variables are
quantised in integer units. As mentioned above, the $K$-gap
solutions discussed here correspond to motions of a classical
string on an $\mathbb{R}\times S^{3}$ submanifold of
$AdS_{5}\times S^{5}$. A leading-order semiclassical quantisation
of these solutions, valid for $\lambda \gg 1$, would also involve
applying the Bohr-Sommerfeld quantisation condition on periodic
orbits of the string \cite{DHN}. This should coincide with the
semiclassical quantisation of the integrable system described
above\footnote{Note that the familiar sum over the zero point
energies of small fluctuations around a classical solution appears
at the {\em next} order in the semiclassical expansion. Obviously
this includes fluctuations of the target space coordinates outside
the $\mathbb{R}\times S^3$ submanifold and would not be captured
by a quantisation of the integrable system proposed here.} and
therefore lead to a discrete spectrum of string states where the
filling fractions are quantised in integer units. This would be
very natural in the context of the AdS/CFT correspondence where
the filling fractions correspond to the number of Bethe roots of
each mode number in the dual spin chain description.

As it stands the Hamiltonian description of the dynamics of the
data given above is not quite complete. In particular, we have
only discussed the $(\sigma,\tau)$-evolution of the data on the
reduced moduli space $\mathcal{M}^{(2g)}_{\mathbb{R}}$ and not on
the full moduli space $\mathcal{M}_{\mathbb{R}}$. The difference
is the additional $S^{1}$ parametrised by the angle $\bar{\theta}$
whose linear evolution is given in \eqref{linbar}. As mentioned
above, the evolution of $\bar{\theta}$ corresponds to a global
$U(1)_R$ rotation of the string with constant angular velocity. As
in the study of rotating rigid bodies, we can remove this motion
by working in an appropriate reference frame which rotates with
the string. As we discuss in subsection \ref{section:
Hamiltonian}, the motion on the reduced moduli space
$\mathcal{M}^{(2g)}_{\mathbb{R}}$ naturally corresponds to the
motion of the string in this rotating frame.



\section{The $\sigma$-model on $\mathbb{R} \times S^3$}
\label{section: sigma model}

The action for a string on $\mathbb{R} \times S^3$ in conformal
gauge reads
\begin{equation*}
S = \frac{\sqrt{\lambda}}{4 \pi} \int d\sigma d\tau \left[ \sum_i
\partial_a X_i \partial^a X_i - \partial_a X_0
\partial^a X_0 + \Lambda \left( \sum_i X_i^2 - 1 \right)\right]
\end{equation*}
where $X_0$ is the time coordinate, $X_i$ are Cartesian
coordinates on $\mathbb{R}^4$ and $\Lambda$ is a Lagrange
multiplier constraining the string to the unit sphere $S^3 \subset
\mathbb{R}^4$. In terms of worldsheet lightcone coordinates
\begin{equation} \label{LC coords}
\sigma^{\pm} = \frac{1}{2}(\tau \pm \sigma) = \frac{1}{2}(\sigma^0
\pm \sigma^1), \quad \partial_{\pm} = \partial_0 \pm \partial_1,
\end{equation}
the equations of motion derived from this action read
\begin{equation*}
\partial_+ \partial_- X_i + \left( \sum_j \partial_+ X_j \partial_-
X_j \right) X_i = 0, \quad \partial_+ \partial_- X_0 = 0,
\end{equation*}
which must be supplemented by the Virasoro constraints
\begin{equation*}
\sum_i (\partial_{\pm} X_i)^2 = (\partial_{\pm} X_0)^2.
\end{equation*}
The equation for $X_0$ being decoupled from the other fields it
can be solved separately and has the general solution $X_0 =
\kappa \tau + f_+(\sigma^+) + f_-(\sigma^-)$. Using the residual
gauge symmetry $\sigma^{\pm} \rightarrow F_{\pm}(\sigma^{\pm})$
one can gauge away $f_{\pm}$, leaving $X_0 = \kappa \tau$. The
Virasoro constraint becomes
\begin{equation*}
\sum_i (\partial_{\pm} X_i)^2 = \kappa^{2}.
\end{equation*}
In this
gauge the spacetime energy of the string is given as,
\begin{equation*}
\Delta=\frac{\sqrt{\lambda}}{2\pi}\int_{0}^{2\pi}\, d\sigma\,
\partial_{\tau}X_{0}=\sqrt{\lambda}\kappa
\end{equation*}
and we therefore identify the constant $\kappa$ as
$\Delta/\sqrt{\lambda}$.

\subsection{$SU(2)$ principal chiral model}

Since the sphere $S^3$ is the group $SU(2)$, the $\sigma$-model in
question can be reformulated as an $SU(2)$ principal chiral model
for a field $g$ taking values in $SU(2)$ by defining
\begin{equation} \label{SU(2) sigma-model field}
g = \left( \begin{array}{cc} X_1 + i X_2 & X_3 + i X_4 \\ -X_3 + i
X_4 & X_1 - i X_2 \end{array}\right) \in SU(2).
\end{equation}
In terms of the current\footnote{Note the sign difference in the
definition of $j$ with \cite{Kazakov:2004qf}. The choice of sign
here is to agree with the literature on algebro-geometric methods
and finite-gap integration \cite{Babelon, Krichever1, Krichever2,
Belokolos}.} $j = -g^{-1} dg$, the action may be rewritten as
follows
\begin{equation} \label{action}
S = - \frac{\sqrt{\lambda}}{4 \pi} \int \left[ \frac{1}{2}
\text{tr}(j \wedge \ast j) + dX_0 \wedge \ast dX_0 \right].
\end{equation}
The current $j$ is identically flat from its definition and the
equations of motion are equivalent to its conservation
\begin{equation} \label{flatness & conservation}
dj - j \wedge j = 0, \quad d \ast j = 0.
\end{equation}

In static gauge, $X_{0}=\kappa\tau$, the conserved Noether charges
associated with translations of the world-sheet coordinates $\sigma$
and $\tau$ are $\mathcal{P}$ and
$\mathcal{E}-\sqrt{\lambda}\kappa^{2}/2$ where,
\begin{equation} \label{defem}
\begin{split}
\mathcal{P} & = -\frac{\sqrt{\lambda}}{4\pi}\, \int_0^{2\pi}\,
d\sigma\, {\rm tr}\left[ j_0 j_1 \right] \\
\mathcal{E} & = -\frac{\sqrt{\lambda}}{4\pi}\, \int_0^{2\pi}\,
d\sigma\, \frac{1}{2}{\rm tr}\left[ j_0^2 + j_1^2\right]
\end{split}
\end{equation}
are the two-dimensional momentum and energy of the principal
chiral model respectively.

In terms of the lightcone components $j_{\pm} = j_0 \pm j_1$ of
the current $j$, the Virasoro constraints read,
\begin{equation} \label{Virasoro}
\frac{1}{2} \text{tr} j_{\pm}^2 = -\kappa^2.
\end{equation}
For later convenience, it will be useful to split these constraints
into two parts. The first set of constraints read,
\begin{equation} \label{V1}
\frac{1}{2} \text{tr} j_{\pm}^2 = -\kappa_{\pm}^2.
\end{equation}
where, for the moment, $\kappa_{\pm}$ are undetermined constants.
The world-sheet momentum and energy defined above then become,
\begin{equation} \label{EP1}
\begin{split}
\mathcal{P} & = \frac{\sqrt{\lambda}}{4}(\kappa^2_+ - \kappa^2_-),\\
\mathcal{E} & = \frac{\sqrt{\lambda}}{4}(\kappa^2_+ + \kappa^2_-).
\end{split}
\end{equation}
The remaining content of the Virasoro constraint \eqref{Virasoro} is
the vanishing of the total momentum $\mathcal{P} = 0$ which implies
$\kappa_+^2 = \kappa_-^2 = \kappa^2$ and the string mass-shell
condition,
\begin{equation}\label{massshell}
\mathcal{E}=\frac{\sqrt{\lambda}}{2}\kappa^2 =
\frac{\Delta^2}{2\sqrt{\lambda}}.
\end{equation}

It will also be useful to consider the Hamiltonian description of
string dynamics in static gauge. The string action \eqref{action}
leads to a set of non-trivial equal-$\tau$ Poisson brackets for
the current components, $j_0(\sigma)$ and $j_1(\sigma)$. Writing
the currents as $j_0 = j_0^a t^a$, $j_1 = j_1^a t^a$ in terms of
$SU(2)$ generators $t^a$ satisfying,
\begin{equation*}
[t^a,t^b] = f^{abc} t^c, \quad \text{tr}(t^a t^b) = - \delta^{ab}.
\end{equation*}
the Poisson brackets are given as,
\begin{equation} \label{PB}
\begin{split}
\left\{ j_1^a(\sigma), j_1^b(\sigma')\right\} &=  0, \\
\frac{\sqrt{\lambda}}{4 \pi} \left\{ j_0^a(\sigma),
j_1^b(\sigma')\right\} &=  - f^{abc} j_1^c(\sigma) \delta(\sigma -
\sigma') - \delta^{ab}
\delta'(\sigma - \sigma'), \\
\frac{\sqrt{\lambda}}{4 \pi} \left\{ j_0^a(\sigma),
j_0^b(\sigma')\right\} &= - f^{abc} j_0^c(\sigma) \delta(\sigma -
\sigma').
\end{split}
\end{equation}
The resulting symplectic structure is non-ultra-local reflecting the
presence of the term proportional to $\delta'(\sigma-\sigma')$ in
the second bracket. This term leads to some ambiguities in the
Poisson brackets of the monodromy matrix of the model. However, the
brackets lead (unambiguously) to a Hamiltonian version of the
equations of motion \eqref{flatness & conservation} which read,
\begin{equation}
\partial_{\sigma} j = \{\mathcal{P}, j\}, \quad \partial_{\tau} j =
\{\mathcal{E}, j\}
\end{equation}
where $\mathcal{P}$ and $\mathcal{E}$ are the world-sheet momentum
and energy defined above. We will not make any further use of these
Poisson brackets in this paper other than to note  the fact that the
$\sigma$ and $\tau$ evolution of the current $j$ are generated by
$\mathcal{P}$ and $\mathcal{E}$ respectively. A more complete
discussion of the Hamiltonian formalism for this model including the
issues raised by non-ultra-locality will appear in a forthcoming
paper \cite{BVII}.

\subsection{Symmetries} \label{section: symmetries}

\begin{figure}
\centering \psfrag{g1}{\footnotesize $\gamma_1$}
\psfrag{g2}{\footnotesize $\gamma_2$} \psfrag{D}{$\Bigg\} \footnotesize
D$}
\includegraphics[height=40mm,width=32mm]{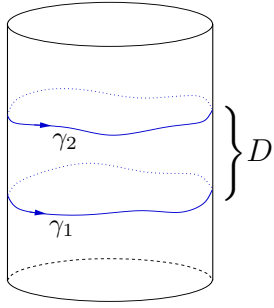}
\caption{Conservation of $Q_R$ and $Q_L$.} \label{Noether charge figure}
\end{figure}

The action \eqref{action} has a global $SU(2)_L \times SU(2)_R$
symmetry
\begin{equation*}
g \rightarrow U_L g U_R,
\end{equation*}
where $U_L$ and $U_R$ are constant matrices. The Noether current
corresponding to $SU(2)_R$ is the current $j = -g^{-1} dg$ introduced
above
whereas the
Noether current for the $SU(2)_L$ symmetry is $l = - dg \, g^{-1} = g
j g^{-1}$. The corresponding Noether charges are
\begin{align*}
&SU(2)_R : \quad Q_R = \frac{\sqrt{\lambda}}{4 \pi} \int_{\gamma} \ast
  j = - \frac{\sqrt{\lambda}}{4 \pi} \int_0^{2 \pi} d\sigma j_0, \\
&SU(2)_L : \quad Q_L = \frac{\sqrt{\lambda}}{4 \pi} \int_{\gamma} \ast
  l = - \frac{\sqrt{\lambda}}{4 \pi} \int_0^{2 \pi} d\sigma l_0,
\end{align*}
where $\gamma$ is any curve winding once around the world-sheet,
expressing the conservation of these Noether charges (see Figure
\ref{Noether charge figure}), e.g.
\begin{equation*}
\int_{\gamma_2} \ast j - \int_{\gamma_1} \ast j = \int_{\partial D}
\ast j = \int_D d \ast j = 0.
\end{equation*}
Notice that the $SU(2)_R$ current $j$ which appears in the action
\eqref{action} is invariant under the action of $SU(2)_L$.
On the other hand the $SU(2)_R$ symmetry acts non-trivially on the current
as
\begin{equation} \label{R symmetry on j}
j \rightarrow U_R^{-1} j U_R.
\end{equation}

\section{Classical integrability} \label{section: classical integrability}

The equations \eqref{flatness & conservation} of the $SU(2)$
principal chiral model can be reformulated as a $1$-parameter
family of zero-curvature equations which is one of the most
general representations known to give rise to classical integrable
systems and constitutes the starting point for constructing
solutions by algebro-geometric methods \cite{Babelon, Krichever1,
Krichever2, Belokolos}.

\subsection{Lax connection and Monodromy} \label{section: Lax & monodromy}

We search for the $1$-parameter family of flat currents in the
form $J(x) = \alpha(x) j + \beta(x) \ast j$. Using \eqref{flatness
& conservation} we find $dJ(x) - J(x) \wedge J(x) = -(\alpha(x)^2
- \alpha(x) - \beta(x)^2) j \wedge j$ so that $J(x)$ is flat
provided $\alpha(x)^2 - \alpha(x) - \beta(x)^2 = 0$ which is
solved by $\alpha(x) = \frac{1}{1 - x^2} ,\; \beta(x) =
\frac{-x}{1 - x^2}$. Therefore
\begin{equation} \label{Lax connection}
J(x) = \frac{1}{1 - x^2} (j - x \ast j),
\end{equation}
or in components $J(x) = J_0 d\tau + J_1 d\sigma$ (with the
convention $\epsilon_{01} = 1$, $(\ast j)_{\mu} = \epsilon_{\nu
\mu}j^{\nu}$)
\begin{align*}
J_0(x,\sigma,\tau) &= \frac{1}{2} \left(\frac{j_+}{1-x} + \frac{j_-}{1+x}\right)\\
J_1(x,\sigma,\tau) &= \frac{1}{2} \left(\frac{j_+}{1-x} -
\frac{j_-}{1+x}\right).
\end{align*}

\begin{figure}
\centering \psfrag{g}{\footnotesize $\tilde{\gamma}$}
\psfrag{(s')}{\footnotesize $(\sigma',\tau')$}
\psfrag{(s)}{\footnotesize $(\sigma,\tau)$}
\psfrag{g1}{\footnotesize
  $\gamma_{(\sigma',\tau')}$} \psfrag{g2}{\footnotesize $\gamma_{(\sigma,\tau)}$}
\includegraphics[height=40mm,width=37mm]{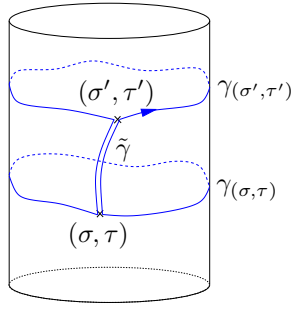}
\caption{$(\sigma,\tau)$-evolution of the monodromy matrix
$\Omega(x,\sigma,\tau)$.} \label{Monodromy figure}
\end{figure}

By construction, the equation
\begin{equation} \label{zero curvature equation}
d J(x,\sigma,\tau) - J(x,\sigma,\tau) \wedge J(x,\sigma,\tau) = 0
\end{equation}
holds identically in $x$ and is equivalent to the pair of
equations in \eqref{flatness & conservation}. Given such a family
of flat connections $J(x,\sigma,\tau)$ on the world-sheet it is
natural to consider the corresponding parallel transporters along
any given curve $\gamma$. By virtue of the flatness condition
\eqref{zero curvature equation} the non-Abelian version of
Stokes's theorem implies that for any simply connected domain $D$
(so that $\partial D$ is a single closed curve) one has
\begin{equation} \label{non-Abelian Stokes}
P \overleftarrow{\exp} \left\{ \int_{\partial D} J(x,\sigma,\tau)
\right\} = {\bf 1}.
\end{equation}
It follows that the parallel transporter along any given curve
$\gamma$ will only depend on the homotopy class $[\gamma]$ of
$\gamma$ with fixed end points. Of particular interest is the
monodromy matrix defined as the parallel transporter along a
closed loop $\gamma(\sigma,\tau)$ bound at $(\sigma,\tau)$ and
winding once around the world-sheet
\begin{equation} \label{monodromy definition}
\Omega(x,\sigma,\tau) = P \overleftarrow{\exp}
\int_{[\gamma(\sigma,\tau)]} J(x,\sigma,\tau) = P
\overleftarrow{\exp} \int_{\sigma}^{\sigma + 2 \pi} d\sigma'
\frac{1}{2} \left(\frac{j_+(\sigma',\tau)}{1-x} -
\frac{j_-(\sigma',\tau)}{1+x}\right).
\end{equation}
Equation \eqref{non-Abelian Stokes} implies that monodromy
matrices based at different points $(\sigma,\tau)$ and
$(\sigma',\tau')$ are related by conjugation (see Figure \ref{Monodromy figure})
\begin{equation} \label{monodromy conjugation}
\Omega(x,\sigma',\tau') = U \Omega(x,\sigma,\tau) U^{-1}, \quad
\text{with } U = P \overleftarrow{\exp} \int_{\widetilde{\gamma}}
J(x,\sigma,\tau),
\end{equation}
where $\widetilde{\gamma}$ is any curve connecting the two points
$(\sigma,\tau)$ and $(\sigma',\tau')$ in question.

\subsection{Analyticity}

Since the dependence of $\Omega(x,\sigma,\tau)$ on $(\sigma,\tau)$
is governed by
\begin{equation} \label{monodromy evolution}
\left[ d - J(x,\sigma,\tau), \Omega(x,\sigma,\tau) \right] = 0,
\end{equation}
it follows from Poincar\'e's theorem on holomorphic differential
equations that $\Omega(x,\sigma,\tau)$ is holomorphic in $x$
except at the singular points $x=\pm 1$ of the Lax connection
$J(x,\sigma,\tau)$ where it has essential singularities. Also,
since $\text{tr} j_{\pm} = 0$ it follows that the monodromy matrix
is unimodular, i.e. $\text{det } \Omega(x,\sigma,\tau) = 1$. So at
generic values of $x$ where it can be diagonalised it takes on the
following diagonal form
\begin{equation} \label{monodromy diagonalised}
u(x,\sigma,\tau) \Omega(x,\sigma,\tau) u(x,\sigma,\tau)^{-1} =
\text{diag}\left( e^{i p(x)}, e^{-i p(x)} \right),
\end{equation}
where the \textit{quasi-momentum} $p(x)$ has simple poles at $x=\pm 1$. But
even though $\Omega(x,\sigma,\tau)$ is meromorphic in $x \in
\mathbb{CP}^1 \setminus \{ x = \pm 1 \}$, the quasi-momentum
$p(x)$ is only well defined on the $x$-plane with branch cuts
\cite{Beisert:2004ag, Beisert:2005bm} since going around a simple
zero of $\Delta(x) = \left(e^{i p(x)} - e^{-i p(x)}\right)^2$ is
easily seen to interchange $e^{i p(x)}$ with $e^{-i p(x)}$.

The asymptotic behaviour of the monodromy matrix near $x = \pm 1$
follows directly from \eqref{monodromy definition}
\begin{equation*}
\Omega(x,\sigma,\tau) = P
\overleftarrow{\exp} \int_{\sigma}^{\sigma + 2 \pi} d\sigma'
\left[ - \frac{1}{2} \frac{j_{\pm}(\sigma',\tau)}{x \mp 1} +
O\left( (x \mp 1)^0 \right) \right], \quad \text{as } x \rightarrow \pm 1
\end{equation*}
It is diagonalised to leading order in $(x \mp 1)$ by the
eigenvectors of $j_{\pm}$ using the relation $u(\sigma,\tau)
j_{\pm}(\sigma,\tau) u(\sigma,\tau)^{-1} = i\kappa \sigma_3$,
where $\sigma_3 = \text{diag}(1,-1)$ is the third Pauli matrix.
Subleading orders are diagonalised by adding to $u(\sigma,\tau)$
higher orders in $(x \mp 1)$ to obtain a matrix $u(x,\sigma,\tau)
= u(\sigma,\tau) + O(x \mp 1)$ analytic in $x$ such that
\begin{equation*}
u(x,\sigma,\tau) \Omega(x,\sigma,\tau) u(x,\sigma,\tau)^{-1}
= \exp \left[ - \frac{i \pi \kappa}{x \mp 1} \sigma_3 + O\left( (x \mp
  1)^0 \right) \right], \quad \text{as } x \rightarrow \pm 1.
\end{equation*}
In the definition \eqref{monodromy diagonalised} of the
quasi-momentum $p(x)$ there is an ambiguity coming from the
freedom of swapping the two sheets. We fix this ambiguity by
choosing the asymptotic behaviour of $p(x)$ at $x = \pm 1$ to be
\begin{equation} \label{p asymptotics at pm 1}
p(x) = - \frac{\pi \kappa}{x \mp 1} + O\left( (x \mp 1)^0 \right),
\quad \text{as } x \rightarrow \pm 1.
\end{equation}
The other possible choice for the asymptotics of the quasi-momentum
near $x = \pm 1$ would be to take opposite relative signs for the pole
terms at $x = 1$ and $x = -1$, namely to replace $p(x)$ by the
function $q(x)$ with the following asymptotics
\begin{equation} \label{q asymptotics at pm 1}
q(x) = \mp \frac{\pi \kappa}{x \mp 1} + O\left( (x \mp 1)^0 \right),
\quad \text{as } x \rightarrow \pm 1.
\end{equation}
This function will play an important role in the analysis with
\eqref{p asymptotics at pm 1} as quasi-momentum and so we will refer
to it as the \textit{quasi-energy}.

\subsection{Asymptotics}

The asymptotic expansion of the connection \eqref{Lax connection}
at $x = \infty$
\begin{equation} \label{connection asymptotics}
J(x) = \frac{1}{x} \ast j + O\left( \frac{1}{x^2} \right),
\end{equation}
leads to the following asymptotic expansion of the monodromy
matrix at $x = \infty$
\begin{equation} \label{monodromy asymptotics at infty}
\begin{split}
\Omega(x,\sigma,\tau) &= P \overleftarrow{\exp}
\int_{[\gamma(\sigma,\tau)]} \left( \frac{1}{x}
\ast j + O\left( \frac{1}{x^2} \right) \right) \\
&= {\bf 1} - \frac{1}{x} \int_{\sigma}^{\sigma + 2 \pi} d\sigma'
j_0 + O\left( \frac{1}{x^2} \right), \quad \text{as } x
\rightarrow \infty \\
&= {\bf 1} + \frac{1}{x} \frac{4 \pi Q_R}{\sqrt{\lambda}} +
O\left( \frac{1}{x^2} \right),
\quad \text{as } x \rightarrow \infty.
\end{split}
\end{equation}
Since the Noether charges $Q_R$ and $Q_L$ are conserved classically,
we may fix them to lie in a particular direction of $\mathfrak{su}(2)$
and take them for example to be proportional to the third Pauli matrix
$\sigma_3 = \text{diag}(1,-1)$
\begin{equation*}
Q_R = \frac{1}{2i}\, R\sigma_3, \; Q_L =\frac{1}{2i}\, L\sigma_3 ,
\quad R,L \in \mathbb{R}_+.
\end{equation*}
where $R$ and $L$ are constants of the motion. By restricting the
Noether charges in this way we will only obtain the restricted set
of `highest weight' solutions to the equations of motion, but all
other solutions can be obtained from these by applying a
combination of $SU(2)_R$ and $SU(2)_L$ to such a `highest weight'
solution. With this restriction the asymptotic expansion of the
monodromy matrix at $x = \infty$ reduces to
\begin{equation} \label{monodromy asymptotics at infty 2}
\Omega(x,\sigma,\tau) = {\bf 1} - \frac{1}{x} \frac{2 \pi i
R}{\sqrt{\lambda}} \sigma_3 + O\left( \frac{1}{x^2} \right), \quad
\text{as } x \rightarrow \infty.
\end{equation}
We have the freedom of choosing the branch of the logarithm in the
definition of the quasi-momentum and we do so such that $p(x) \sim
O(1/x)$ as $x \rightarrow \infty$, in other words
\begin{equation} \label{p asymptotics at infty}
p(x) = -\frac{1}{x} \frac{2 \pi R}{\sqrt{\lambda}} + O\left(
\frac{1}{x^2} \right), \quad \text{as } x \rightarrow \infty.
\end{equation}

The asymptotics of the connection at $x = 0$ is $J(x) = j - x \ast
j + O\left( x^2 \right)$, so that
\begin{align*}
d - J(x) &= d - j + x \ast j + O\left( x^2 \right),\\
&= g^{-1} \left( d + x \ast l + O\left( x^2 \right) \right) g,
\end{align*}
where $l = - dg \, g^{-1} = g j g^{-1}$. Now because the field
$g(\sigma,\tau)$ is periodic in $\sigma$ it follows that the
asymptotic expansion of the monodromy matrix near $x = 0$ is given
by
\begin{equation} \label{monodromy asymptotics at zero}
\begin{split}
g(\sigma,\tau) \Omega(x,\sigma,\tau) g^{-1}(\sigma,\tau) &= P
\overleftarrow{\exp} \left( \int_{[\gamma(\sigma,\tau)]} - x
\ast l + O\left( x^2 \right) \right),\\
&= {\bf 1} + x \int_{\sigma}^{\sigma + 2 \pi} d\sigma' l_0 +
O\left( x^2 \right), \quad \text{as } x \rightarrow 0 \\
&= {\bf 1} - x \frac{4 \pi Q_L}{\sqrt{\lambda}} + O\left( x^2 \right),
\quad \text{as } x \rightarrow 0. \\
&= {\bf 1} + x \frac{2 \pi i L}{\sqrt{\lambda}} \sigma_3 + O\left( x^2 \right),
\quad \text{as } x \rightarrow 0,
\end{split}
\end{equation}
which implies the following asymptotic expansion of the quasi-momentum
at $x = 0$
\begin{equation} \label{p asymptotics at zero}
p(x) = 2 \pi m +x \frac{2 \pi L}{\sqrt{\lambda}} + O\left( x^2
\right), \quad \text{as } x \rightarrow 0,
\end{equation}
where $m \in \mathbb{Z}$. The integer $m$ corresponds to the winding
number of the string around an equator of $S^{3}$. Although
$\pi_{1}(S^{3})=0$, a non-trivial winding number emerges when we
restrict our attention to highest weight solutions as described in the
previous subsection. For convenience we will focus on the sector with
$m=0$, which includes the pointlike string solution. Our results can
easily be generalised to the other sectors.

\subsection{Reality conditions}

The monodromy matrix also satisfies a reality condition coming from
the property $j_{\pm}^{\dag} = - j_{\pm}$ of the matrices $j_{\pm}
\in \mathfrak{su}(2)$, namely
\begin{equation} \label{monodromy reality condition}
\Omega(x,\sigma,\tau)^{\dag} = \Omega(\bar{x},\sigma,\tau)^{-1},
\end{equation}
so in particular, for $x \in \mathbb{R}$ we have
$\Omega(x,\sigma,\tau) \in SU(2)$. Since the eigenvalues of
$\Omega(x,\sigma,\tau)$ are $\{ e^{i p(x)}, e^{-i p(x)} \}$ it
follows that for $x \in \mathbb{R}$ in the cut plane, where $p(x)$
is well defined, we have $p(x) \in \mathbb{R}$. Using
\eqref{monodromy reality condition} the eigenvalues $\{ e^{i
p(\bar{x})}, e^{-i p(\bar{x})} \}$ of $\Omega(\bar{x},\sigma,\tau)$
can be also written as $\{ e^{i \overline{p(x)}}, e^{-i
\overline{p(x)}} \}$, but by the previous argument $p(\bar{x}) =
\overline{p(x)}$ when $x \in \mathbb{R}$ in the cut plane and so by
continuity $e^{i p(\bar{x})} = e^{i \overline{p(x)}}$ for all $x$ in
the cut plane. In other words $p(\bar{x}) = \overline{p(x)} + 2 \pi
k, \; k \in \mathbb{Z}$, but then $p(x) = \overline{p(\bar{x})} + 2
\pi k = p(x) + 4 \pi k$ so that $k=0$. Hence for all $x$ where
$p(x)$ is defined it satisfies the reality condition
\begin{equation} \label{reality of p}
\overline{p(x)} = p(\bar{x}).
\end{equation}
Thus, in particular, $p(x)$ is real on the real axis.

\subsection{Gauge redundancy}

The zero-curvature equation \eqref{zero curvature equation} is
invariant under gauge transformations
\begin{equation} \label{gauge redundancy}
J(x,\sigma,\tau) \rightarrow \tilde{g} J(x,\sigma,\tau)
\tilde{g}^{-1} + d\tilde{g} \tilde{g}^{-1},
\end{equation}
where $\tilde{g} = \tilde{g}(\sigma,\tau)$ is periodic in $\sigma$
and under which $\Omega(x,\sigma,\tau) \rightarrow \tilde{g}
\Omega(x,\sigma,\tau) \tilde{g}^{-1}$. The gauge transformation
parameter $\tilde{g}$ must be independent of $x$ so as to preserve
the analytical properties in $x$ of the connection
$J(x,\sigma,\tau)$ and of $\Omega(x,\sigma,\tau)$ (and thus of the
spectral curve, introduced in section \ref{section: spectral
curve}). Preserving also the reality condition \eqref{monodromy
reality condition} on $\Omega(x,\sigma,\tau)$ coming from
$j_{\pm}^{\dag} = - j_{\pm}$ leads to the further restriction
$\tilde{g}(\sigma,\tau) \in SU(2)$. By equation \eqref{monodromy
asymptotics at infty 2} the matrix of eigenvectors
$u^{-1}(x,\sigma,\tau)$ of $\Omega(x,\sigma,\tau)$ introduced in
\eqref{monodromy diagonalised} is diagonal above $x = \infty$,
\begin{equation*}
u^{-1}(x,\sigma,\tau) = {\bf 1} + O\left( \frac{1}{x} \right) \quad
\text{as } x \rightarrow \infty,
\end{equation*}
and in order to preserve this condition one has to further restrict
$\tilde{g}(\sigma,\tau) \in SU(2)$ to be diagonal. In other words,
the level set $Q_R = \frac{1}{2 i} R \sigma_3$ is preserved only by
a residual $U(1)_R$ action of the full $SU(2)_R$ gauge group.

\section{The spectral curve} \label{section: spectral curve}

Monodromy matrices at different points $(\sigma,\tau)$ are related
by similarity transformations \eqref{monodromy conjugation} and so
the eigenvalues $\{ e^{i p(x)}, e^{-i p(x)} \}$ of
$\Omega(x,\sigma,\tau)$ are independent of $(\sigma,\tau)$. This
motivates the definition of an invariant curve in $\mathbb{C}^2$
as the characteristic equation of $\Omega(x,\sigma,\tau)$, namely
\begin{equation} \label{spectral curve}
\Gamma : \quad \Gamma(x,y) = \text{det}(y \mathbf{1} -
\Omega(x,\sigma,\tau)) = 0,
\end{equation}
which is independent of $(\sigma, \tau)$. In the present case $\Gamma$
is represented as a 2-sheeted ramified cover of the $x$-plane and as
such it has a natural hyperelliptic holomorphic involution which
exchanges the two sheets
\begin{equation} \label{holomorphic involution}
\hat{\sigma} : \Gamma \rightarrow \Gamma, (x,y) \mapsto
(x,y^{-1}).
\end{equation}

\subsection{The algebraic curve} \label{section: algebraic curve}

The problem with the curve in \eqref{spectral curve} is that it
doesn't define an algebraic curve in $\mathbb{C}^2$ since the
eigenvalues $\{ e^{i p(x)}, e^{-i p(x)} \}$ have essential
singularities at $x = \pm 1$; so in particular the curve $\Gamma$
has infinitely many singular points $\{ x_k \}$ at which the two
sheets meet $e^{i p(x_k)} = e^{-i p(x_k)}$ and which accumulate at
$x = \pm 1$ (since the function $e^{2 i p(x)}$ must take the value
1 infinitely many times in any neighbourhood of $x = \pm 1$). This
can however be remedied \cite{Beisert:2004ag} by defining a new
matrix $L(x,\sigma,\tau)$ from $\Omega(x,\sigma,\tau)$
\begin{equation*}
u(x,\sigma,\tau) L(x,\sigma,\tau) u(x,\sigma,\tau)^{-1} = -i \frac{\partial}{\partial x}
\text{log}\left( u(x,\sigma,\tau) \Omega(x,\sigma,\tau) u(x,\sigma,\tau)^{-1}\right),
\end{equation*}
which has the same eigenvectors as $\Omega(x,\sigma,\tau)$ but the
corresponding eigenvalues $\{ p'(x), -p'(x) \}$ are meromorphic in
$x$ with only double poles at $x = \pm 1$. Using this new matrix
$L(x,\sigma,\tau)$ we can now define an algebraic curve in
$\mathbb{C}^2$ by
\begin{equation} \label{pre-desingularised curve}
\widehat{\Sigma} : \quad \widehat{\Sigma}(x,y) = \text{det}(y
\mathbf{1} - L(x,\sigma,\tau)) = 0.
\end{equation}
A complete discussion of the properties of this algebraic curve
for strings on $\mathbb{R} \times S^p$ can be found in
\cite{Beisert:2004ag} and those of the full
algebraic curve for superstrings on $AdS_5 \times S^5$ in
\cite{Beisert:2005bm}; see also \cite{Kazakov:2004nh,
Schafer-Nameki:2004ik, Alday:2005gi}. We show in Appendix \ref{section:
desingularised curve} that by performing a birational transformation
on the variable $y$ to remove the (finitely many) singular points it
can be turned into the following hyperelliptic form
\begin{equation} \label{desingularised curve}
\Sigma : y^2 = \prod_{I=1}^K (x - u_I)(x - v_I),
\end{equation}
where $K$ is the number of branch cuts $\mathcal{C}_I$
which connect the branch points $u_I,v_I, I = 1, \ldots,
K$. A solution for which the corresponding curve has finitely many
branch points and hence finite genus $g = K - 1$ is called a
\textit{finite-gap solution}. Let $\hat{\pi} : \Sigma \rightarrow
\mathbb{CP}^1$ denote the hyperelliptic projection of $\Sigma$ onto
the $x$-plane. We shall refer to the sheet of this double cover
corresponding to $p(x)$ as the physical sheet, and for any value $x
\in \mathbb{CP}^1$ we shall denote the pair of points in
$\hat{\pi}^{-1}(x)$ as $x^{\pm}$ (related to each other by $x^+ =
\hat{\sigma}(x^-)$), with $x^+$ being the point on the physical
sheet.

\begin{figure}
\centering \psfrag{a}{$\mathcal{A}_I$} \psfrag{b}{$\mathcal{B}_I$}
\psfrag{c}{$\mathcal{C}_I$} \psfrag{pinf}{$\infty^+$}
\psfrag{minf}{$\infty^-$} \psfrag{pp}{$p(x)$} \psfrag{mp}{$-p(x)$}
\includegraphics{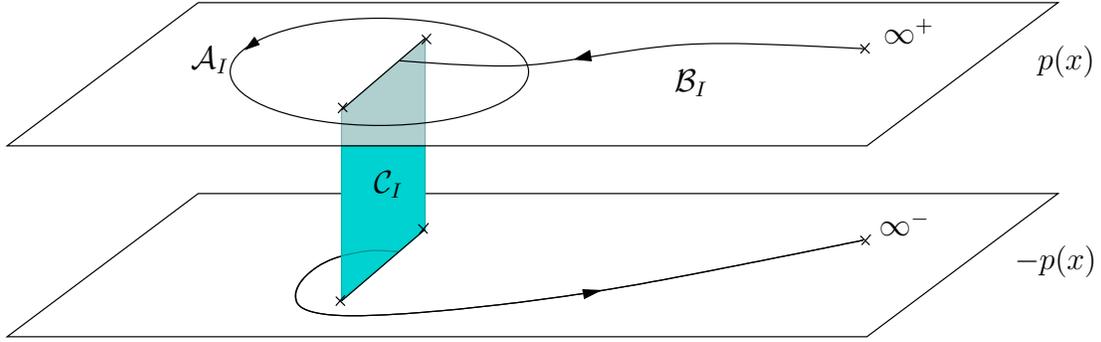}
\caption{The cycle $\mathcal{A}_I$ and path $\mathcal{B}_I$ for
the cut $\mathcal{C}_I$.} \label{AnBcycles}
\end{figure}

For each cut $\mathcal{C}_I$ we define a cycle $\mathcal{A}_I$
encircling the cut on the physical sheet and a path
$\mathcal{B}_I$ joining the points $x = \infty$ on both sheets
through the cut $\mathcal{C}_I$ as shown in Figure
\ref{AnBcycles}.
Let us also define a basis $\{ a_i, b_i \}_{i=1}^g$ for the first
homology group $H_1(\Sigma,\mathbb{R})$ of the curve $\Sigma$ by
setting for example $a_i = \mathcal{A}_i$ and $b_i = \mathcal{B}_i -
\mathcal{B}_{g+1}$. This basis has the following canonical
intersection matrix
\begin{equation*}
a_i \cap a_j = b_i \cap b_j = 0, \; a_i \cap b_j = \delta_{ij}.
\end{equation*}
Define also the dual basis $\{ \omega_i \}_{i=1}^g$ of holomorphic
differentials on $\Sigma$ normalised such that $\int_{a_i} \omega_j =
\delta_{ij}$. The matrix of $b$-periods of these holomorphic
differentials defines a symmetric matrix with positive imaginary part
called the period matrix $\Pi$ of the Riemann surface $\Sigma$
\begin{equation} \label{period matrix}
\Pi_{ij} = \int_{b_i} \omega_j.
\end{equation}
Using this data of $\Sigma$ one can define an important object
associated with any Riemann surface known as the Jacobian variety and
defined as the following $g$-dimensional complex torus
\begin{equation*}
J(\Sigma) = \mathbb{C}^g / (2\pi \mathbb{Z}^g + 2\pi\Pi \mathbb{Z}^g).
\end{equation*}
Now since by definition $p(x)$ is single valued on the cut $x$-plane
it follows that the $\mathcal{A}$-periods of $dp$ must vanish, and
using $\Omega(x) \sim_{x \rightarrow \infty} {\bf 1}$ from
\eqref{monodromy asymptotics at infty} we find that the
$\mathcal{B}$-periods of $dp$ take values in $2 \pi \mathbb{Z}$
\begin{equation} \label{AnB periods of dp}
\int_{\mathcal{A}_I} dp = 0, \quad \int_{\mathcal{B}_I} dp = 2 \pi
n_I, \; n_I \in \mathbb{Z}.
\end{equation}
The integers $n_I$ appearing in the $\mathcal{B}_I$ periods
are precisely the mode numbers introduced in \eqref{mode} above.
A normalised (i.e. vanishing $\mathcal{A}$-periods) meromorphic
differential on $\Sigma$ is uniquely specified by its singular
parts, and so equation \eqref{p asymptotics at pm 1} along with
the $\mathcal{A}$-cycle conditions in \eqref{AnB periods of dp}
uniquely specify $dp$ as a meromorphic differential on $\Sigma$
with singular parts
\begin{equation} \label{dp asymptotics at pm 1}
\begin{split}
dp(x^{\pm}) = \mp d\left(\frac{\pi \kappa}{x - 1}\right) + O\left( (x -
1)^0 \right) \quad \text{as } x \rightarrow +1, \\
dp(x^{\pm}) = \mp d\left(\frac{\pi \kappa}{x + 1}\right) + O\left( (x +
1)^0 \right) \quad \text{as } x \rightarrow -1.
\end{split}
\end{equation}
Because the normalised meromorphic differential $dp$ on $\Sigma$
is uniquely specified by its singular parts \eqref{dp asymptotics
at pm 1} it follows that
\begin{equation*}
\hat{\sigma}^{\ast} dp = -dp.
\end{equation*}

\begin{figure}
\centering \psfrag{x=p1}{\tiny $x = 1$}
\psfrag{x=m1}{\tiny $x = - 1$}
\psfrag{p(x)}{\footnotesize $p(x)$} \psfrag{-p(x)}{\footnotesize $-p(x)$}
\includegraphics[height=46mm,width=157mm]{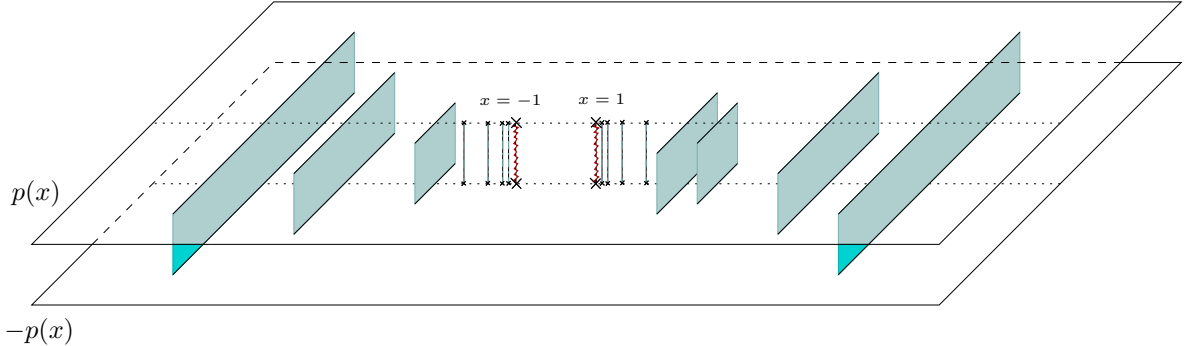}
\caption{The algebraic curve $\Sigma$ with its finitely many branch
cuts and infinitely many marked points accumulating at $x = \pm 1$.}
\label{Riemann Surface Sigma}
\end{figure}

The infinite set of singular points $\{ x_k \}$ of the original
curve $\Gamma$ become marked points on the desingularised curve
$\Sigma$ in \eqref{desingularised curve} and can be characterised
by the following condition
\begin{equation} \label{singular points}
p(x_k) = n_k \pi,  \qquad n_k \in \mathbb{Z}.
\end{equation}
Singular points can otherwise be seen as degenerated branch cuts,
the two branch points of which sit on top of each other. From this
point of view, the condition \eqref{singular points} corresponds
to the degeneration of the $\mathcal{B}$-period condition
$\int_{\mathcal{B}_k} dp = 2 \pi n_k$ of the meromorphic
differential $dp$ for this particular cut. Indeed, given the
normalised Abelian differential $dp$ we can define its Abelian
integral as
\begin{equation*}
p(P) = \int_{\infty^+}^P dp, \quad P \in \Sigma,
\end{equation*}
If we restrict the integration path to lie on the physical sheet
then this function is just the single valued function $p(x)$
itself, i.e. $p(x) = p(x^+)$. The $\mathcal{B}$-period condition
on $dp$ now reads (where $Q^+ \in \Sigma$ is a point on the
physical sheet arbitrarily close to the cut $\mathcal{C}_k$)
\begin{equation*}
\begin{split}
2 \pi n_k &= \int_{\infty^+}^{Q^+} dp + \int_{Q^+}^{\infty^-} dp =
\int_{\infty^+}^{Q^+} dp - \int_{Q^+}^{\infty^-} \hat{\sigma}^{\ast} dp \\
&= \int_{\infty^+}^{Q^+} dp - \int_{\hat{\sigma} Q^+}^{\infty^+}
dp = \int_{\infty^+}^{Q^+} dp + \int_{\infty^+}^{\hat{\sigma} Q^+} dp \\
\text{i.e. } 2 \pi n_k &= p(Q^+) + p(\hat{\sigma} Q^+).
\end{split}
\end{equation*}
Thus if the cut $\mathcal{C}_k$ degenerates to a singular point
then $Q^+ \in \Sigma$ becomes that singular point which satisfies
$\hat{\sigma} Q^+ = Q^+$ so that
\begin{equation*}
p(Q^+) = n_k \pi.
\end{equation*}
A typical algebraic curve $\Sigma$ for a finite-gap solution of genus
$g$ is represented schematically in Figure \ref{Riemann Surface Sigma}
with its $g+1$ cuts and marked points (representing singular points of
$\Gamma$) accumulating at $x = \pm 1$.

As suggested in \cite{Beisert:2005bm}, the spectral curve can thus
be understood as the analogue of a Fourier mode decomposition of
the string, in the sense that to every integer mode number $n_i
\in \mathbb{Z}$ there corresponds either a branch cut (when the
mode is excited) with corresponding $\mathcal{B}$-period
$\int_{\mathcal{B}_i} dp = 2 \pi n_i$ or a singular point (when
the mode is switched off) at $x_i$ such that $p(x_i) = n_i \pi$.
Turning on a mode, with mode number $m \in \mathbb{Z}$ say,
corresponds to opening up the singular point at $p(x) = m \pi$
into a full grown cut, which in turn increases the genus of the
curve by one requiring the introduction of a new moduli as we will
shortly argue, i.e. the amplitude of the mode $m$. Now
since the algebraic curve is $(\sigma,\tau)$-invariant, its moduli
gives a set of action variables of the theory, which by the previous
reasoning forms a complete set. For a general solution we expect the
corresponding curve to have infinite genus, but we can view such a
solution as a limit of finite-gap solutions.

Finally, we note that the spectral curve $\widehat{\Sigma}$
\eqref{pre-desingularised curve} of \cite{Beisert:2004ag,
Beisert:2005bm} can be uniquely recovered from the algebraic curve
$\Sigma$ \eqref{desingularised curve} of \cite{Kazakov:2004qf}
equipped with the meromorphic differential $dp$ simply by keeping
the same spectral parameter $x$ but redefining the $y$ coordinate
by
\begin{equation*}
dp = y dx.
\end{equation*}
This definition makes sense away from branch points of $\Sigma$ where $x$
can be taken as a local parameter, and can be extended by continuity to branch
points where $dx$ has zeroes so that $y$ picks up a pole at branch points.
Therefore the specification of the pair $(\Sigma,dp)$ is equivalent to
the specification of the curve $\widehat{\Sigma}$.


\subsection{The moduli space of the spectral curve} \label{section: moduli}

The curve $\widehat{\Sigma}$, or equivalently the pair
$(\Sigma,dp)$, uniquely determine the spectrum of conserved
charges for any finite-gap solution. In the following, we will
define a moduli space for this holomorphic data which, when
subjected to the appropriate reality conditions, corresponds to
the space of allowed values for the conserved charges. In this
section it will be convenient to work initially with
configurations where only the first set of Virasoro constraints
\eqref{V1} are imposed. The two remaining conditions, which are
the momentum constraint and the string mass shell condition
\eqref{massshell}, will be imposed at a later stage in our
discussion. As discussed above, we will also restrict our
attention to variations of the spectral curve for which the left
and right Noether charges $L$ and $R$ remain constant.

The required moduli space is a special case of a very general
construction due to Krichever and Phong \cite{Krichever+Phong}
which is reviewed in Appendix \ref{section: The moduli space}.
Following \cite{Krichever+Phong}, the starting point of the
construction is a much larger {\em universal moduli space},
$\mathcal{U}$, of smooth Riemann surfaces of genus $g$, with $N$
punctures $P_{\alpha}$, equipped with two Abelian integrals $E$
and $Q$ which have poles of order $n_{\alpha}$ and $m_{\alpha}$
respectively at the punctures. This space has complex dimension
$5g-3+3N+\sum_{i=1}^N (n_{\alpha} + m_{\alpha})$. Krichever and
Phong define a set of local holomorphic coordinates on
$\mathcal{U}$ and prove that level sets of certain subsets of
these coordinates define smooth foliations of $\mathcal{U}$. As
shown in \cite{Krichever+Phong}, the moduli spaces of the spectral
data for a large class of integrable systems can be each described
as leaves of such a foliation. Remarkably, the relevant moduli
space for the holomorphic data $(\Sigma,dp)$ also admits precisely
such a description.

In the present case the Riemann surface is identified with the
algebraic curve $\Sigma$ of genus $g=K-1$ defined above and the
punctures $P_{\alpha}$ correspond to the eight points
$\{\infty^{\pm}, 0^{\pm}, (+1)^{\pm}, (-1)^{\pm}\}$ on $\Sigma$.
The first Abelian integral $E$ is identified with the
quasi-momentum $p(P)$ defined in the previous section. As in
\eqref{p asymptotics at pm 1}, $p(P)$ has simple poles at the
points $\{ (+1)^{\pm}, (-1)^{\pm}\}$ and no other singularities.
To make contact with \cite{Krichever+Phong} we are led to
introduce a second Abelian integral $z(P)$ on $\Sigma$, identified
with $Q$ in the general construction, which has simple poles at
the remaining punctures $\{\infty^{\pm}, 0^{\pm}\}$ and no other
singularities. Thus we are considering the $N=8$ case of the
construction of \cite{Krichever+Phong} with $m_{\alpha} +
n_{\alpha} = 1$ at each of the eight punctures $P_{\alpha}$. In
this case, the universal moduli space $\mathcal{U}$ has complex
dimension $5g + 29 = 5K + 24$.

As in the previous section $\Sigma$ and $p$ are subject to several
constraints. These include,
\begin{itemize}
\item[$\bullet$] The $a$- and $b$-periods \eqref{AnB periods of
dp} of the differential $dp$. \item[$\bullet$] The asymptotics of
$dp$ near the points $\{(+1)^{\pm},(-1)^{\pm}\}$. As explained
above, at this stage we are only imposing the first set of
Virasoro constraints \eqref{V1}. In this case the asymptotics
\eqref{dp asymptotics at pm 1} are replaced by,
\begin{equation} \label{weak dp asymptotics at pm 1}
\begin{split}
dp(x^{\pm}) = \mp d\left(\frac{\pi \kappa_+}{x - 1}\right) + O\left( (x -
1)^0 \right) \quad \text{as } x \rightarrow +1, \\
dp(x^{\pm}) = \mp d\left(\frac{\pi \kappa_-}{x + 1}\right) + O\left( (x +
1)^0 \right) \quad \text{as } x \rightarrow -1.
\end{split}
\end{equation}
\item[$\bullet$] The asymptotics \eqref{p asymptotics at infty},
\label{weak dp asymptotics at zero} of $p$ near
the points $\{\infty^{\pm},0^{\pm}\}$.
\end{itemize}
In Appendix \ref{section: The moduli space}, we show that the set of
constraints described above is equivalent to a collection of
level-set conditions for a subset of the holomorphic coordinates on
$\mathcal{U}$ introduced in \cite{Krichever+Phong}. Further
level-set conditions are provided  by specifying appropriate
residues and periods for the second Abelian integral $z(P)$. In
total, we find level-set conditions for exactly $4g+29$ of the
$5g+29$ holomorphic coordinates on $\mathcal{U}$. This defines a
leaf $\mathcal{L}$ in a smooth $g$-dimensional foliation of the
universal moduli space.

In the above discussion we did not assume the hyperelliptic form
\eqref{desingularised curve} for the curve $\Sigma$. In fact, as we
show in Appendix \ref{section: The moduli space}, the hyperelliptic
property of $\Sigma$ is actually a consequence of the level-set
conditions described above and therefore leads to no further
constraints.  Finally we can also derive an explicit expression for
the second Abelian integral $z(P)$ in terms of the hyperelliptic
coordinate $x(P)$. Specifically, we find the formula,
\begin{equation}\label{zdef}
z(P) = x(P) + \frac{1}{x(P)}.
\end{equation}

The remaining holomorphic coordinates which are not fixed by the
level-set conditions provide holomorphic coordinates on the leaf
${\cal L}$. According to \cite{Krichever+Phong}, these coordinates
correspond to certain periods (and residues) of the meromophic
differential
\begin{equation} \label{symplectic 1-form}
\alpha = \frac{\sqrt{\lambda}}{4 \pi} z dp,
\end{equation}
where we have chosen an appropriate normalisation. The $g$ independent
coordinates parameterising the leaf are \cite{Krichever+Phong},
\begin{equation} \label{independent moduli}
S_i = \frac{1}{2\pi i} \int_{a_i} \alpha = \frac{1}{2\pi
i}\frac{\sqrt{\lambda}}{4 \pi} \int_{a_i}
\left(x+\frac{1}{x}\right)dp, \quad i = 1, \ldots, g,
\end{equation}
where $a_{i}$ are the canonical one-cycles on $\Sigma$ defined in
section \ref{section: algebraic curve} and we have used
\eqref{zdef}. Equivalently one can work with the periods of
$\alpha$ around the $K=g+1$ cycles, $\mathcal{A}_I$, also defined
in section \ref{section: algebraic curve}, and choose coordinates,
\begin{equation} \label{filling fractions}
\mathcal{S}_I = \frac{1}{2\pi i} \int_{\mathcal{A}_I} \alpha =
\frac{1}{2\pi i} \frac{\sqrt{\lambda}}{4 \pi} \int_{\mathcal{A}_I}
\left(x+\frac{1}{x}\right)dp, \quad I = 1, \ldots, K.
\end{equation}
subject to the constraint,
\begin{equation}\label{sumrule}
\sum_{I=1}^K {\cal S}_I = \text{Res}_{\infty^{+}} \alpha + \text{Res}_{0^{+}}
\alpha =\frac{1}{2}(L - R).
\end{equation}
Thus we have one coordinate $\mathcal{S}_I$ associated with each
cut $\mathcal{C}_I$ in the $x$-plane. The relation between the two
sets of cycles implies that $\mathcal{S}_I = S_i$ for
$I=i=1,\ldots,g$.

The variables $\mathcal{S}_I$ are known as {\em filling fractions}
\cite{Kazakov:2004qf, Beisert:2004ag} and they play an important
role in the context of the AdS/CFT correspondence. As in
\eqref{sumrule}, their sum is the global charge $J=(L-R)/2$. In
the dual spin-chain description, each magnon carries one unit of
the charge $J$ and also corresponds to a single Bethe root in the
complex rapidity plane. The Bethe roots condense to form cuts in
an appropriate thermodynamic limit. The quantity $\mathcal{S}_I$
corresponds to the total $U(1)_J$ charge associated with the cut
$\mathcal{C}_I$ and thus to the total number of Bethe roots with
the corresponding mode number $n_I$. The AdS/CFT correspondence
therefore suggest that the quantities $\mathcal{S}_I$ (and thus
the moduli $S_i$) should be quantised in integer units. In
subsection \ref{section: Hamiltonian}, we will find an intriguing
hint that this quantisation will emerge naturally on the string
theory side of the correspondence.

It is important to emphasise that the quantities $\kappa_{\pm}$
which appear in the Virasoro constraint \eqref{V1} and the
asymptotics \eqref{weak dp asymptotics at pm 1}, are not
independent parameters in the construction described above. Rather
they are non-trivial functions of the coordinates $\{S_i\}$ on the
leaf $\mathcal{L}$. These quantities also determine (via
\eqref{EP1}) the worldsheet momentum, $\mathcal{P}$ and energy,
$\mathcal{E}$, which should likewise be thought of as functions on
$\mathcal{L}$. Fortunately it is possible to describe the
variations of $\mathcal{P}$ and $\mathcal{E}$ with the moduli $\{
S_i \}$ quite explicitly. In Appendix \ref{section: variations of
moduli} we derive the following equations,
\begin{equation} \label{variation2b}
\delta^{\mathcal{L}} \mathcal{P} = -\sum_{i=1}^g k_i
\delta^{\mathcal{L}} S_i, \quad \delta^{\mathcal{L}}\mathcal{E} =
-\sum_{i=1}^g w_i \delta^{\mathcal{L}} S_i,
\end{equation}
using the Riemann bilinear identity. Here $k_i$ and $w_i$ are
defined in terms of the periods of the meromorphic differentials
$dp$ and $dq$ of the quasi-momentum and quasi-energy respectively,
\begin{equation} \label{B periods of dp and dq}
\int_{b_i} dp = 2\pi k_i, \quad \int_{b_i} dq = 2 \pi w_i.
\end{equation}
The variation $\delta^{\mathcal{L}}$, corresponds to the exterior
derivative on the complex leaf $\mathcal{L}$.

As in \eqref{AnB periods of dp} the closed string boundary
condition constrains the $k_i$ to be integers which are related to
the mode numbers $n_I$ as $k_i = n_i - n_K$. Thus the $k_i$ are
constant on the leaf $\mathcal{L}$ and we can integrate the first
equality in \eqref{variation2b} to obtain,
\begin{equation*}
\mathcal{P} = -\sum_{i=1}^g k_i S_ i + {\rm constant}.
\end{equation*}
Using \eqref{sumrule}, we can rewrite $\mathcal{P}$ in terms of
the filling fractions $\mathcal{S}_I$ and fix the integration
constant by comparison to the pointlike string solution with
$\mathcal{S}_I = 0$ for all $I$. The final result is,
\begin{equation}\label{mcons2}
\mathcal{P} = -\sum_{I=1}^K n_I \mathcal{S}_I.
\end{equation}

Before imposing the momentum constraint and the string mass shell
condition, the moduli space of admissible spectral curves $\Sigma$
is precisely the leaf $\mathcal{L}$ described above. From
\eqref{mcons2} , the momentum constraint implies a further linear
condition on the filling fractions,
\begin{equation}\label{constraint}
\sum_{I=1}^K n_I \mathcal{S}_I = 0.
\end{equation}
As before the string mass shell condition identifies the string
energy $\Delta$ as $\sqrt{2\sqrt{\lambda}\mathcal{E}}$ The second
equality in \eqref{variation2b} then provides a set of first order
differential equations for $\Delta$ as a function of the moduli
$\{S_{i}\}$,
\begin{equation*}
\Delta \frac{\partial \Delta}{\partial S_i} =
-\frac{\sqrt{\lambda}}{2 \pi} \int_{b_i}\, dq.
\end{equation*}
In contrast to the integers $k_i$, the $b$-periods $w_i$ of $dq$
are not constant along the leaf $\mathcal{L}$ and integration of
these equations is non-trivial.

\subsection{Reality conditions} \label{section: reality of curve}

The reality condition \eqref{monodromy reality condition} on the
monodromy matrix implies that the curve $\Gamma$ is invariant under
the following anti-holomorphic involution
\begin{equation} \label{anti-holomorphic involution}
\hat{\tau} : \Gamma \rightarrow \Gamma, (x,y) \mapsto
(\bar{x},\bar{y}^{-1}).
\end{equation}
In terms of the representation of $\Gamma$ as a 2-sheeted ramified
cover of the $x$-plane the effect of this anti-holomorphic
involution is simply to map both sheets to themselves by $x \mapsto
\bar{x}$. The two involutions \eqref{anti-holomorphic involution}
and \eqref{holomorphic involution} together generate a $\mathbb{Z}_2
\times \mathbb{Z}_2$ group of involutions on $\Gamma$ such that
$\hat{\sigma} \hat{\tau} = \hat{\tau} \hat{\sigma}$. Since in going
from the curve $\Gamma$ in \eqref{spectral curve} to the fully
desingularised curve $\Sigma$ in \eqref{desingularised curve} we
haven't modified the spectral parameter $x$, the desingularised
curve $\Sigma$ is still invariant under the anti-holomorphic
involution which takes both sheets to themselves by $x \mapsto
\bar{x}$, so that the branch points $u_I,v_I$ defined in
\eqref{desingularised curve} should either be both real ($u_I =
\bar{u}_I$, $v_I = \bar{v}_I$) or form a complex conjugate pair
($u_I = \bar{v}_I$).

It follows from its definition that the homology basis $\{ a_i, b_i
\}_{i=1}^g$ has the following property under the action of the
anti-holomorphic involution $\hat{\tau}$
\begin{equation} \label{a,b cycles reality}
\hat{\tau} a_i = - a_i, \quad \hat{\tau} b_i = b_i,
\end{equation}
where the equalities in these expressions are to be understood
modulo cycles homotopic to zero. Note also that the basis of
holomorphic differentials of $\Sigma$ is such that
$\overline{\hat{\tau}^{\ast} \omega_i} = - \omega_i$ since
$\overline{\hat{\tau}^{\ast} \omega_i}$ are holomorphic $1$-forms
(because the local parameter $x$ is such that $\hat{\tau}^{\ast} x =
\bar{x}$) and
\begin{equation*}
\delta_{ij} = \int_{a_i} \omega_j = \int_{\hat{\tau} a_i}
\hat{\tau}^{\ast} \omega_j = - \int_{a_i} \hat{\tau}^{\ast}
\omega_j.
\end{equation*}
Thus it follows that the periodic matrix $\Pi_{ij} = \int_{b_i}
\omega_j$ of $\Sigma$ is pure imaginary $\overline{\Pi} = - \Pi$
because
\begin{equation*}
\overline{\Pi_{ij}} = \int_{b_i} \overline{\omega_j} =
\int_{\hat{\tau} b_i} \overline{\hat{\tau}^{\ast} \omega_j} = -
\int_{b_i} \omega_j = - \Pi_{ij}.
\end{equation*}

The reality condition on the differential introduced in
\eqref{symplectic 1-form} is $\overline{\hat{\tau}^{\ast} \alpha} = \alpha$, from
which the reality condition on the filling fractions
\eqref{filling fractions} follows
\begin{equation*}
\overline{\mathcal{S}_I} = - \frac{1}{2 \pi i} \int_{\mathcal{A}_I}
\bar{\alpha} = - \frac{1}{2 \pi i} \int_{\mathcal{A}_I}
\hat{\tau}^{\ast} \alpha = - \frac{1}{2 \pi i} \int_{\hat{\tau}
\mathcal{A}_I} \alpha = \mathcal{S}_I, \quad I = 1, \ldots, K.
\end{equation*}

\section{Finite-gap solutions}

In the previous sections we have seen how from a given solution to
the equations of motion one can construct a flat connection
$J(x,\sigma,\tau)$, and from this obtain a non-singular algebraic
curve carrying the action variables of the solution in question.
In section \ref{section: identifying data} we identify the
remaining algebro-geometric data which uniquely characterises the
dynamics of the solution, i.e. the angle variables. We then
explain in section \ref{section: reconstruction} the method of
finite-gap integration for recovering the explicit solution
corresponding to a given set of algebro-geometric data
\cite{Babelon, Krichever1, Krichever2, Belokolos}. In the
following section we will use various results proved for example
in \cite{Babelon} so to avoid cluttering the main argument we will
simply refer to specific pages in this reference for the proofs.

\subsection{Identifying the algebro-geometric data} \label{section: identifying data}

We start by describing the dynamics of the monodromy matrix which
from the relation \eqref{monodromy diagonalised} is encoded in its
eigenvectors. By definition of the spectral curve there is a
unique eigenvector of $\Omega(x,\sigma,\tau)$ corresponding to any
regular point $P \in \Gamma$, and it can be shown (see the
Proposition p131-132 in \cite{Babelon} and especially the remarks
following it) that this is also the case when $P$ is a branch
point. Thus the desingularised curve $\Sigma$ has a unique
eigenvector $\bm{h}(P,\sigma,\tau)$ at each point $P \in \Sigma$,
which we normalise by setting its first component to one for
convenience, i.e. $h_1(P,\sigma,\tau) = 1$. The second component
$h_2$ on $\Sigma$ can be shown to have exactly $g + 1$ poles in
the present case (see the Proposition p134 in \cite{Babelon}). At
the points $\infty^+, \infty^- \in \Sigma$ above $x_0 = \infty$
the eigenvector is proportional to ${\tiny \left( \begin{array}{c} 1\\
0 \end{array} \right), \left( \begin{array}{c} 0\\ 1 \end{array}
\right)}$ respectively, so the second component
$h_2(P,\sigma,\tau)$ of the normalised eigenvector has a (simple)
zero at $\infty^+$ and one of its $g + 1$ poles is at $\infty^-$.
The remaining $g$ poles define the \textit{dynamical divisor}
which we denote by $\gamma(\sigma,\tau) = \prod_{i=1}^g
\gamma_i(\sigma,\tau)$. It is easy to see that the
algebro-geometric data $\{ \Sigma, dp, \gamma(\sigma,\tau),
\infty^{\pm} \}$ is invariant under the residual gauge group
(section \ref{section: classical integrability}) and so we have a
well defined map
\begin{equation} \label{Omega algebro-geometric data}
[\Omega(x,\sigma,\tau)] \mapsto \{ \Sigma, dp,
\gamma(\sigma,\tau), \infty^{\pm} \},
\end{equation}
where $[\cdot]$ denotes the equivalence class under residual gauge
transformations. But now by the Riemann-Roch theorem there is up
to multiplication by a function independent of $P$ a unique
meromorphic function on $\Sigma$ with pole divisor in
$\gamma(\sigma,\tau) \infty^-$ and with a zero at $\infty^+$. Thus
the divisor $\left( \gamma(\sigma,\tau) \infty^- \right)^{-1}
\infty^+$ on $\Sigma$ uniquely specifies the normalised
eigenvector $\bm{h}(P,\sigma,\tau)$ up to left multiplication by a
diagonal matrix $\text{diag}(1,d(\sigma,\tau))$ and hence also
completely specifies the monodromy matrix up to a similarity
transformation $\Omega(x,\sigma,\tau) \rightarrow
\text{diag}(1,d(\sigma,\tau)) \Omega(x,\sigma,\tau)
\text{diag}(1,d(\sigma,\tau))^{-1}$. Therefore the map
\eqref{Omega algebro-geometric data} is injective.

To determine the dynamics of the connection $J(x,\sigma,\tau)$ we
first notice that the zero-curvature equation \eqref{zero
curvature equation} is equivalent to the consistency condition for
the following linear system
\begin{equation} \label{auxiliary linear system}
\left( d - J(x,\sigma,\tau) \right) \bm{\psi} = 0,
\end{equation}
The Lax connection $J(x,\sigma,\tau)$ can be recovered at a
generic value of $x$ from the solution $\bm{\psi}(P,\sigma,\tau)$
to the auxiliary linear problem by
\begin{equation} \label{reconstruction formula for U,V}
J(x,\sigma,\tau) = \left(d \Psi(x,\sigma,\tau)\right)
\Psi(x,\sigma,\tau)^{-1},
\end{equation}
where $\Psi(x,\sigma,\tau) = (\bm{\psi}(x^+), \bm{\psi}(x^-))$,
and thus we can focus on describing the dynamics of the solution
$\bm{\psi}(P,\sigma,\tau)$ to the auxiliary linear problem. Now by
virtue of the relation \eqref{monodromy evolution}, the operators
$(d - J(x,\sigma,\tau))$ and $\Omega(x,\sigma,\tau)$ can be
simultaneously diagonalised, and hence the vector
$\bm{\psi}(P,\sigma,\tau)$ in \eqref{auxiliary linear system} can
be taken as a multiple of the normalised eigenvector
\begin{equation} \label{psi factored scalar}
\bm{\psi}(P,\sigma,\tau) = \varphi(P,\sigma,\tau) \bm{h}(P,\sigma,\tau).
\end{equation}
Taking the first component of the auxiliary linear problem
\eqref{auxiliary linear system} yields a system of equations for
the scalar function $\varphi(P,\sigma,\tau)$, namely
\begin{equation} \label{scalar function equation}
d \varphi(P,\sigma,\tau) = \big( J(x,\sigma,\tau)
\bm{h}(P,\sigma,\tau) \big)_1 \varphi(P,\sigma,\tau),
\end{equation}
Now since the Lax connection $J(x,\sigma,\tau)$ has poles at the
constant values $x = \pm 1$ and the normalised eigenvector is
holomorphic in a neighbourhood of every point above $x = \pm 1$,
equation \eqref{scalar function equation} implies that the scalar
function $\varphi(P,\sigma,\tau)$ has essential singularities at
every point on $\Sigma$ above $x = \pm 1$ which are shown in
Appendix \ref{section: Singular parts} to be of the form
\begin{equation} \label{BA singular parts}
\begin{split}
\varphi(x^{\pm},\sigma,\tau) &= O(1) \; \exp \left\{ \mp
\frac{i\kappa_+}{2} \frac{\sigma + \tau}{x - 1} \right\} , \quad
\text{as}\; x \rightarrow 1, \\
\varphi(x^{\pm},\sigma,\tau) &= O(1) \; \exp \left\{ \mp
\frac{i\kappa_-}{2} \frac{\sigma - \tau}{x + 1} \right\} , \quad
\text{as}\; x \rightarrow -1,
\end{split}
\end{equation}
where $O(1)$ denotes a function holomorphic in a neighbourhood of
$x = \pm 1$. Because of \eqref{connection asymptotics} and the
gauge redundancy \eqref{gauge redundancy}, we have in a general
residual gauge $J(\infty,\sigma,\tau) = d \tilde{g}
\tilde{g}^{-1}$ where $\tilde{g}(\sigma,\tau)$ is diagonal, so
that the constant pole of $h_2$ at $\infty^-$ does not give rise
to any essential singularities in $\varphi(P,\sigma,\tau)$. The
function $\varphi(P,\sigma,\tau)$ also has zeroes at the dynamical
divisor $\gamma(\sigma,\tau)$ (see \cite{Babelon} p151-152) and if
we consider the solution to \eqref{auxiliary linear system} with
initial condition\footnote{Normalising $\varphi$ such that
$\varphi(P,0,0) = 1$ is always possible because one can divide the
solution $\varphi$ to the auxiliary linear problem
\eqref{auxiliary linear system} by an arbitrary function $f(P)$ of
$P \in \Sigma$ only. Note that this scaling by $f(P)$ has no
effect on the current $J(x,\sigma,\tau)$ reconstructed using
\eqref{reconstruction formula for U,V}. Choosing $f(P) =
\varphi(P,0,0)$ yields the required initial condition on
$\varphi$.} $\varphi(P,0,0) = 1$ then $\varphi(P,\sigma,\tau)$ has
poles at $\gamma(0,0)$. But a meromorphic function on $\Sigma
\setminus \{ x = \pm 1 \}$ with pole divisor in $\gamma(0,0)$ and
essential singularities at the punctures $x = \pm 1$ of the
prescribed form \eqref{BA singular parts} is
unique\footnote{Indeed, suppose there are two such functions, then
their quotient is meromorphic on the genus $g$ Riemann surface
$\Sigma$ and has at most $g = \text{deg}\, \gamma(0,0)$ poles
which must be independent of $P \in \Sigma$ by the Riemann-Roch
theorem.} up to multiplication by a function independent of $P \in
\Sigma$. Therefore the first component $\psi_1 = \varphi$ of
$\bm{\psi}(P,\sigma,\tau)$ is uniquely specified up to a
multiplicative $c(\sigma,\tau)$ by the divisor $\gamma(0,0)$ and
its behaviour \eqref{BA singular parts} at $x = \pm 1$. Likewise,
the second component $\psi_2 = \varphi h_2$ is uniquely specified
up to a multiplicative $d(\sigma,\tau)$ by: the behaviour
\eqref{BA singular parts} of $\varphi$ at $x = \pm 1$, the fact
that $\psi_2$ has a zero at $\infty^+$ and that its pole divisor
is contained in $\gamma(0,0) \infty^-$. Therefore the set of
algebro-geometric data for the solution $\bm{\psi}(P,\sigma,\tau)$
to the auxiliary linear problem \eqref{auxiliary linear system}
can be taken to be $\{ \Sigma, dp, \gamma(0,0), \infty^{\pm},
S_{\pm}(P,\sigma,\tau) \}$ where $S_{\pm}(P,\sigma,\tau)$ are the
singular parts defined in \eqref{Singular parts} and it is easily
seen to be invariant under residual gauge transformations
($\bm{\psi} \rightarrow \tilde{g} \bm{\psi}$). Hence we have a
well defined injective map
\begin{equation} \label{U,V algebro-geometric data full}
[J(x,\sigma,\tau)] \mapsto \{ \Sigma, dp, \gamma(0,0),
\infty^{\pm}, S_{\pm}(P,\sigma,\tau) \}.
\end{equation}
Note that the singular parts are specific to the choice of
analytic behaviour of $J(x,\sigma,\tau)$ in $x$ and that the pair
of punctures $\infty^{\pm} \in \Sigma$ is specific to the gauge
fixing choice \eqref{monodromy asymptotics at infty 2} and neither
of them depends on the solution considered. So since this extra
data is common to all solutions we may omit it when specifying the
algebro-geometric data so that the injective map \eqref{U,V
algebro-geometric data full} boils down to
\begin{equation} \label{U,V algebro-geometric data}
[J(x,\sigma,\tau)] \mapsto \{ \Sigma, dp, \gamma(0,0) \}.
\end{equation}
The divisor $\gamma(0,0)$ being of degree $g$ it lives in the $g$-th
symmetric product $S^g(\Sigma) = \Sigma^g / S_g$ of the curve
$\Sigma$ which is in bijection with the Jacobian $J(\Sigma)$ of
$\Sigma$ by means of the Abel map
\begin{equation} \label{Abel map on divisors}
\begin{split}
\bm{\mathcal{A}} : S^g(\Sigma) &\rightarrow J(\Sigma) \\
\prod_{i=1}^g P_i &\mapsto 2 \pi \sum_{i=1}^g \int_{\infty^+}^{P_i} \bm{\omega}.
\end{split}
\end{equation}
Therefore the algebro-geometric data $\{ \Sigma, dp, \gamma(0,0)
\}$ corresponding to a generic solution specifies a point on the
Jacobian bundle $\mathcal{M}^{(2g)}_{\mathbb{C}}$ over the leaf
$\mathcal{L}$ which we defined in section \ref{section: spectral
curve}
\begin{equation} \label{Jacobian bundle}
J(\Sigma) \rightarrow \mathcal{M}^{(2g)}_{\mathbb{C}} \rightarrow
\mathcal{L},
\end{equation}
where the fibre over any point $\Sigma$ in $\mathcal{L}$ is the
Jacobian $J(\Sigma)$ of the corresponding curve $\Sigma$. So in
effect the map \eqref{U,V algebro-geometric data} constructed in
this section takes the orbit under residual gauge transformations
of any solution of the equations of motion \eqref{flatness &
conservation} (note that not every point on this orbit is a
solution) to a single point on $\mathcal{M}^{(2g)}_{\mathbb{C}}$.

\subsection{Reconstruction formulae} \label{section: reconstruction}

In the previous section we have identified the set of
algebro-geometric data which uniquely specifies (up to residual
gauge transformations) a solution $J(x,\sigma,\tau)$ to the
zero-curvature equations, as exhibited by the injective map
\eqref{U,V algebro-geometric data}. In this section we want to
construct the (left) inverse of this map \eqref{U,V
algebro-geometric data}, namely
\begin{equation} \label{left inverse map}
\{ \Sigma, dp, \gamma(0,0) \} \mapsto \left[ J(x,\sigma,\tau)
\right].
\end{equation}
In other words, starting from a given set of admissible
algebro-geometric data (i.e. a point on
$\mathcal{M}^{(2g)}_{\mathbb{C}}$) - a curve $\Sigma$ of the form
\eqref{desingularised curve} equipped with an Abelian differential
$dp$ and a constant divisor $\gamma(0,0)$ of degree $g$, along
with the pair of punctures $\infty^{\pm} \in \Sigma$ above $x_0 =
\infty \in \mathbb{CP}^1$ and the singular parts
$S_{\pm}(P,\sigma,\tau)$ - we want to recover the Lax connection
$J(x,\sigma,\tau)$ corresponding to this set of data modulo
residual gauge transformations.

The components $\psi_1, \psi_2$ of $\bm{\psi}$ are meromorphic
functions on $\Sigma \setminus \{ x = \pm 1 \}$ which were shown in
the previous section to be uniquely specified (up to multiplicative
functions independent of $P \in \Sigma$) by their respective
divisors $(\psi_1) \geq \gamma(0,0)^{-1}$ and $(\psi_2) \geq \left(
\gamma(0,0) \infty^- \right)^{-1} \infty^+$ and by their common
behaviour \eqref{BA singular parts} near the essential singularities
at $x = \pm 1$. We can scale the eigenvector $\bm{\psi}$ by a
function $k_-(P)$ independent of $\sigma$ and $\tau$ so we let
$\hat{\gamma}$ be a positive divisor of degree $\text{deg }
\hat{\gamma} = g + 1$ equivalent\footnote{Two divisors $D_1,D_2$ are
said to be equivalent and we write $D_1 \sim D_2$ if there exists a
meromorphic function $f$ with divisor $(f) = D_1 D_2^{-1}$.} to
$\gamma(0,0) \infty^-$ and define $k_-(P)$ as the unique function
with divisor $\left( \gamma(0,0) \infty^- \right) \hat{\gamma}^{-1}$
normalised by $k_-(\infty^+) = 1$. After scaling by $k_-(P)$ the
divisors of the components of $\bm{\psi}$ satisfy
\begin{equation} \label{psi divisor}
(\psi_1) \geq \hat{\gamma}^{-1} \infty^-, \quad (\psi_2) \geq
\hat{\gamma}^{-1} \infty^+,
\end{equation}
whereas their behaviour near the essential singularities at $x = \pm
1$ is unchanged. Functions with such defining properties are known
as Baker-Akhiezer functions and can be constructed on the Riemann
surface $\Sigma$ of genus $g$ with the help of Riemann
$\theta$-functions \cite{Babelon, Belokolos}, so by uniqueness these
constructed functions will have to be equal to $\psi_1, \psi_2$ up
to a multiplicative function independent of $P \in \Sigma$; we fix
these multiplicative functions of $\sigma, \tau$ in the definitions
of $\psi_1, \psi_2$ by choosing the following normalisation
$\psi_1(\infty^+) = \psi_2(\infty^-) = 1$. Using the standard
expressions for Baker-Akhiezer functions in terms of Riemann
$\theta$-functions (see for instance section 2.7 in
\cite{Belokolos}), the components of $\bm{\psi}(P,\sigma,\tau)$ are
given by
\begin{subequations} \label{reconstruction formula for psi}
\begin{multline} \label{reconstruction formula for psi_1}
\psi_1(P,\sigma,\tau) = k_-(P) \frac{\theta \left(
\bm{\mathcal{A}}(P) + \bm{k} \sigma + \bm{w} \tau -
\bm{\zeta}_{\gamma(0,0)} \right) \theta \left(
\bm{\mathcal{A}}(\infty^+) - \bm{\zeta}_{\gamma(0,0)} \right)}{\theta
\left( \bm{\mathcal{A}}(P) - \bm{\zeta}_{\gamma(0,0)} \right) \theta
\left( \bm{\mathcal{A}}(\infty^+) + \bm{k} \sigma +
\bm{w} \tau - \bm{\zeta}_{\gamma(0,0)} \right)} \\ \times
\exp \left( \frac{i \sigma}{2 \pi} \int_{\infty^+}^P dp + \frac{i
\tau}{2 \pi} \int_{\infty^+}^P dq \right),
\end{multline}
\begin{multline} \label{reconstruction formula for psi_2}
\psi_2(P,\sigma,\tau) = k_+(P) \frac{\theta \left(
\bm{\mathcal{A}}(P) + \bm{k} \sigma + \bm{w} \tau -
\bm{\zeta}_{\gamma'(0,0)} \right) \theta \left(
\bm{\mathcal{A}}(\infty^-) - \bm{\zeta}_{\gamma'(0,0)} \right)}{\theta
\left( \bm{\mathcal{A}}(P) - \bm{\zeta}_{\gamma'(0,0)} \right) \theta \left(
\bm{\mathcal{A}}(\infty^-) + \bm{k} \sigma + \bm{w} \tau -
\bm{\zeta}_{\gamma'(0,0)} \right)} \\ \times \exp \left( \frac{i
  \sigma}{2 \pi} \int_{\infty^-}^P dp + \frac{i \tau}{2 \pi}
\int_{\infty^-}^P dq \right),
\end{multline}
\end{subequations}
where the ingredients for these formulae are defined as follows:
\begin{itemize}
\item[$\bullet$] The function $\theta : \mathbb{C}^g \rightarrow
\mathbb{C}$ is the Riemann $\theta$-function associated with
$\Sigma$ defined for $\bm{z} \in \mathbb{C}^g$ by
\begin{equation*}
\theta(\bm{z}) = \sum_{\bm{m} \in \mathbb{Z}^g} \exp \left\{ i
\langle \bm{m}, \bm{z} \rangle + \pi \, i \langle \Pi \bm{m}, \bm{m}
\rangle \right\},
\end{equation*}
where $\langle \bm{x}, \bm{y} \rangle = \sum_{i=1}^g x_i y_i$ and
$\Pi$ is the period matrix \eqref{period matrix}. It has the
following important property under translation by vectors $2\pi \bm{n} +
2\pi \Pi \bm{m} \in 2\pi \mathbb{Z}^g + 2\pi \Pi \mathbb{Z}^g$
\begin{equation} \label{automorphy theta}
\theta(\bm{z} + 2\pi \bm{n} + 2\pi \Pi \bm{m}) = \exp \left\{ - i
\langle \bm{m}, \bm{z} \rangle - \pi i \langle \Pi \bm{m}, \bm{m}
\rangle \right\} \theta(\bm{z}).
\end{equation}

\item[$\bullet$] The divisor $\gamma'(0,0)$ is the unique positive
divisor such that
\begin{equation}
\gamma'(0,0) \infty^+ \sim \gamma(0,0) \infty^- \sim \hat{\gamma}.
\end{equation}
Then $k_+(P)$ is the unique function with divisor $(k_+) = \left(
\gamma'(0,0) \infty^+ \right) \hat{\gamma}^{-1}$ normalised by the
condition $k_+(\infty^-) = 1$, just as $k_-(P)$ is the unique
function with divisor $(k_-) = \left( \gamma(0,0) \infty^- \right)
\hat{\gamma}^{-1}$ normalised by $k_-(\infty^+) = 1$.

\item[$\bullet$] The function $\bm{\mathcal{A}} : \Sigma \rightarrow
J(\Sigma)$ is the Abel map with base point $P_0$
\begin{align*}
\bm{\mathcal{A}} : \Sigma &\rightarrow J(\Sigma) \\
P &\mapsto 2\pi \int_{P_0}^P \bm{\omega}.
\end{align*}
This map is an injective holomorphic map by Abel's theorem. As in
section \ref{section: identifying data} it can be extented
\eqref{Abel map on divisors} to act on divisors of degree $g$ by
requiring that it be a morphism. This extended map is still
injective by Abel's theorem but is also surjective by Jacobi's
inversion theorem, and so the map \eqref{Abel map on divisors}
defines a bijection between positive divisors of degree $g$ and
points on $J(\Sigma)$ as already pointed out in section
\ref{section: identifying data}.

We define $\zeta_D = \bm{\mathcal{A}}(D) + \bm{\mathcal{K}}$ for any
divisor of degree $\text{deg } D = g$ where $\bm{\mathcal{K}}$ is the
vector of Riemann's constants. Since $\zeta_{\gamma'(0,0)} =
\zeta_{\gamma(0,0)} - 2\pi \int_{\infty^-}^{\infty^+} \bm{\omega}$, the
Abel map only enters in \eqref{reconstruction formula for psi} through
the expression $\bm{\mathcal{A}}(P) - \zeta_{\gamma(0,0)}$ and it
follows that the formulae \eqref{reconstruction formula for psi} are
base-point independent: changing the base point of the Abel map
$\bm{\mathcal{A}}$ can be absorbed in a change of the initial
conditions $\gamma(0,0)$ of the solution and so without loss of
generality we choose $P_0 = \infty^+$.
We also note that the vector $\zeta_{\gamma(0,0)}$ is almost
arbitrary, the only requirement being that the function $\theta \left(
\bm{\mathcal{A}}(P) - \zeta_{\gamma(0,0)} \right)$ should not vanish
identically, which is equivalent to the requirement that the divisor
$\gamma(0,0)$ be non-special \cite{Belokolos}.

\item[$\bullet$]
Apart from the holomorphic differentials $\bm{\omega}$ the
expressions in \eqref{reconstruction formula for psi} involve
various other Abelian differentials. Firstly, $dp$ is the normalised
second kind Abelian differential of the quasi-momentum and is
uniquely specified by its pole structure \eqref{weak dp asymptotics at pm
1} at $x = \pm 1$. Secondly, $dq$ is the normalised second kind
Abelian differential of the quasi-energy $q(x)$ uniquely specified
by its poles at $x = \pm 1$. As for the differential $dp$ in
\eqref{weak dp asymptotics at pm 1}, imposing only the first set of
Virasoro constraints \eqref{V1}, the asymptotics of $dq$ read,
\begin{equation} \label{dq asymptotics at pm 1}
\begin{split}
dq(x^{\pm}) = \mp d\left(\frac{\pi \kappa_+}{x - 1}\right) + O\left( (x -
1)^0 \right) \quad \text{as } x \rightarrow +1, \\
dq(x^{\pm}) = \pm d\left(\frac{\pi \kappa_-}{x + 1}\right) + O\left( (x +
1)^0 \right) \quad \text{as } x \rightarrow -1.
\end{split}
\end{equation}
The relation among the pole part of the quasi-energy \eqref{dq
asymptotics at pm 1}, the pole part of the quasi-momentum in
\eqref{weak dp asymptotics at pm 1} and the singular parts $S_{\pm}$ of
the problem in \eqref{Singular parts} is
\begin{equation} \label{dp, dq and dS relation}
\sigma dp + \tau dq \underset{x \rightarrow \pm 1}\sim 2 \pi i
dS_{\pm}.
\end{equation}

The vectors in $\mathbb{C}^g$ of $b$-periods of these Abelian
differentials
\begin{equation*}
\bm{k} = \frac{1}{2\pi} \int_{\bm{b}} dp, \quad
\bm{w} = \frac{1}{2\pi} \int_{\bm{b}} dq
\end{equation*}
being generically non-zero, the Abelian integrals $\int^P dp$ and
$\int^P dq$ define multi-valued functions on
$\Sigma$. Nevertheless it is straightforward to check using the
property \eqref{automorphy theta} of $\theta$-functions that the
formulae in \eqref{reconstruction formula for psi} do define
single-valued functions $\psi_1$ and $\psi_2$ on $\Sigma$: if $P \in
\Sigma$ goes around an arbitrary cycle $\gamma = \sum_{i=1}^g [N_i a_i
+ M_i b_i] \in H_1(\Sigma,\mathbb{Z})$, $\bm{N},\bm{M} \in
\mathbb{Z}^g$ then the expressions in \eqref{reconstruction formula
for psi} are unchanged.

\item[$\bullet$] Since Abelian integrals are multi-valued it is
preferable to work with specific branches of these functions by
cutting up the Riemann surface $\Sigma$ along the homology basis
$a_i,b_i, i = 1, \ldots, g$ to form a simply connected domain
$\Sigma_{\text{cut}}$ called the normal form of $\Sigma$, see Figure
\ref{RS normal form figure}. The Abelian integrals
$\bm{\mathcal{A}}(P)$, $\int^P dp$, and $\int^P dq$ are
now well defined on $\Sigma_{\text{cut}}$ where all the paths of
integration should lie in the interior of $\Sigma_{\text{cut}}$.
\begin{figure}
\centering \psfrag{a1}{\footnotesize $a_1$}
\psfrag{ag}{\footnotesize $a_g$} \psfrag{b1}{\footnotesize $b_1$}
\psfrag{bg}{\footnotesize $b_g$} \psfrag{m}{\footnotesize
$\infty^-$} \psfrag{p}{\footnotesize $\infty^+$}
\psfrag{Q0}{\footnotesize $P$}
\includegraphics[height=50mm,width=50mm]{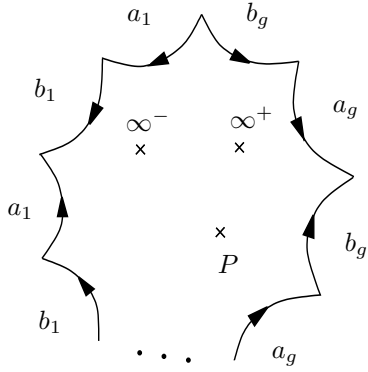}
\caption{Normal form $\Sigma_{\text{cut}}$ of the Riemann surface
$\Sigma$ of genus $g$.} \label{RS normal form figure}
\end{figure}

\item[$\bullet$] At this stage the vector $\bm{\psi}$ is determined
only up to a residual gauge transformation $\bm{\psi} \rightarrow
\tilde{g} \bm{\psi}$ where $\tilde{g} = \text{diag }
(c(\sigma,\tau),d(\sigma,\tau))$, but we will show in the remainder
of this section that no further residual gauge transformation is
actually required for the reconstruction of the current $j$.

\end{itemize}

Since $\gamma(\sigma,\tau) \infty^-$ is the divisor of zeroes of the
first component $\psi_1$ it follows from the formula
\eqref{reconstruction formula for psi_1} for $\psi_1$ and using
Riemann's theorem on the zeroes of the $\theta$-function that
\begin{equation} \label{linear motion on Jac}
\bm{\mathcal{A}}(\gamma(\sigma,\tau)) -
\bm{\mathcal{A}}(\gamma(0,0)) = - \bm{k} \sigma - \bm{w} \tau.
\end{equation}
In other words, the motion of the dynamical divisor
$\gamma(\sigma,\tau)$ is linearised on the Jacobian $J(\Sigma)$ by
the Abel map. Notice that since the right hand side of
\eqref{linear motion on Jac} is uniquely determined by the
singular parts $S_{\pm}(P)$ of the problem, it follows that the
set of algebro-geometric data \eqref{Omega algebro-geometric data}
for $\Omega(x,\sigma,\tau)$ is actually equivalent to the set of
algebro-geometric data \eqref{U,V algebro-geometric data full} for
$J(x,\sigma,\tau)$.

The connection $J(x,\sigma,\tau)$ can now be reconstructed from
the formula \eqref{reconstruction formula for U,V} using the
expressions \eqref{reconstruction formula for psi} for the vector
$\bm{\psi}$ up to a residual gauge \eqref{gauge redundancy} with
diagonal parameter $\tilde{g}(\sigma,\tau)$. In order to obtain
expressions for the components $j_0, j_1$ of the current $j$ we
must first show that the reconstructed connection
$J(x,\sigma,\tau)$ in \eqref{reconstruction formula for U,V} takes
the original form \eqref{Lax connection} for some current $j$. In
section \ref{section: identifying data} the behaviour of the
eigenvector $\bm{\psi}$ near the essential singularities at $x =
\pm 1$ was found to be
\begin{equation*}
\bm{\psi}(x^{\pm}) \underset{x \rightarrow 1}\sim O(1) e^{\mp
\frac{i \kappa_+ \sigma^+}{x - 1}}, \quad \bm{\psi}(x^{\pm})
\underset{x \rightarrow -1}\sim O(1) e^{\pm \frac{i \kappa_-
\sigma^-}{x + 1}},
\end{equation*}
from which we may write the behaviour of the matrix $\Psi(x) =
(\bm{\psi}(x^+), \bm{\psi}(x^-))$ near $x = \pm 1$ as follows
\begin{equation} \label{Psi essential singularity}
\begin{split}
\Psi(x,\sigma,\tau) &= \left( \Psi_0(\sigma,\tau) + \sum_{s =
1}^{\infty} \Psi_s(\sigma,\tau) (x - 1)^s \right) e^{\frac{i
\kappa_+ \sigma^+}{1 - x} \sigma_3} \quad \text{as } x
\rightarrow 1, \\
\Psi(x,\sigma,\tau) &= \left( \Phi_0(\sigma,\tau) + \sum_{s =
1}^{\infty} \Phi_s(\sigma,\tau) (x + 1)^s \right) e^{\frac{i
\kappa_- \sigma^-}{1 + x} \sigma_3}  \quad \text{as } x \rightarrow
-1,
\end{split}
\end{equation}
where $\sigma_3 = \text{diag}(1,-1)$ is the third Pauli matrix and
$\sigma^{\pm}$ are the light-cone coordinates on the string
worldsheet introduced in \eqref{LC coords}. It is straightforward
to derive from these expansions the following asymptotics near $x
= 1$
\begin{equation*}
\left\{
\begin{split}
\left( \partial_+ \Psi \right) \Psi^{-1} &= \frac{i \kappa_+}{1-x}
\left( \Psi_0 \sigma_3 \Psi_0^{-1} \right) + O(1) \\
\left( \partial_- \Psi \right) \Psi^{-1} &= O(1)
\end{split}
\right. \quad \text{as } x \rightarrow 1
\end{equation*}
and likewise near $x = -1$,
\begin{equation*}
\left\{
\begin{split}
\left( \partial_+ \Psi \right) \Psi^{-1} &= O(1) \\
\left( \partial_- \Psi \right) \Psi^{-1} &= \frac{i \kappa_-}{1 + x}
\left( \Phi_0 \sigma_3 \Phi_0^{-1} \right) + O(1)
\end{split}
\right. \quad \text{as } x \rightarrow -1.
\end{equation*}
However we also find from \eqref{reconstruction formula for psi}
that $\Psi = {\bf 1} + O\left( \frac{1}{x} \right)$ as $x
\rightarrow \infty$ due to the special normalisation of the vector
$\bm{\psi}$ defined in \eqref{reconstruction formula for psi}, so
that
\begin{equation*}
\left( \partial_{\pm} \Psi \right) \Psi^{-1} = O\left( \frac{1}{x}
\right) \quad \text{as } x \rightarrow \infty.
\end{equation*}
Thus the above asymptotics at $x = \pm 1, \infty$ take the following
form
\begin{subequations} \label{f_pm essential singularities}
\begin{gather}
\left( \partial_+ \Psi \right) \Psi^{-1} = J_+(x) + O(1), \quad
\left( \partial_- \Psi \right) \Psi^{-1} = J_-(1)
+ O(1) \quad \text{as } x \rightarrow 1 \label{f_pm at p1} \\
\left( \partial_+ \Psi \right) \Psi^{-1} = J_+(-1) + O(1), \quad
\left( \partial_- \Psi \right) \Psi^{-1} = J_-(x) + O(1) \quad
\text{as } x
\rightarrow -1 \label{f_pm at m1} \\
\left( \partial_{\pm} \Psi \right) \Psi^{-1} = J_{\pm}(\infty) +
O\left( \frac{1}{x} \right) \quad \text{as } x \rightarrow \infty,
\label{f_pm at inf}
\end{gather}
\end{subequations}
where the matrices $J_{\pm}(x)$ here have been defined as
\begin{equation} \label{reconstruction formula for j}
\begin{split}
J_+(x) &= \frac{i \kappa_+}{1 - x} \left( \Psi_0 \sigma_3 \Psi_0^{-1} \right), \\
J_-(x) &= \frac{i \kappa_-}{1 + x} \left( \Phi_0 \sigma_3 \Phi_0^{-1}
\right).
\end{split}
\end{equation}
To show that these are in fact the light-cone components $J_0 \pm
J_1$ of the Lax connection $J(x)$ we use a standard argument based
on the uniqueness of the Baker-Akhiezer function (see for instance
\cite{Belokolos}) and consider the following vector-valued
functions on $\Sigma$
\begin{equation*}
\bm{f}_{\pm}(P) = \left( \partial_{\pm} - J_{\pm}(x) \right)
\bm{\psi}(P) = \left[ \left(\partial_{\pm} \Psi(x) \right)
\Psi(x)^{-1} - J_{\pm}(x) \right] \bm{\psi}(P),
\end{equation*}
where $\hat{\pi}(P) = x$. By their definition, on $\Sigma
\setminus \{ x = \pm 1 \}$ the components of the vectors
$\bm{f}_{\pm}(P)$ have exactly the same \textit{constant} poles as
$\bm{\psi}(P)$ at $\hat{\gamma}$ as well as the same
\textit{constant} zeroes as the components of $\bm{\psi}(P)$ at
$\infty^{\pm}$ (see \eqref{psi divisor}) using the fact that
$J_{\pm}(\infty) = 0$. Also from their definition and using the
asymptotics at $x = \pm 1$ in \eqref{f_pm at p1} and \eqref{f_pm
at m1}, these vectors have essential singularities at $x = \pm 1$
of exactly the same form as that of the vector $\bm{\psi}$. These
were the properties which determined the Baker-Akhiezer vector
$\bm{\psi}(P)$ uniquely on $\Sigma$ up to multiplication by a
diagonal matrix independent of $P \in \Sigma$, and thus by the
uniqueness of the Baker-Akhiezer vector the following must be true
\begin{equation*}
\bm{f}_{\pm}(P) = D(\sigma,\tau) \bm{\psi}(P),
\end{equation*}
for some diagonal matrix $D(\sigma,\tau)$ independent of $P \in
\Sigma$. But the asymptotics at $x = \infty$ in \eqref{f_pm
at inf} now show that in fact $D(\sigma,\tau)$
must be zero, so we conclude
\begin{equation*}
\bm{f}_{\pm}(P) \equiv 0.
\end{equation*}
Going back to the definition of these vectors this implies
\begin{equation} \label{reconstructed Lax connection}
J_{\pm}(x) = \left(\partial_{\pm} \Psi(x) \right) \Psi(x)^{-1}.
\end{equation}
In other words the connection defined in \eqref{reconstruction
formula for j} is exactly the reconstructed Lax connection, and so
the latter is indeed of the form \eqref{Lax connection} once we
have completely fixed the residual gauge by imposing the final
condition $J(\infty) = 0$ (obtained from \eqref{connection
asymptotics}). We can now read off from \eqref{reconstruction
formula for j} the reconstruction formula for the light-cone
components of the current $j$
\begin{equation} \label{(pre)reconstruction formula for j components}
\begin{split}
j_+(\sigma,\tau) &= i \kappa_+ \left(\Psi_0 \sigma_3 \Psi_0^{-1} \right), \\
j_-(\sigma,\tau) &= i \kappa_- \left(\Phi_0 \sigma_3 \Phi_0^{-1}
\right),
\end{split}
\end{equation}
where $\Psi_0$ and $\Phi_0$ are the matrices defined in terms of
the reconstructed matrix $\Psi$ through equation \eqref{Psi
essential singularity}. The components of the reconstructed
current \eqref{(pre)reconstruction formula for j components} can
alternatively be written more compactly as follows
\begin{equation} \label{reconstruction formula for j components}
j_{\pm}(\sigma,\tau) = i \kappa_{\pm} \lim_{x \rightarrow \pm 1}
\left(\Psi(x,\sigma,\tau) \sigma_3 \Psi(x,\sigma,\tau)^{-1} \right).
\end{equation}
One can easily check that these reconstructed currents satisfy the
first set of Virasoro constraints \eqref{V1} since $j_{\pm}^2 = -
\kappa_{\pm}^2 {\bf 1}$ so that
\begin{equation*}
\text{tr } j_{\pm}^2 = - \kappa_{\pm}^2 \, \text{tr } {\bf 1} = - 2
\kappa_{\pm}^2.
\end{equation*}
Also, before having imposed any reality conditions on the
algebro-geometric data the reconstructed current
\eqref{reconstruction formula for j components} takes values in
$\mathfrak{sl}(2, \mathbb{C})$ since it is obviously invertible
and traceless,
\begin{equation*}
\text{tr } j_{\pm} = i \kappa_{\pm} \, \text{tr } \sigma_3 = 0.
\end{equation*}
We can write out the components of the reconstructed current
\eqref{reconstruction formula for j components} even more explicit
in terms of $\theta$-functions starting from the expression
\eqref{reconstruction formula for psi} for the matrix of
Baker-Akhiezer vectors $\Psi(x,\sigma,\tau)$ in terms of
$\theta$-functions. The final expression for the light-cone
components of the reconstructed $SU(2)_R$ current $j$ then reads
\begin{equation} \label{explicit reconstruction of j}
j_{\pm}(\sigma,\tau) = e^{\big( \frac{i}{2} \bar{\theta}_0 -
\frac{i}{2} \int_{\infty^-}^{\infty^+} d\mathcal{Q} \big)
\sigma_3} \Theta(\pm 1,\sigma,\tau) \left( i \kappa_{\pm} \sigma_3
\right) \Theta(\pm 1,\sigma,\tau)^{-1} e^{- \big(
\frac{i}{2}\bar{\theta}_0 - \frac{i}{2} \int_{\infty^-}^{\infty^+}
d\mathcal{Q} \big) \sigma_3},
\end{equation}
where the notation in this expression is as follows:
\begin{itemize}
\item[$\bullet$] After reconstructing the current
\eqref{reconstruction formula for j components}, the only part of
the residual gauge transformation that is still unfixed is
conjugation by constant diagonal matrices corresponding precisely
to the $U(1)_R$ subgroup of the physical $SU(2)_R$ symmetry
\eqref{R symmetry on j} of the action that preserves the level set
$Q_R = \frac{1}{2 i} R \sigma_3$ (although at this stage, before
having imposed reality conditions, we are really dealing with a
$\mathbb{C}^{\ast}$ subgroup of an $SL(2,\mathbb{C})_R$
transformation acting on complexified currents $j$). All the other
spurious gauge redundancy introduced by the zero-curvature
formulation \eqref{zero curvature equation} of the equations of
motion has been fixed. This undetermined $\mathbb{C}^{\ast}$
conjugation matrix can be expressed in terms of a single arbitrary
constant $\bar{\theta}_0 \in \mathbb{C}$ as $e^{\frac{i}{2}
\bar{\theta}_0 \sigma_3}$.

\item[$\bullet$] We can combine the Abelian differentials $dp$ and
$dq$ into the following second kind Abelian differential
\begin{equation*}
d\mathcal{Q}(\sigma,\tau) = \frac{1}{2 \pi} \left( \sigma dp +
\tau dq \right),
\end{equation*}
which plays a special role in the dynamics of finite-gap
solutions. Referring to equation \eqref{dp, dq and dS relation},
it can otherwise be defined as the unique normalised second kind
Abelian differential with double poles at $x = \pm 1$ of the form
\begin{equation*}
d \mathcal{Q} \underset{x \rightarrow \pm 1}\sim i dS_{\pm}.
\end{equation*}
Notice that the only $(\sigma,\tau)$-dependence of the matrix
$\Theta(x,\sigma,\tau)$ defined below enters through the
$b$-periods of $d\mathcal{Q}(\sigma,\tau)$ in the expression
\begin{equation*}
\bm{\zeta}_{\gamma(\sigma,\tau)} = \bm{\zeta}_{\gamma(0,0)} -
\int_{\bm{b}} d\mathcal{Q}(\sigma,\tau),
\end{equation*}
and that the quantity entering in the exponents of expression
\eqref{explicit reconstruction of j} for $j_{\pm}$ is just (minus)
the $\mathcal{B}_{g+1}$-period of $d\mathcal{Q}(\sigma,\tau)$,
namely
\begin{equation*}
\int_{\infty^-}^{\infty^+} d\mathcal{Q}(\sigma,\tau) =
-\int_{\mathcal{B}_{g+1}} d\mathcal{Q}(\sigma,\tau).
\end{equation*}

\item[$\bullet$] The reconstruction formulae \eqref{reconstruction
formula for psi} for the matrix $\Psi(x,\sigma,\tau) = \left(
\bm{\psi}(x^+,\sigma,\tau), \bm{\psi}(x^-,\sigma,\tau) \right)$
can be conveniently factored as follows
\begin{equation} \label{Psi factored}
\Psi(x,\sigma,\tau) = e^{\frac{i}{2} \int_{\infty^-}^{\infty^+}
d\mathcal{Q}(\sigma,\tau)} C(\sigma,\tau) \Theta(x,\sigma,\tau)
e^{\Omega_S(x,\sigma,\tau)},
\end{equation}
where $C(\sigma,\tau)$ and $\Omega_S(x,\sigma,\tau)$ are diagonal
matrices defined as
\begin{equation*}
C(\sigma,\tau) = e^{- \frac{i}{2} \int_{\infty^-}^{\infty^+}
d\mathcal{Q}(\sigma,\tau) \sigma_3}, \quad \Omega_S(x,\sigma,\tau)
= \text{diag} \left( i \int_{\infty^+}^{x^+}
d\mathcal{Q}(\sigma,\tau), i \int_{\infty^+}^{x^-}
d\mathcal{Q}(\sigma,\tau) \right),
\end{equation*}
and the matrix $\Theta(x,\sigma,\tau)$ of $\theta$-functions is
defined by
\begin{multline*}
\Theta(x,\sigma,\tau) = \\
\left( \begin{array}{cc} k_-(x^+) \frac{\theta \left(
\bm{\mathcal{A}}(x^+) - \bm{\zeta}_{\gamma(\sigma,\tau)} \right)
\theta \left( \bm{\mathcal{A}}(\infty^+) -
\bm{\zeta}_{\gamma(0,0)} \right)}{\theta \left(
\bm{\mathcal{A}}(x^+) - \bm{\zeta}_{\gamma(0,0)} \right) \theta
\left( \bm{\mathcal{A}}(\infty^+) -
\bm{\zeta}_{\gamma(\sigma,\tau)} \right)} & k_-(x^-) \frac{\theta
\left( \bm{\mathcal{A}}(x^-) - \bm{\zeta}_{\gamma(\sigma,\tau)}
\right) \theta \left( \bm{\mathcal{A}}(\infty^+) -
\bm{\zeta}_{\gamma(0,0)} \right)}{\theta \left(
\bm{\mathcal{A}}(x^-) - \bm{\zeta}_{\gamma(0,0)} \right) \theta
\left( \bm{\mathcal{A}}(\infty^+) -
\bm{\zeta}_{\gamma(\sigma,\tau)}
\right)} \\
k_+(x^+) \frac{\theta \left( \bm{\mathcal{A}}(x^+) -
\bm{\zeta}_{\gamma'(\sigma,\tau)} \right) \theta \left(
\bm{\mathcal{A}}(\infty^-) - \bm{\zeta}_{\gamma'(0,0)}
\right)}{\theta \left( \bm{\mathcal{A}}(x^+) -
\bm{\zeta}_{\gamma'(0,0)} \right) \theta \left(
\bm{\mathcal{A}}(\infty^-) - \bm{\zeta}_{\gamma'(\sigma,\tau)}
\right)} & k_+(x^-) \frac{\theta \left( \bm{\mathcal{A}}(x^-) -
\bm{\zeta}_{\gamma'(\sigma,\tau)} \right) \theta \left(
\bm{\mathcal{A}}(\infty^-) - \bm{\zeta}_{\gamma'(0,0)}
\right)}{\theta \left( \bm{\mathcal{A}}(x^-) -
\bm{\zeta}_{\gamma'(0,0)} \right) \theta \left(
\bm{\mathcal{A}}(\infty^-) - \bm{\zeta}_{\gamma'(\sigma,\tau)}
\right)}
\end{array} \right).
\end{multline*}
Substituting \eqref{Psi factored} into \eqref{reconstruction
formula for j components} yields the required expression for
$j_{\pm}$ expressed in terms of $\theta$-functions through
$\Theta(\pm 1,\sigma,\tau)$.

\end{itemize}

We can now also reconstruct the monodromy matrix corresponding to
the reconstructed current \eqref{reconstruction formula for j
components} that would be obtained by substituting
\eqref{reconstruction formula for j components} directly into the
definition \eqref{monodromy definition} of $\Omega(x)$. For this
we note that the monodromy matrix satisfies the differential
equation \eqref{monodromy evolution} with the Lax connection now
being given by \eqref{reconstructed Lax connection}, i.e.
\begin{equation*}
\left[ \partial_{\pm} - (\partial_{\pm} \Psi(x)) \Psi(x)^{-1}, \Omega(x) \right] = 0.
\end{equation*}
This equation is solved by $\Omega(x) = \Psi(x) \Lambda(x) \Psi(x)^{-1}$
where $\Lambda(x)$ is a constant diagonal matrix, $\partial_{\pm}
\Lambda(x) = 0$. The requirement that $\Omega(x)$ be the monodromy
matrix fixes these integration constants to be the eigenvalues of the
monodromy matrix, namely
\begin{equation*}
\Lambda(x) = \text{diag } (e^{i p(x)}, e^{- i p(x)}),
\end{equation*}
so that the monodromy matrix of the reconstructed current
\eqref{reconstruction formula for j components} reads
\begin{equation*}
\Omega(x) = \Psi(x) \left( \begin{array}{cc} e^{i p(x)} & 0 \\ 0 &
  e^{- i p(x)} \end{array} \right) \Psi(x)^{-1},
\end{equation*}
where $\Psi(x) = (\bm{\psi}(x^+), \bm{\psi}(x^-))$ and $\bm{\psi}$ is
the reconstructed Baker-Akhiezer vector in \eqref{reconstruction
formula for psi}.

The original field $g$ of the $SU(2)$ principal chiral model
defined in \eqref{SU(2) sigma-model field} can also be recovered
now that the current $j = - g^{-1} dg$ is known. It is the unique
solution to
\begin{equation*}
dg + g j = 0,
\end{equation*}
up to left multiplication by an arbitrary constant $U_L \in
SL(2,\mathbb{C})$ matrix. In particular it is given by
\begin{equation} \label{reconstruction of original field g}
g(\sigma,\tau) = U_L \cdot P \overleftarrow{\exp} \int_{(\sigma,\tau)} j,
\end{equation}
where the integral runs from $(\sigma,\tau)$ on the worldsheet to an
arbitrary but fixed point. For the moment though, before having
imposed any reality conditions, this field takes values in
$SL(2,\mathbb{C})$.

We conclude this section with a summary of the results obtained so
far. In section \ref{section: identifying data} we constructed a map
\eqref{U,V algebro-geometric data} associating to an equivalence class
$[J(x,\sigma,\tau)]$ a set of algebro-geometric data $\{ \Sigma, \ dp,
\gamma(0,0)\} \in \mathcal{M}_{\mathbb{C}}^{(2g)}$. This map is
injective, essentially by the Riemann-Roch theorem, and thus admits a
(left) inverse \eqref{left inverse map} which is also injective. For
any class $[J(x,\sigma,\tau)]$, only certain particular
representatives, determined by the requirement $J(\infty) = 0$, are
genuine solutions of the equations of motion. As was shown above, the
condition $J(\infty) = 0$ picks out a family of solutions related by
residual gauge transformations consisting of constant diagonal
matrices $\tilde{g}$. This constant diagonal residual degree of
freedom, corresponding to the $U(1)_R$ symmetric group preserving the
`highest weight' condition, was parametrised by introducing a new
complex variable $W = \exp(i\bar{\theta}_0) \in \mathbb{C}^{\ast}$. So
amending the algebro-geometric data $\mathcal{M}_{\mathbb{C}}^{(2g)}$
with this extra piece of data to form $\mathcal{M}_{\mathbb{C}} =
\mathcal{M}_{\mathbb{C}}^{(2g)} \times \mathbb{C}^{\ast}$ we can
describe the reconstruction of the current \eqref{explicit
reconstruction of j} by an injective map, which we call the
\textit{geometric map},
\begin{equation} \label{geometric map}
\mathcal{G} : \mathcal{M}_{\mathbb{C}} \rightarrow
\mathcal{S}_{\mathbb{C}} \subset \mathcal{M}^{\infty}_{\mathbb{C}},
\end{equation}
from the \textit{moduli space} $\mathcal{M}_{\mathbb{C}} =
\mathcal{M}_{\mathbb{C}}^{(2g)} \times \mathbb{C}^{\ast}$
describing the extended algebro-geometric data $\{ \Sigma, \ dp,
\gamma(0,0), \bar{\theta}_0\}$ into the space
$\mathcal{S}_{\mathbb{C}}$ of complexified solutions $j \in
\mathfrak{sl}(2,\mathbb{C})$ of the equations of motion
\eqref{flatness & conservation} and the Virasoro constraints
\eqref{Virasoro} (after imposing the final Virasoro constraint
$\kappa_+ = \kappa_-$), which is a subspace of the space
$\mathcal{M}^{\infty}_{\mathbb{C}}$ of all pairs $j =
(j_0(\sigma,\tau), j_1(\sigma,\tau))$. The main feature of the
geometric map specified by \eqref{explicit reconstruction of j}
which we wish to emphasise is that it linearises the motion on the
moduli space $\mathcal{M}_{\mathbb{C}}$. In other words, once the
initial data has been specified as a point on
$\mathcal{M}_{\mathbb{C}}$, the subsequent evolution of the system
is described by a linear flow on the corresponding
$J(\Sigma)$-fibre of $\mathcal{M}_{\mathbb{C}}^{(2g)}$ as well as
on the extra $\mathbb{C}^{\ast}$ factor of
$\mathcal{M}_{\mathbb{C}}$\\
\\
\begin{equation} \label{linear motion of system}
\begin{split}
\bm{\theta}(\sigma,\tau) &= \bm{\theta}(0,0) - \bm{k} \sigma - \bm{w} \tau, \\
\bar{\theta}(\sigma,\tau) &= \bar{\theta}_0 - k_{\bar{\theta}} \sigma
- w_{\bar{\theta}} \tau,
\end{split}
\end{equation}
where $\bm{\theta}(\sigma,\tau) =
\bm{\mathcal{A}}(\gamma(\sigma,\tau)) + \bm{\theta}_0$ and $W=\exp(i\bar{\theta})$ is
the coordinate along the $\mathbb{C}^{*}$ factor of
$\mathcal{M}_{\mathbb{C}}$. The vector $\bm{\theta}_0 \in \mathbb{C}^g$
is a constant vector whose specific value will be chosen in section
\ref{section: reality of divisor} to make the reality conditions on
$\bm{\theta}$ simple. The velocities have been identified in the
remarks following \eqref{explicit reconstruction of j} as
being linear combinations of all $K = g + 1$ of the
$\mathcal{B}$-periods of the spectral curve $\Sigma$ defined in
section \ref{section: spectral curve}, specifically
\begin{gather*}
\bm{k} = \frac{1}{2\pi} \int_{\bm{b}} dp, \quad
\bm{w} = \frac{1}{2\pi} \int_{\bm{b}} dq, \\
k_{\bar{\theta}} = -\frac{1}{2 \pi} \int_{\mathcal{B}_{g+1}} dp,
\quad w_{\bar{\theta}} = -\frac{1}{2\pi} \int_{\mathcal{B}_{g+1}}
dq.
\end{gather*}
The first equation in \eqref{linear motion of system} is a rewriting
of \eqref{linear motion on Jac} and the second equation for
$\bar{\theta}$ comes from observing the reconstruction formula
\eqref{explicit reconstruction of j} written explicitly in terms of
all the data.


\subsection{The dual linear system} \label{section: dual}

In section \ref{section: reality of divisor} we will show that the
reconstructed current \eqref{reconstruction formula for j components}
corresponding to real algebro-geometric data is
$\mathfrak{su}(2)$ valued, but in order to do so we will need the
concept of the dual Baker-Akhiezer vector which we now introduce. We
will derive a useful formula for the inverse matrix
$\Psi(x,\sigma,\tau)^{-1}$ appearing in most of the reconstruction
formulae, in particular \eqref{reconstruction formula for j
components}, expressing it as a matrix of row vectors
$\left(\bm{\psi}^+(x^+)^{\text{T}},\bm{\psi}^+(x^-)^{\text{T}}
\right)^{\text{T}}$ where the row-vector $\bm{\psi}^+$ is the
\textit{dual Baker-Akhiezer vector} obeying
\begin{equation} \label{dual vector ortho condition}
\bm{\psi}^+(x^{\pm}) \cdot \bm{\psi}(x^{\pm}) = 1, \quad
\bm{\psi}^+(x^{\pm}) \cdot \bm{\psi}(x^{\mp}) = 0,
\end{equation}
which can be constructed as follows \cite{Babelon}: consider a
meromorphic differential $\widetilde{\Omega}$ with double poles at
$\infty^{\pm}$ and $g+1$ specified zeroes at $\hat{\gamma}$. The set
of $g+1$ remaining zeroes of $\widetilde{\Omega}$ define the dual
divisor $\hat{\gamma}^+$ up to equivalence; indeed the differential
$\widetilde{\Omega}$ is not unique, but the divisors
$\hat{\gamma}^+, \hat{\gamma}'^+$ defined from two such
differentials $\widetilde{\Omega}, \widetilde{\Omega}'$ are
equivalent $\hat{\gamma}^+ \sim \hat{\gamma}'^+$. Then the dual
Baker-Akhiezer vector $\bm{\psi}^+$ is defined as
\begin{equation*}
\psi_i^+(P) = \chi(P) \widetilde{\psi}_i^+(P), \quad \text{with } \chi(P) =
\frac{\widetilde{\Omega}(P)}{dx},
\end{equation*}
where $\widetilde{\Omega}$ is normalised by $\chi(\infty^+) = 1$ say,
and where the components of the vector $\widetilde{\bm{\psi}}^+$ are
Baker-Akhiezer functions specified by their respective divisors
\begin{equation*}
(\widetilde{\psi}^+_1) \geq \left(\hat{\gamma}^+\right)^{-1} \infty^-, \quad
(\widetilde{\psi}^+_2) \geq \left(\hat{\gamma}^+\right)^{-1} \infty^+,
\end{equation*}
as well as singular parts this time of opposite sign $-S_{\pm}$ at
$x = \pm 1$ and are normalised such that
\begin{equation} \label{dual BA normalisation}
\psi_1^+(\infty^+) = \psi_2^+(\infty^-) = 1.
\end{equation}
So just as the components of $\bm{\psi}$ were constructed in
\eqref{reconstruction formula for psi}, those of
$\widetilde{\bm{\psi}}^+$ can be constructed explicitly using a
reconstruction formula analogous to \eqref{reconstruction formula
for psi} but with the replacements $\hat{\gamma} \rightarrow
\hat{\gamma}^+$, $k_{\pm} \rightarrow h_{\pm}$ where $(h_-) = \gamma^+(0,0)
\infty^- (\hat{\gamma}^+)^{-1}$ and $(h_+) = \gamma'^+(0,0)
\infty^+ (\hat{\gamma}^+)^{-1}$, as well as $dp \rightarrow -dp$,
$dq \rightarrow -dq$ (and hence also $\bm{k} \rightarrow -\bm{k}$,
$\bm{w} \rightarrow -\bm{w}$), namely we have
\begin{subequations} \label{reconstruction formula for psi^+}
\begin{multline} \label{reconstruction formula for psi^+_1}
\widetilde{\psi}^+_1(P,\sigma,\tau) = h_-(P) \frac{\theta \left(
\bm{\mathcal{A}}(P) - \bm{k} \sigma - \bm{w} \tau -
\bm{\zeta}_{\gamma^+(0,0)} \right) \theta \left(
\bm{\mathcal{A}}(\infty^+) - \bm{\zeta}_{\gamma^+(0,0)}
\right)}{\theta \left( \bm{\mathcal{A}}(P) -
\bm{\zeta}_{\gamma^+(0,0)} \right) \theta \left(
\bm{\mathcal{A}}(\infty^+) - \bm{k} \sigma - \bm{w} \tau -
\bm{\zeta}_{\gamma^+(0,0)} \right)} \\ \times \exp \left( - \frac{i
\sigma}{2 \pi} \int_{\infty^+}^P dp - \frac{i \tau}{2 \pi}
\int_{\infty^+}^P dq \right),
\end{multline}
\begin{multline} \label{reconstruction formula for psi^+_2}
\widetilde{\psi}^+_2(P,\sigma,\tau) = h_+(P) \frac{\theta \left(
\bm{\mathcal{A}}(P) - \bm{k} \sigma - \bm{w} \tau -
\bm{\zeta}_{\gamma'^+(0,0)} \right) \theta \left(
\bm{\mathcal{A}}(\infty^-) - \bm{\zeta}_{\gamma'^+(0,0)}
\right)}{\theta \left( \bm{\mathcal{A}}(P) -
\bm{\zeta}_{\gamma'^+(0,0)} \right) \theta \left(
\bm{\mathcal{A}}(\infty^-) - \bm{k} \sigma - \bm{w}
\tau - \bm{\zeta}_{\gamma'^+(0,0)} \right)} \\
\times \exp \left( - \frac{i \sigma}{2 \pi} \int_{\infty^-}^P dp -
\frac{i \tau}{2 \pi} \int_{\infty^-}^P dq \right),
\end{multline}
\end{subequations}
where $\gamma'^+(0,0) \infty^+ \sim \gamma^+(0,0) \infty^- \sim
\hat{\gamma}^+$. That the vector $\bm{\psi}^+$ thus constructed
indeed satisfies the orthogonality condition \eqref{dual vector
ortho condition} and hence can be used to express the inverse matrix
$\Psi(x,\sigma,\tau)^{-1} =
\left(\bm{\psi}^+(x^+)^{\text{T}},\bm{\psi}^+(x^-)^{\text{T}}
\right)^{\text{T}}$ is shown for example in \cite{Babelon}. Using
the orthogonality relation \eqref{dual vector ortho condition} it
is straightforward to show that $\bm{\psi}^+(P)$ is a left row
eigenvector of the monodromy matrix $\Omega(x,\sigma,\tau)$ with the
same eigenvalue as the right column eigenvector $\bm{\psi}(P)$ and
furthermore that it is a solution to the \textit{dual auxiliary linear
system}, namely (c.f. the auxiliary linear system \eqref{auxiliary
linear system})
\begin{equation*}
d \bm{\psi}^+ + \bm{\psi}^+ J(x,\sigma,\tau) = 0.
\end{equation*}

Defining the \textit{dual dynamical divisor}
$\gamma^+(\sigma,\tau)$ such that $\gamma^+(\sigma,\tau) \infty^-$
is the divisor of zeroes of $\widetilde{\psi}_1^+$, then for the
same reasons that lead to \eqref{linear motion on Jac}, it follows
from the analogue \eqref{reconstruction formula for psi^+_1} of
equation \eqref{reconstruction formula for psi_1} for the first
component of the un-normalised dual Baker-Akhiezer vector
$\widetilde{\bm{\psi}}^+$ that the motion of the dual divisor is
also linearised on $J(\Sigma)$ by the Abel map
\begin{equation} \label{linear motion on Jac for dual}
\bm{\mathcal{A}}(\gamma^+(\sigma,\tau)) -
\bm{\mathcal{A}}(\gamma^+(0,0)) = \bm{k} \sigma + \bm{w} \tau,
\end{equation}
but the motion is in the opposite direction to that of the dynamical
divisor $\gamma(\sigma,\tau)$ in \eqref{linear motion on Jac}.

\subsection{Singular points and the dynamical divisor} \label{section:
sklyanin}

Following a general procedure developed by Sklyanin
(\cite{Sklyanin:1995bm} and references therein) we show that the
points of the dynamical divisor $\gamma(\sigma,\tau)$ along with
the set of singular points $\{ x_k \}$ of $\Gamma$ can be
conveniently characterised as the zeroes of the particular
component $\mathcal{B}(x)$ of the monodromy matrix
\begin{equation} \label{Omega components}
\Omega(x) = \left(
\begin{array}{cc}
\mathcal{A}(x) & \mathcal{B}(x) \\
\mathcal{C}(x) & \mathcal{D}(x)
\end{array}
\right).
\end{equation}
To see this note that provided $x$ is not a branch point, so that
both eigenvectors are linearly independent, the monodromy matrix
can be diagonalised by its matrix of (normalised)
eigenvectors\footnote{This is simply a rewriting of equation
\eqref{monodromy diagonalised}, with $u(x,\sigma,\tau) =
D(x,\sigma,\tau) \left(
\begin{array}{cc}
1 & 1 \\
h_2(x^+) & h_2(x^-)
\end{array}
\right)^{-1}$ where $D(x,\sigma,\tau)$ is a diagonal matrix whose
effect is simply to change the norms of the individual
eigenvectors.}
\begin{equation} \label{Omega}
\Omega(x) = \left(
\begin{array}{cc}
1 & 1 \\
h_2(x^+) & h_2(x^-)
\end{array}
\right) \left(
\begin{array}{cc}
\Lambda(x^+) & 0 \\
0 & \Lambda(x^-)
\end{array}
\right) \left(
\begin{array}{cc}
1 & 1 \\
h_2(x^+) & h_2(x^-)
\end{array}
\right)^{-1}
\end{equation}
where $\Lambda(Q) = e^{i p(Q)} = e^{i \int^Q_{\infty^+} dp}$ is
the function on $\Sigma$ which at $x^+,x^-$ gives the two
eigenvalues $e^{i p(x)}, e^{-i p(x)}$ of $\Omega(x)$ since
\begin{equation*}
\Lambda(x^+) = e^{i \int^{x^+}_{\infty^+} dp} = e^{i p(x)},
\end{equation*}
\begin{equation*}
\Lambda(x^-) = e^{i \int^{x^-}_{\infty^+} dp} = e^{- i
\int^{x^+}_{\infty^+} dp} = e^{- i p(x)},
\end{equation*}
where we have used the relation $\int^{x^+}_{\infty^+} dp +
\int^{\hat{\sigma} x^+}_{\infty^+} dp \in 2 \pi \mathbb{Z}$ along
with $\hat{\sigma} x^+ = x^-$. Now because the function $h_2(P)$ is
meromorphic on the algebraic curve $\Sigma : y^2 = \prod_{i = 1}^{2
g + 2} (x - x_i)$ where the $\{x_i\}_{i=1}^{2g +2} = \{
u_I,v_I\}_{I=1}^{g+1}$ denote the branch points, it can be written
as a rational function in $x$ and $y$. Its pole structure implies
the following form
\begin{equation*}
h_2(x^{\pm}) = \frac{\pm y_+ + h(x)}{\tilde{B}(x)},
\end{equation*}
where $\tilde{B}(x) = C \prod_{i=1}^g (x - x_{\gamma_i})$ with
$x_{\gamma_i} = \hat{\pi}(\gamma_i)$ and $C = C(\sigma,\tau)$,
$y_+ = y(x^+)$ is the value of $y$ on the physical sheet and
$h(x)$ is a polynomial of degree $g + 1$.

The function $\mathcal{B}(x)$ can now be found from \eqref{Omega}
to be
\begin{equation*}
\mathcal{B}(x) = \left( \Lambda(x^+) - \Lambda(x^-) \right)
\frac{\tilde{B}(x)}{2 y_+}.
\end{equation*}
But the function $\Delta(x) = \left( \Lambda(x^+) - \Lambda(x^-)
\right) ^2 = \left( e^{i p(x)} - e^{-i p(x)} \right)^2$ is the
discriminant of the quadratic polynomial $\Gamma(x,y)$ in $y$
which has simple zeroes at the branch points, double zeroes at the
singular points of the curve $\Gamma$ and no other zeroes. Thus we
see that the function $\mathcal{B}(x)$ has simple zeroes at the
dynamical divisor and at the singular points of $\Gamma$. We shall
therefore denote the singular points more naturally by $\gamma_i,
i = -\infty, \ldots, 0$. The relation between singular points and
points of the dynamical divisor is even stronger, as we show in
Appendix \ref{section: motion of divisor}, since the singular
points can be thought of as `trapped points' of the dynamical
divisor in the sense that if we increase the genus of the curve
$\Sigma$ by blowing up certain singular points into genuine
handles, then the number of points of the dynamical divisor
increases by the same amount, as if the singular points in
question had become dynamical.

When the zero $x_{\gamma_i}$ of $\mathcal{B}(x)$ corresponds to a
point $\gamma_i \in \Sigma$ of the dynamical divisor then using
equation \eqref{Omega} we easily obtain the following asymptotics
for the monodromy matrix (note that $\Lambda(P^+) =
\Lambda(P^-)^{-1}$)
\begin{equation*}
\Omega(x) \underset{x \rightarrow x_{\gamma_i}}\longrightarrow
\left( \begin{array}{cc} \Lambda(\gamma_i)^{-1} & 0\\
\star & \Lambda(\gamma_i) \end{array} \right),
\end{equation*}
from which it follows that $\mathcal{A}(x_{\gamma_i}) =
\Lambda(\gamma_i)^{-1}$. This result also holds when the zero
$x_{\gamma_i}$ of $\mathcal{B}(x)$ corresponds to a singular point
of $\Gamma$ since in this case we have $\mathcal{A}(x_{\gamma_i})
= \Lambda(\gamma_i)^{-1} = \Lambda(\gamma_i)$.

Therefore the set of points $\gamma_i \in \Gamma, i = -\infty,
\ldots, g$ comprising of the points $\gamma_i, i = 1, \ldots, g$
of the dynamical divisor and the singular points $\gamma_i, i =
-\infty, \ldots, 0$ is uniquely characterised by the following
conditions in terms of the two components $\mathcal{A}(x)$ and
$\mathcal{B}(x)$ of $\Omega(x)$
\begin{equation} \label{A and B functions}
\mathcal{B}(x_{\gamma_i}) = 0, \qquad \Lambda(\gamma_i)^{-1} =
\mathcal{A}(x_{\gamma_i}).
\end{equation}

By the exact same reasoning one can also argue from
\begin{equation*}
\Omega(x) = \left(
\begin{array}{cc}
1 & k_2(x^+) \\
1 & k_2(x^-)
\end{array}
\right)^{-1} \left(
\begin{array}{cc}
\Lambda(x^+) & 0 \\
0 & \Lambda(x^-)
\end{array}
\right) \left(
\begin{array}{cc}
1 & k_2(x^+) \\
1 & k_2(x^-)
\end{array}
\right),
\end{equation*}
where $k_2(P) = \psi_2^+(P) / \psi_1^+(P)$ has divisor $(k_2) \geq
\left( \gamma^+(\sigma,\tau) \infty^- \right)^{-1} \infty^+$, that
the set of points $\gamma_i^+, i = 1, \ldots, g$ of the dual
divisor $\gamma^+(\sigma,\tau)$ along with the infinite set of
singular points $\gamma_i^+ \equiv \gamma_i, i = -\infty, \ldots,
0$ of $\Gamma$ can be uniquely characterised by
\begin{equation} \label{C and D functions}
\mathcal{C}(x_{\gamma_i^+}) = 0, \qquad \Lambda(\gamma_i^+) =
\mathcal{D}(x_{\gamma_i^+}).
\end{equation}

Besides restricting the moduli of the algebraic curve, the reality
condition \eqref{monodromy reality condition} also imposes
constraints on the dynamical divisor $\gamma(\sigma,\tau)$.
The reality condition \eqref{monodromy reality condition} on
$\Omega(x)$ reads in components
\begin{equation*}
\mathcal{D}(\bar{x}) = \overline{\mathcal{A}(x)}, \quad
\mathcal{C}(\bar{x}) = - \overline{\mathcal{B}(x)}.
\end{equation*}
Using these relations the equation \eqref{A and B functions}
characterising the dynamical divisor (and singular points) can be
rewritten as
\begin{equation*} \mathcal{C}(x_{\hat{\tau} \gamma_i}) = 0,
\qquad \Lambda(\hat{\tau} \gamma_i) = \mathcal{D}(x_{\hat{\tau}
\gamma_i}), \quad i = -\infty,\ldots,g
\end{equation*}
where we have used the fact that the anti-holomorphic involution
$\hat{\tau}$ defined in \eqref{anti-holomorphic involution} maps the
point $\gamma_i = (x_{\gamma_i}, \Lambda(\gamma_i))$ to $\hat{\tau}
\gamma_i = (x_{\hat{\tau} \gamma_i}, \Lambda(\hat{\tau} \gamma_i)) =
\left(\bar{x}_{\gamma_i}, \overline{\Lambda(\gamma_i)}^{\;
-1}\right)$. But this last equation expresses the fact that the
image $\hat{\tau} \gamma(\sigma,\tau)$ of the dynamical divisor along
with the images $\hat{\tau} \gamma_i, i = -\infty,\ldots,0$ of the
singular points are precisely the dual dynamical divisor and the
singular points themselves by comparison with \eqref{C and D
functions}, i.e.
\begin{equation} \label{reality of divisor}
\hat{\tau} \gamma(\sigma,\tau) = \gamma^+(\sigma,\tau), \qquad \hat{\tau}
\gamma_i = \gamma_i, \quad i = -\infty,\ldots,0.
\end{equation}
Thus we conclude that the singular points all lie along the real
axis, which as we already know accumulate at $x = \pm 1$, and the
dual dynamical divisor $\gamma^+(\sigma,\tau)$ is nothing but the
reflection $\hat{\tau} \gamma(\sigma,\tau)$ of the dynamical divisor
$\gamma(\sigma,\tau)$ through the real axis.

\subsection{Reality conditions} \label{section: reality of divisor}

We can describe the reality conditions \eqref{reality of divisor} on
the dynamical divisor more explicitly. By definition of the dual
dynamical divisor we know that $\gamma(0,0) \cdot \gamma^+(0,0) \sim Z
\cdot (\infty^+)^2$ where $Z = (\widetilde{\Omega})$ is the canonical
class and so
\begin{equation*}
\gamma(\sigma,\tau) \cdot \left(\hat{\tau}\gamma(\sigma,\tau)\right)
\sim \gamma(0,0) \cdot \left(\hat{\tau}\gamma(0,0)\right) \sim
\hat{\gamma} \cdot \left(\hat{\tau} \hat{\gamma}\right) \cdot (\infty^-)^{-2}
\sim Z \cdot (\infty^+)^2,
\end{equation*}
where the first equivalence follows from the fact that $k_-^{-1}
h_-^{-1} \psi_1 \widetilde{\psi}_1^+$ is meromorphic or equivalently
from equations \eqref{linear motion on Jac} and \eqref{linear motion
on Jac for dual} which together imply
\begin{equation} \label{real divisor on Jac}
\bm{\mathcal{A}}(\gamma(\sigma,\tau)) +
\bm{\mathcal{A}}(\hat{\tau}\gamma(\sigma,\tau)) =
\bm{\mathcal{A}}(\gamma(0,0)) +
\bm{\mathcal{A}}(\hat{\tau}\gamma(0,0)),
\end{equation}
and then invoking Abel's theorem. Now to obtain the induced action
of the anti-holomorphic involution $\hat{\tau}$ on the Jacobian
$J(\Sigma)$ consider a positive divisor $D = P_1 \ldots P_g$ of
degree $\text{deg } D = g$, then
\begin{equation*}
\bm{\mathcal{A}}(\hat{\tau} D) = 2\pi \sum_{i=1}^g
\int_{\infty^+}^{\hat{\tau} P_i} \bm{\omega} = 2\pi \sum_{i=1}^g
\int_{\infty^+}^{P_i} \hat{\tau}^{\ast} \bm{\omega} = - 2\pi
\sum_{i=1}^g \int_{\infty^+}^{P_i} \overline{\bm{\omega}} = -
\overline{\bm{\mathcal{A}}(D)}.
\end{equation*}
It follows then from \eqref{real divisor on Jac} that the dynamical
divisor $\gamma(\sigma,\tau)$ corresponding to real solutions evolves
such that
\begin{equation*}
2 \, \text{Im } \bm{\mathcal{A}}(\gamma(\sigma,\tau)) = 2 \,
\text{Im } \bm{\mathcal{A}}(\gamma(0,0)) = \bm{\mathcal{A}}(Z
\cdot (\infty^+)^2).
\end{equation*}
In other words, the dynamical divisor $\gamma(\sigma,\tau)$
corresponding to real solutions is constrained to move on the real
$g$-dimensional sub-torus $T^g$ of the complex $g$-dimensional
Jacobian $J(\Sigma)$ defined by
\begin{equation} \label{real slice of Jac}
T^g = \left\{ \bm{X} \in J(\Sigma) \quad | \quad 2 \, \text{Im }
\bm{X} = \bm{\mathcal{A}}(Z \cdot (\infty^+)^2) \right\} \subset
J(\Sigma).
\end{equation}

The reality of the singular parts at $x = \pm 1$ of both $dp$ and
$dq$ in \eqref{weak dp asymptotics at pm 1} and \eqref{dq
asymptotics at pm 1} following from the reality condition on the
quasi-momentum \eqref{reality of p} implies that (using also the
fact that the local parameter $x$ is such that $\hat{\tau}^{\ast}
x = \bar{x}$, by definition of $\hat{\tau}$)
\begin{equation*}
\overline{\hat{\tau}^{\ast} dp} = dp, \qquad
\overline{\hat{\tau}^{\ast} dq} = dq,
\end{equation*}
because a normalised (vanishing $a$-periods) Abelian differential
is uniquely specified by its singular parts. It follows from this
and \eqref{a,b cycles reality} that $\bm{k},\bm{w} \in
\mathbb{R}^g$ since
\begin{equation} \label{reality of U}
\begin{split}
\overline{k_i} &= \frac{1}{2\pi} \int_{b_i} \overline{dp} =
\frac{1}{2\pi} \int_{\hat{\tau} b_i} \overline{\hat{\tau}^{\ast}
dp} = \frac{1}{2\pi} \int_{b_i} dp
= k_i, \\
\overline{w_i} &= \frac{1}{2\pi} \int_{b_i} \overline{dq} =
\frac{1}{2\pi} \int_{\hat{\tau} b_i} \overline{\hat{\tau}^{\ast}
dq} = \frac{1}{2\pi} \int_{b_i} dq = w_i.
\end{split}
\end{equation}
Therefore we find again that once the initial divisor $\gamma(0,0)$, which
is part of the algebro-geometric data, is chosen to live on $T^g$, the
dynamical divisor $\gamma(\sigma,\tau)$ is constrained to move on the
$g$-dimensional real sub-torus $T^g \subset J(\Sigma)$. Thus the
restriction on the algebro-geometric data corresponding to real
solutions can be summarised as
\begin{equation} \label{real subtorus of Jac}
\bm{\mathcal{A}} \left(\gamma(0,0)\right) \in T^g.
\end{equation}
Moreover, the reality condition $j_{\pm}^{\dag} = - j_{\pm}$ has
the effect of reducing the $\mathbb{C}^{\ast}$ action on the space
of complexified solutions $\mathcal{S}_{\mathbb{C}}$ discussed at
the end of section \ref{section: reconstruction} to a $U(1)_R$
action on the space of real solutions $\mathcal{S}_{\mathbb{R}}$.
This yields the simple reality condition $\bar{\theta}_0 \in
\mathbb{R}/2 \pi \mathbb{Z}$ on the $\bar{\theta}_0$ component of
the extended algebro-geometric data. Furthermore, the
$g$-dimensional real sub-torus \eqref{real subtorus of Jac} of
$J(\Sigma)$ can be parametrised by the $g$ component vector
$\bm{\theta} = (\theta_1, \ldots, \theta_g)$ defined for every
point $\bm{X} \in T^g$ by
\begin{equation*}
\bm{\theta} = \bm{X} + \bm{\theta}_0,
\end{equation*}
which can be made \textit{real} provided we choose $-
\bm{\theta}_0 \in T^g$, so we set for instance
\begin{equation} \label{reality of theta}
\bm{\theta}_0 = - \frac{1}{2} \bm{\mathcal{A}}(Z \cdot
(\infty^+)^2).
\end{equation}
Through these coordinates the sub-torus $T^g \subset J(\Sigma)$ is
identified with the real torus $\mathbb{R}^g/2 \pi \mathbb{Z}^g$
which $\bm{\theta}$ takes values in.

Conversely we now show that if the algebro-geomtric data is
\textit{real} so that the initial divisor satisfies $\hat{\tau}
\gamma(0,0) = \gamma^+(0,0)$ then the current $j$ given by the
reconstruction formula \eqref{reconstruction formula for j
components} is real, i.e. $j^{\dag} = -j$. Since $\hat{\tau}
\gamma(0,0) = \gamma^+(0,0)$ we have $\hat{\tau} \hat{\gamma} \sim
\hat{\gamma}^+$ and so we choose the form $\widetilde{\Omega}$
such that equality actually holds, namely $\hat{\tau} \hat{\gamma}
= \hat{\gamma}^+$. Now consider the functions
\begin{equation*}
f_i(P,\sigma,\tau) = \widetilde{\psi}_i^+(P,\sigma,\tau) \Big/
\overline{\psi_i(\hat{\tau} P,\sigma,\tau)}.
\end{equation*}
These are meromorphic functions with at most $g$ poles and hence
are constant by the Riemann-Roch theorem. These constants are all
fixed by the normalisation conditions \eqref{dual BA
normalisation} to be equal to $1$ since $f_i(\infty^+,\sigma,\tau)
= 1$ and so we obtain the following reality condition on the
Baker-Akhiezer vector
\begin{equation} \label{BA reality}
\bm{\psi}(\hat{\tau} P,\sigma,\tau)^{\dag} =
\widetilde{\bm{\psi}}^+(P,\sigma,\tau).
\end{equation}
Now using the dual Baker-Akhiezer vector to construct the inverse
matrix $\Psi(x,\sigma,\tau)^{-1}$ it can be written as follows
\begin{equation*}
\Psi(x,\sigma,\tau)^{-1} =
\left(\bm{\psi}^+(x^+)^{\text{T}},\bm{\psi}^+(x^-)^{\text{T}}
\right)^{\text{T}} = \text{diag} \left( \chi(x^+), \chi(x^-) \right)
\left(\widetilde{\bm{\psi}}^+(x^+)^{\text{T}},
\widetilde{\bm{\psi}}^+(x^-)^{\text{T}} \right)^{\text{T}},
\end{equation*}
where $\chi(P) = \frac{\widetilde{\Omega}(P)}{dx}$ is meromorphic
with zeroes at $\hat{\gamma}$ and $\hat{\tau} \hat{\gamma}$ and
poles at the $2g + 2$ branch points which we denote by the divisor
$B$, i.e.
\begin{equation*}
(\chi) = \hat{\gamma} \cdot \left( \hat{\tau} \hat{\gamma} \right)
\cdot B^{-1}.
\end{equation*}
Using the reality condition on the algebraic curve $\Sigma$ we can
split the divisor $B$ of branch points into two divisors $B_+,B_-$
such that
\begin{equation*}
B = B_+ \cdot B_-, \quad B_+ = \hat{\tau} B_-.
\end{equation*}
Thus we can write the meromorphic function $\chi$ as a product
$\chi = \chi_+ \chi_-$ of two meromorphic functions $\chi_{\pm}$
with respective divisors
\begin{equation*}
(\chi_+) = \hat{\gamma} \cdot B_+^{-1}, \quad (\chi_-) =
\hat{\gamma}^+ \cdot B_-^{-1},
\end{equation*}
and which are therefore related by $\chi_+(P) =
\overline{\chi_-(\hat{\tau} P)}$. The reconstructed current
\eqref{reconstruction formula for j components} now takes the form
\begin{equation} \label{j reconstruction reality}
\begin{split}
j_+(\sigma,\tau) &= i \kappa_+ \left(\Psi_0 \text{ diag} \left(
\chi_+(1^+), \chi_+(1^-) \right) \sigma_3 \text{ diag} \left(
\chi_-(1^+), \chi_-(1^-) \right) \widetilde{\Psi}_0^+ \right), \\
j_-(\sigma,\tau) &= i \kappa_- \left(\Phi_0 \text{ diag} \left(
\chi_+(1^+), \chi_+(1^-) \right)\sigma_3 \text{ diag} \left(
\chi_-(1^+), \chi_-(1^-) \right) \widetilde{\Phi}_0^+
\right),
\end{split}
\end{equation}
where $\widetilde{\Psi}_0^+, \widetilde{\Phi}_0^+$ are the leading
terms in the expansion of $\left(
\widetilde{\bm{\psi}}^+(x^+)^{\text{T}},
\widetilde{\bm{\psi}}^+(x^-)^{\text{T}} \right)^{\text{T}}$ near
$x = \pm 1$ just as $\Psi_0, \Phi_0$ were the leading terms for
$\Psi(x,\sigma,\tau)$ in the expansion \eqref{Psi essential
singularity}. Using the reality condition \eqref{BA reality} on
the Baker-Akiezer vector we find $\widetilde{\Psi}_0^+ =
\Psi_0^{\dag}, \widetilde{\Phi}_0^+ = \Phi_0^{\dag}$ and since
also
\begin{equation*}
\text{diag} \left(\chi_-(1^+), \chi_-(1^-)\right) = \text{diag}
\left(\chi_+(1^+), \chi_+(1^-)\right)^{\dag},
\end{equation*}
it follows from the reconstruction formulae written in the form
\eqref{j reconstruction reality} that the reconstructed currents
are anti-hermitian $j_{\pm}^{\dag} = -j_{\pm}$, i.e. $j_{\pm} \in
\mathfrak{su}(2)$. Now with the extra reality condition
$\bar{\theta}_0 \in \mathbb{R}/2 \pi \mathbb{Z} = S^1$ on the
extension of the algebro-geometric data, the reconstruction formula
\eqref{reconstruction formula for j components} is also easily checked
to give an anti-hermitian current so that $j_{\pm} \in
\mathfrak{su}(2)$.
It therefore follows that as a result of imposing the reality conditions, the
reconstructed field $g$ in \eqref{reconstruction of original field
g} becomes $SU(2)$-valued, and in particular, the original fields
$X_1, \ldots,  X_4$ describing the embedding of the string into
the $S^3 \subset \mathbb{R}^4$ part of the target space are real
valued as required.

To summarise the discussion of reality conditions, the
\textit{real} algebro-geometric data which gives rise to real
solutions through the geometric map \eqref{geometric map} can be
identified with a sub-bundle $\mathcal{M}_{\mathbb{R}} =
\mathcal{M}_{\mathbb{R}}^{(2g)} \times S^1$ of the extended bundle
$\mathcal{M}_{\mathbb{C}} = \mathcal{M}_{\mathbb{C}}^{(2g)} \times
\mathbb{C}^{\ast}$ introduced in section \ref{section: reconstruction}, namely
\begin{equation} \label{real extended bundle}
T^g \rightarrow \mathcal{M}_{\mathbb{R}}^{(2g)}
\rightarrow \mathcal{L}_{\mathbb{R}} \, ,
\end{equation}
where $\mathcal{L}_{\mathbb{R}}$ is the real part of the leaf
$\mathcal{L}$ parametrised by real values of the filling fractions
\eqref{filling fractions}. The restriction
$\mathcal{G}_{\mathbb{R}} = \mathcal{G}
|_{\mathcal{M}_{\mathbb{R}}}$ of the geometric map
\eqref{geometric map} to the real bundle $\mathcal{M}_{\mathbb{R}}
\subset \mathcal{M}_{\mathbb{C}}$ is an injective map from
\textit{real} algebro-geometric data to \textit{real} solutions in
$\mathcal{S}_{\mathbb{R}}$,
\begin{equation*}
\xymatrix{ \mathcal{M}_{\mathbb{C}} \ar[rr]^{\mathcal{G}} & &
\mathcal{S}_{\mathbb{C}} \subset \mathcal{M}^{\infty}_{\mathbb{C}} \\
\mathcal{M}_{\mathbb{R}} \ar@{^{(}->}[u]
\ar[rr]^{\mathcal{G_{\mathbb{R}}}} & & \mathcal{S}_{\mathbb{R}}
\subset \mathcal{M}^{\infty}_{\mathbb{R}} \ar@{^{(}->}[u] }
\end{equation*}

\subsection{Periodicity conditions}

Since the configuration of a finite-gap solution is specified by the
position of the point $\mathcal{A}(\gamma(\sigma,\tau)) \in
J(\Sigma)$ on the Jacobian of $\Sigma$, a necessary condition for
the solution to be periodic is that the motion of this point be
periodic on $J(\Sigma)$. But by \eqref{linear motion on Jac} we know
that $\mathcal{A}(\gamma(\sigma,\tau))$ moves linearly on
$J(\Sigma)$ in both $\sigma$ and $\tau$ so that periodic motion
under $\sigma \rightarrow \sigma + 2\pi$ or $\tau \rightarrow \tau +
T$ can occur respectively only when the following conditions are met
\begin{equation} \label{periodicity conditions}
\begin{split}
2 \pi \bm{k} &= 2\pi \bm{n} + 2\pi \Pi \bm{m}, \qquad
\bm{n},\bm{m} \in \mathbb{Z}^g, \\
T \bm{w} &= 2\pi \bm{n}' + 2\pi \Pi \bm{m}', \qquad \bm{n}',\bm{m}'
\in \mathbb{Z}^g.
\end{split}
\end{equation}
Combining these periodicity conditions with the reality conditions
\eqref{reality of U} on the vectors $\bm{k}$ and
$\bm{w}$ as well as the fact that the components of $\Pi$ are
pure imaginary we see that the motion is periodic under $\sigma
\rightarrow \sigma + 2\pi$ and $\tau \rightarrow \tau + T$
respectively when
\begin{equation} \label{periodicity conditions 1}
\begin{split}
\bm{k} \in \mathbb{Z}^g, \\
\frac{T}{2\pi} \bm{w} \in \mathbb{Z}^g.
\end{split}
\end{equation}
Note that the $\sigma$-periodicity condition is precisely
equivalent to the integrality of the $b$-period of the
differential $dp$ in \eqref{AnB periods of dp}.

However these conditions are not sufficient. If we look at the
explicit reconstruction formula \eqref{explicit reconstruction of
j} for $j_{\pm}$ we find that the periodicity in $\sigma
\rightarrow \sigma + 2 \pi$ and $\tau \rightarrow \tau + T$ of the
similarity transformation corresponding to the $S^1$ factor of the
algebro-geometric data $\mathcal{M}_{\mathbb{R}}$ requires also
that the $\mathcal{B}_{g+1}$-period of the differentials $dp$ and
$dq$ be integer valued respectively,
\begin{equation} \label{periodicity conditions 2}
\begin{split}
k_{\bar{\theta}} = \frac{1}{2 \pi} \int_{\infty^-}^{\infty^+} dp \in \mathbb{Z}, \\
T w_{\bar{\theta}} = \frac{T}{2 \pi} \int_{\infty^-}^{\infty^+} dq \in \mathbb{Z}.
\end{split}
\end{equation}
To show that the conditions \eqref{periodicity conditions 1} and
\eqref{periodicity conditions 2} are sufficient to ensure
periodicity of the reconstructed $j(\sigma,\tau)$ we note from
\eqref{reconstruction formula for psi} and using the property
\eqref{automorphy theta} of $\theta$-functions that
\begin{align*}
\bm{\psi}(P,\sigma + 2 \pi,\tau) &= \exp \left\{ i \int_{\infty^+}^P
dp \right\} \bm{\psi}(P,\sigma,\tau), \\
\bm{\psi}(P,\sigma,\tau + T) &= \exp \left\{ \frac{i T}{2 \pi}
\int_{\infty^+}^P dq \right\} \bm{\psi}(P,\sigma,\tau).
\end{align*}
Since the exponent in these expressions are
$(\sigma,\tau)$-independent it follows that the matrix
$\Psi(x,\sigma,\tau) = \left( \bm{\psi}(P^+,\sigma,\tau),
\bm{\psi}(P^-,\sigma,\tau) \right)$ gets multiplied on the right by
a $(\sigma,\tau)$-independent diagonal matrix under $\sigma
\rightarrow \sigma + 2 \pi$ or $\tau \rightarrow \tau + T$, and
hence \eqref{reconstruction formula for j components} is indeed
periodic
\begin{align*}
j_{\pm}(x,\sigma + 2 \pi,\tau) &= j_{\pm}(x,\sigma,\tau),\\
j_{\pm}(x,\sigma,\tau + T) &= j_{\pm}(x,\sigma,\tau).
\end{align*}
But periodicity of the current $j = -g^{-1}dg$, although
necessary, is not sufficient to ensure periodicity of the original
field $g$. The extra condition necessary for periodicity of
$g(\sigma,\tau)$ in $\sigma$ can be obtained as follows. Starting
from the formula \eqref{reconstruction of original field g} for
the reconstruction of $g$
\begin{equation*}
g(\sigma,\tau) = U_L \cdot P \overleftarrow{\exp}
\int_{(\sigma,\tau)} j,
\end{equation*}
we compare this expression to the same expression translated by
$\sigma \rightarrow \sigma + 2 \pi$, namely
\begin{equation*}
g(\sigma + 2 \pi,\tau) = U_L \cdot P \overleftarrow{\exp}
\int_{(\sigma + 2 \pi,\tau)} j.
\end{equation*}
Its inverse is given by $g^{-1}(\sigma + 2 \pi,\tau) = \left( P
\overleftarrow{\exp} \int^{(\sigma + 2 \pi,\tau)} j \right) \cdot
U_L^{-1}$ so that
\begin{equation*}
g^{-1}(\sigma + 2 \pi,\tau) g(\sigma,\tau) = P
\overleftarrow{\exp} \int^{(\sigma + 2 \pi,\tau)}_{(\sigma,\tau)}
j = \Omega(0,\sigma,\tau).
\end{equation*}
The last equality comes from the definition of the monodromy
matrix \eqref{monodromy definition}. Periodicity of
$g(\sigma,\tau)$ under $\sigma \rightarrow \sigma + 2 \pi$ is
therefore guaranteed provided
\begin{equation*}
\Omega(0,\sigma,\tau) = {\bf 1}.
\end{equation*}
This condition however is equivalent to the condition that $p(0) =
2 \pi m \in 2 \pi \mathbb{Z}$ or stated as a condition on a
certain period of $dp$, as for the other periodicity conditions
\eqref{periodicity conditions 1} and \eqref{periodicity conditions
2}
\begin{equation} \label{periodicity conditions 3}
\frac{1}{2 \pi} \int_{\infty^+}^{0^+} dp \in \mathbb{Z}.
\end{equation}

Therefore if the real geometric map $\mathcal{G}_{\mathbb{R}}$ is
restricted to a sub-bundle of the algebro-geometric data
$\mathcal{M}_{\mathbb{R}}$ corresponding to data satisfying the
periodicity conditions \eqref{periodicity conditions 1},
\eqref{periodicity conditions 2} and \eqref{periodicity conditions
3} then its image will consist of real periodic solutions and the
corresponding reconstructed field $g(\sigma,\tau)$ will be
$SU(2)$-valued and $\sigma$-periodic, as required.

\subsection{String motion as rigid, linear motion on a torus}

The reconstructed solutions described in the previous sections
have the important property that all
dependence on the worldsheet coordinates $\sigma$ and $\tau$ is
contained in the linear evolution \eqref{linear motion of system}
of the Jacobian coordinates $\bm{\theta}(\sigma,\tau)$ and of the
$U(1)_{R}$ coordinate $\bar{\theta}(\sigma,\tau)$,
\begin{equation*}
j(\sigma,\tau) = j[{\bm
\theta}(\sigma,\tau),\bar{\theta}(\sigma,\tau)].
\end{equation*}
For real solutions, $\bar{\theta}$ and each component $\theta_i$
of $\bm{\theta}$ are real and have period $2\pi$. Together these
angular variables parametrise a torus $T^K = T^g \times S^1$ in
$\mathcal{M}_{\mathbb{R}}$. It is convenient to consider a
slightly different choice of basis cycles on $T^K$ by defining new
angular coordinates, $\vec{\varphi} = (\varphi_1, \dots,
\varphi_K)$ via,
\begin{equation}
\varphi_I = \theta_i - \bar{\theta}\quad\quad {\rm for}\,\, I=i =
1, \ldots, g = K-1,\qquad{} \varphi_K = -\bar{\theta}.
\end{equation}
In these variables, the linear evolution takes the very simple form
\begin{equation}\label{lin2}
\varphi_I(\sigma,\tau) = \varphi_I(0,0) - n_I \sigma - v_I \tau
\end{equation}
where,
\begin{equation} \label{B periods of dp and dq 2}
\int_{\mathcal{B}_I} dp = 2\pi n_I, \quad \int_{\mathcal{B}_I} dq
= 2 \pi v_I.
\end{equation}

At fixed $\tau$, the linear evolution \eqref{lin2} corresponds to
a configuration of a wrapped string on the real torus $T^K$. Any
finite gap solution is effectively described by a motion of this
string on $T^K$ which has the following features:

\begin{itemize}
\item[$\bullet$] The string wraps a non-contractible cycle on the
torus. Representing $T^K$ as $\mathbb{R}^K/2\pi\mathbb{Z}^K$ the
cycle corresponds to the lattice vector $-\vec{n}$ where
$\vec{n}=(n_1, \ldots, n_K)$. In other words, the winding numbers
of the string around the fundamental cycles on $T^K$ are
determined by the mode numbers $n_I$ of the finite-gap solution.

\item[$\bullet$] The resulting time evolution corresponds to
rigid, linear motion of the wrapped string on $T^K$ with constant
velocity vector $-\vec{v}$ where $\vec{v}=(v_1, \ldots, v_K)$.

\item[$\bullet$] The filling fractions
$\vec{\mathcal{S}}=(\mathcal{S}_1, \ldots, \mathcal{S}_K)$ play
the role of conjugate momenta to the string coordinates
$\vec{\varphi} = (\varphi_1, \ldots, \varphi_K)$ (this point will
be made more precise in the next subsection). In particular, from
\eqref{mcons2}, the momentum constraint $\mathcal{P}=0$ implies
$\vec{S} \cdot \vec{n}=0$ which means that the momentum carried by
the string is perpendicular to the line on which it lies at fixed
$\tau$.

\end{itemize}

The features described above can be summarised by saying that the
motion resembles that of an infinitely rigid string moving freely
on the torus $T^g$ with winding numbers determined by the mode
numbers of the finite-gap solution and momenta determined by the
filling fractions.

\subsection{The Hamiltonian description} \label{section:
Hamiltonian}

In this final subsection we will show that the $\sigma,\tau$
evolution of the string described above admits a simple
Hamiltonian description in terms of a finite dimensional
integrable system with the moduli space
$\mathcal{M}^{(2g)}_{\mathbb{R}}$ as its phase space. As reviewed
in Appendix \ref{section: The moduli space}, the construction of
\cite{Krichever+Phong} provides a natural symplectic form on
$\mathcal{M}^{(2g)}_{\mathbb{R}}$. In terms of the coordinates
$\{S_i, \theta_i\}$ this takes the form,
\begin{equation*}
\omega_{2g} = \,\sum_{i=1}^g \, \delta S_i \wedge \delta \theta_i
\end{equation*}
where $\delta$ denotes the exterior derivative on the moduli space.
The symplectic form defines a Poisson bracket for the coordinates,
\begin{equation*}
\{S_i,S_j\} = \{\theta_i,\theta_j\} = 0, \quad \{S_i,\theta_j\} =
\delta_{ij}.
\end{equation*}

With the above Poisson bracket any function of the conserved
quantities $\{S_i\}$ generates a linear flow on the Jacobian. To
define the appropriate evolution, we consider the functionals,
 \begin{equation*}
\begin{split}
\mathcal{P} & = -\frac{\sqrt{\lambda}}{4\pi}\, \int_{0}^{2\pi}\,
d\sigma\, {\rm tr}\left[ j_0 j_1\right], \\
\mathcal{E} & = -\frac{\sqrt{\lambda}}{4\pi}\, \int_{0}^{2\pi}\,
d\sigma\, \frac{1}{2}{\rm tr}\left[ j_0^2 + j_1^2\right].
\end{split}
\end{equation*}
These quantities are the Noether charges for rigid translations of the
worldsheet coordinates $\sigma$ and $\tau$ respectively. They are
manifestly independent of $\sigma$ and their independence of $\tau$
follows from the corresponding conservation laws for energy and
momentum on the worldsheet. As $\mathcal{P}$ and $\mathcal{E}$ are
uniquely determined by the time independent data $(\Sigma,dp)$ they
correspond to well-defined functions on the leaf $\mathcal{L}$ which
forms the base of the fibration $\mathcal{M}^{(2g)}_{\mathbb{R}}$.
Explicitly, we define the corresponding Hamiltonian functions,
\begin{equation*}
\begin{split}
H_{\sigma} & = \mathcal{P}[S_1,\ldots, S_g],\\
H_{\tau} & = \mathcal{E}[S_1,\ldots, S_g].
\end{split}
\end{equation*}
We will now consider the linear flows on
$\mathcal{M}^{(2g)}_{\mathbb{R}}$ generated by these functions
with parameters $\sigma$ and $\tau$ respectively. The
corresponding Hamilton equations are,
\begin{equation} \label{Ham1}
\frac{\partial{\theta}_i}{\partial \sigma} = \frac{\partial
\mathcal{P}}{\partial S_i}, \quad
\frac{\partial{\theta}_i}{\partial \tau} = \frac{\partial
\mathcal{E}}{\partial S_i}.
\end{equation}
The variations of $\mathcal{P}$ and $\mathcal{E}$ along the leaf
$\mathcal{L}$ were given in section \ref{section: moduli} (see
also Appendix \ref{section: variations of moduli}). In particular,
the equations \eqref{variation2b} imply,
\begin{equation*}
\frac{\partial \mathcal{P}}{\partial S_i} = -k_{i}, \quad
\frac{\partial \mathcal{E}}{\partial S_i} = -\omega_i,
\end{equation*}
where, as before,
\begin{equation*}
k_i = \frac{1}{2\pi} \int_{b_i} \, dp, \quad w_i = \frac{1}{2\pi}
\int_{b_i} \, dq.
\end{equation*}
Integrating \eqref{Ham1} we immediately obtain the linear flow,
\begin{equation} \label{linb}
\theta_i(\sigma,\tau) = \theta_i(0,0) - k_i \sigma - \omega_i \tau
\end{equation}
which is equivalent to the first equality in \eqref{linear motion
of system}.

To obtain the physical phase space of the model we must impose the
remaining constraint $\mathcal{P}=0$ and then identifying points
related by translations of the worldsheet coordinate $\sigma$. The
combined effect of these steps can be described as a symplectic
reduction of the integrable system
$(\mathcal{M}^{(2g)}_{\mathbb{R}}, \omega_{2g})$. Specifically we
work with the moment map,
\begin{equation*}
\mu = H_{\sigma} = - \sum_{I=1}^K n_I \mathcal{S}_I =
-\sum_{i=1}^g k_i S_i - \frac{n_K}{2}(L-R)
\end{equation*}
where we have used \eqref{mcons2}. We impose the constraint $\mu =
0$ and at the same time gauge the corresponding linear flow,
\begin{equation*}
X_{\mu} = \sum_{i=1}^g \, k_i \frac{\partial}{\partial \theta_i}
\end{equation*}
which defines a $U(1)$ action on $\mathcal{M}^{(2g)}_{\mathbb{R}}$.
A reduced phase-space $\tilde{\mathcal{M}}^{(2g-2)}_{\mathbb{R}}$
with symplectic form $\tilde{\omega}_{2g-2}$ can then be defined
using standard symplectic reduction,
\begin{equation*}
\xymatrix{ \mu^{-1}(0) \ar[d]^{\pi} \ar@{^{(}->}[rr]^{i} & &
\mathcal{M}^{(2g)}_{\mathbb{R}} \\
\tilde{\mathcal{M}}_{\mathbb{R}}^{(2g-2)} & & }
\end{equation*}
where $i$ is the inclusion of $\mu^{-1}(0)$ in
$\mathcal{M}^{(2g)}_{\mathbb{R}}$ and $\pi$ is a projection onto
the base of the corresponding principal $U(1)$ bundle. More
precisely, the resulting quotient is well-defined everywhere
except at the degenerate point where $S_i=0$ for all $i$, which is
a fixed point of the $U(1)$ action.

As mentioned in the Introduction, the striking feature of the
Hamiltonian description given above, is that the canonically
normalised action variables are the filling fractions $S_i$,
defined in \eqref{independent moduli}. A leading-order
semiclassical treatment of the model would therefore yield a
quantisation of the filling fractions in integer units.

So far we have only discussed the Hamiltonian evolution of the
coordinates on the reduced moduli space
$\mathcal{M}^{(2g)}_{\mathbb{R}}$ corresponding to linear motion
of the Jacobian angles $\theta_i$. However, the reconstructed
solution also has additional dependence on $\sigma$ and $\tau$
which originates to linear evolution of the $U(1)_R$ angle
$\bar{\theta}$. To obtain a more complete Hamiltonian description
we should consider an enlarged phase space with two additional
dimensions. In particular, the extra coordinate $\bar{\theta}$
corresponds to a global $U(1)_R$ rotation and is therefore
canonically conjugate to the conserved Noether charge $R$ define
above. Thus we should allow $R$ to vary and include it as one of
the moduli of the solution. In particular we should then consider
a fibration whose base is the moduli space of admissible pairs
$(\Sigma,dp)$ {\em without} fixing the value of $R$. The
corresponding leaf $\hat{\mathcal{L}}$ in the universal moduli
space would have dimension $g+1$ and the resulting moduli space of
dimension $2g+2$ would be a $J(\Sigma)\times\mathbb{C}^{\ast}$
fibration of $\hat{\mathcal{L}}$. Although we will not pursue this
idea here we note in passing that the full $\sigma,\tau$ evolution
of the finite gap solution as given in \eqref{lin2} can easily be
reproduced by considering the Hamiltonians $\mathcal{P}$,
$\mathcal{E}$ defined above as functions of the $K=g+1$ filling
fractions $\mathcal{S}_I$ and working with the symplectic form,
\begin{equation*}
\hat{\omega}_{2K}= \sum_{I=1}^K \delta\mathcal{S}_I \wedge \delta
\varphi_I.
\end{equation*}

Finally we note that the additional $\sigma,\tau$ dependence of
the solution originating from the angle $\bar{\theta}$ can be
eliminated by working in an appropriate reference frame.
Specifically, we can define new coordinates on the target $S^3
\simeq SU(2)$ by setting $\tilde{g} = gh\in SU(2)$ with,
\begin{equation*}
h(\sigma,\tau) = e^{\big( \frac{i}{2} \bar{\theta}_0 - \frac{i}{2}
\int_{\infty^-}^{\infty^+} d\mathcal{Q} \big) \sigma_3} \in SU(2)
\end{equation*}
where, as above,
\begin{equation*}
d\mathcal{Q}(\sigma,\tau) = \frac{1}{2 \pi} \left( \sigma dp + \tau
dq \right),
\end{equation*}
with a corresponding right-current,
\begin{equation*}
\tilde{j} = -\tilde{g} d\tilde{g} = h^{-1}jh - h^{-1}dh
\end{equation*}
which satisfy the equations of motion,
\begin{equation*}
d\tilde{j} - \tilde{j} \wedge \tilde{j} = 0, \quad
\nabla\ast\tilde{j} = d\ast\tilde{j} - [h^{-1}dh,\ast\tilde{j}] =
0.
\end{equation*}
\paragraph{}
The new coordinates correspond to a frame which undergoes the same
constant $SU(2)_R$ rotation as the string itself at each point
along its length. This is analogous to the body-fixed reference
frame used in the standard analysis of a rigid rotation body. The
particular merit of this frame is that, in these coordinates, the
$\sigma,\tau$-dependence of the solution is purely contained in
the internal variables $\bm{\theta}$;
\begin{equation*}
\tilde{j}(\sigma,\tau) = \tilde{j}[\bm{\theta}(\sigma,\tau)].
\end{equation*}
Thus, in this frame, the evolution of the finite-gap solution in
the worldsheet coordinates is precisely equivalent to that of the
integrable system on $\mathcal{M}^{(2g)}_{\mathbb{R}}$ introduced
above.

\section*{Acknowledgements}
The authors acknowledge useful discussions with Gleb Arutyunov and
Konstantin Zarembo. ND is supported by a PPARC Senior Research
Fellowship and BV is supprted by EPSRC.

\appendix

\section{The desingularised curve} \label{section: desingularised curve}

Since the matrix $L(x,\sigma,\tau)$ is
$2 \times 2$ the equation for the curve $\widehat{\Sigma}$ can be written
more explicitly as
\begin{equation} \label{Sigma equation}
\widehat{\Sigma} : \quad \widehat{\Sigma}(x,y) = P_2(x) y^2 +
P_0(x) = 0,
\end{equation}
where $P_0(x), P_2(x)$ are polynomials and the term linear in $y$
is absent because the sum of the eigenvalues $\{ p'(x), -p'(x) \}$
of $L(x,\sigma,\tau)$ is zero. The discriminant for such a
polynomial $\widehat{\Sigma}(x,y)$ in $y$ is given by
\begin{equation} \label{discriminant}
\Delta(x) = -4 \frac{P_0(x)}{P_2(x)}.
\end{equation}
It expresses the square of the difference between the roots of the
quadratic polynomial $\widehat{\Sigma}(x,y)$ in $y$. Now as argued
in \cite{Beisert:2004ag, Beisert:2005bm}, to exclude unphysical
branch points which arise from odd multiplicity zeroes of the
discriminant \eqref{discriminant} but retain the physical branch
points which arise as simple poles of the discriminant
\eqref{discriminant} one should impose the condition
\begin{equation*}
P_2(x) \Delta(x) = \left( Q(x) \right)^2,
\end{equation*}
where $Q(x)$ is some polynomial. This implies that $-4 P_0(x)$ is
a perfect square, and after performing the birational
transformation $y \mapsto y' = \frac{Q(x)}{2 y}$ the equation
\eqref{Sigma equation} for the curve $\widehat{\Sigma}$ turns into
the following standard form for a hyperelliptic curve
\begin{equation*}
y'^2 = P_2(x),
\end{equation*}
which was the curve considered in \cite{Kazakov:2004qf}.

\section{The moduli space} \label{section: The moduli space}

In this appendix we begin by reviewing the construction of
Krichever and Phong described in \cite{Krichever+Phong}. A point
in the universal moduli space $\mathcal{U}$ is specified by a
smooth Riemann surface $\Sigma_{\mathcal{U}}$ of genus $g$ with
$N$ punctures $P_{\alpha}$, together with an $n_{\alpha}$-jet
$[w_{\alpha}]_{n_{\alpha}}$ at each puncture. For any positive
integer $n_{\alpha}\in \mathbb{Z}$, an $n_{\alpha}$-jet is an
equivalence class of local coordinates, $w_{\alpha}$, on
$\Sigma_{\mathcal{U}}$ defined near the puncture $P_{\alpha}$ at
$w_{\alpha}=0$. The equivalence relation is such that two local
coordinates $w_{\alpha}$ and $\tilde{w}_{\alpha}$ are equivalent
if and only if, $\tilde{w}_{\alpha} = w_{\alpha} +
O(w_{\alpha}^{n_{\alpha}+1})$.

In addition, the Riemann surface $\Sigma_{\mathcal{U}}$ is
equipped with two meromorphic Abelian integrals $E$ and $Q$ with
the following specified behaviour near the punctures $P_{\alpha}$,
\begin{equation}\label{conditions}
\begin{split}
E & = w_{\alpha}^{-n_{\alpha}} + c_E + R^E_{\alpha}\log w_{\alpha} +
O(w_{\alpha}) \\
dE &= d\left(w_{\alpha}^{-n_{\alpha}} + O(w_{\alpha})\right) +
R^E_{\alpha} \frac{dw_{\alpha}}{w_{\alpha}} \\
Q & = \sum_{k=1}^{m_{\alpha}} c_{\alpha,k}w_{\alpha}^{-k} + c_Q +
R_{\alpha}^Q \log w_{\alpha} + O(w_{\alpha}) \\
dQ &= d\left(\sum_{k=1}^{m_{\alpha}} c_{\alpha,k}
w_{\alpha}^{-k} + O(w_{\alpha})\right) + R^Q_{\alpha}
\frac{dw_{\alpha}}{w_{\alpha}}
\end{split}
\end{equation}

The universal moduli space is parametrised by the data described
above,
\begin{equation*}
\mathcal{U} = \left\{\Sigma_{\mathcal{U}}, P_{\alpha},
[w_{\alpha}]_{n_{\alpha}}; E,Q\right\}.
\end{equation*}
Its complex dimension is given as,
\begin{equation*}
\text{dim}_{\mathbb{C}} \,\mathcal{U} = 5g - 3 + 3N +
\sum_{\alpha=1}^N (m_{\alpha} + n_{\alpha}).
\end{equation*}

Krichever and Phong introduce a convenient set of holomorphic
coordinates on $\mathcal{U}$. These include the independent
residues of the meromorphic differentials $dE$, $dQ$ and $QdE$.
These are labeled as,
\begin{equation*}
R^E_{\alpha}={\rm Res}_{P_{\alpha}}\,dE, \quad
R^Q_{\alpha}={\rm Res}_{P_{\alpha}}\,dQ, \quad T_{\alpha,0}=
{\rm Res}_{P_{\alpha}}\,\left(QdE\right),
\end{equation*}
for $\alpha = 2, \ldots, N$. They also include the additional
residues,
\begin{equation*}
T_{\alpha,k}=  {\rm
Res}_{P_{\alpha}}\,\left(w_{\alpha}^{k}QdE\right),
\end{equation*}
for $k = 1, \ldots, m_{\alpha}+n_{\alpha}$ and $\alpha = 1,
\ldots, N$. The remaining coordinates are the periods of
meromorphic differentials around canonical basis cycles $a_{i}$
and $b_{i}$, for $i=1,\ldots,g$,
\begin{gather*}
\tau_{a_i,E} = \int_{a_i}\, dE, \quad \tau_{b_i,E} = \int_{b_i}\, dE, \\
\tau_{a_i,Q} = \int_{a_i}\, dQ, \quad \tau_{b_i,Q} = \int_{b_i}\, dQ, \\
s_i = \int_{a_i}\, QdE.
\end{gather*}

Krichever and Phong prove that the $5g-3+3N+\sum_{\alpha=1}^{N}
(m_{\alpha}+n_{\alpha})$ functions,
\begin{equation*}
\{R^{E}_{\alpha}, R^{Q}_{\alpha}, T_{\alpha, 0}, T_{\alpha, k},
\tau_{a_{i},E}, \tau_{b_{i}, E}, \tau_{a_{i}, Q}, \tau_{b_{i}, Q},
s_{i}\}
\end{equation*}
have linearly independent differentials and therefore define a
local holomorphic coordinate system for $\mathcal{U}$. It follows
that the joint level sets of the coordinates,
\begin{equation*}
\{R^{E}_{\alpha}, R^{Q}_{\alpha}, T_{\alpha, 0}, T_{\alpha, k},
\tau_{a_{i},E}, \tau_{b_{i}, E}, \tau_{a_{i}, Q}, \tau_{b_{i}, Q}\}
\end{equation*}
define a smooth $g$-dimensional foliation of $\mathcal{U}$
independent of the choices made in defining the coordinates. The
remaining variables $\{s_{i}\}$ are holomorphic coordinates on a
leaf of this foliation.

We can now analyse the space of holomorphic data for finite-gap
solutions as a special case of the construction described above.
In the present case, the curve $\Sigma_{\mathcal{U}}$ will be
identified with the spectral curve $\Sigma$ of genus $g=K-1$. The
Abelian integrals $E$ and $Q$ will be identified with the
quasi-momentum $p(P)$ and an, as yet unspecified, function $z(P)$
on $\Sigma$. We know that $p(P)$ has only simple poles at
$\{(+1)^{\pm}, (-1)^{\pm}\}$ and no other singularities. We will
choose $z(P)$ to have only simple poles at the points
$\{\infty^{\pm},0^{\pm}\}$ and no other singularities. Thus we are
interested in the $N=8$ case of the construction described above,
where $n_{\alpha}=1$, $m_{\alpha}=0$ at the first four punctures,
and $n_{\alpha}=0$, $m_{\alpha}=1$ at the second four. In this
case the universal moduli space has dimension $5g+29=5K+24$

To make contact with the construction of Krichever and Phong we
must start by assuming nothing about $\Sigma$, $p(P)$ and $z(P)$
other than the existence of eight punctures on $\Sigma$ and the
behaviour of $p(P)$ and $z(P)$ near the punctures implied by
\eqref{conditions} with the identifications $E=p$, $Q=z$ for the
specific values of $n_{\alpha}$ and $m_{\alpha}$ given above. In
particular we do not assume that $\Sigma$ is hyperelliptic. To
properly define the moduli space we must identify each of the
constraints which are imposed on $\Sigma$, $p(P)$ in the text as a
level set condition on the holomorphic coordinates on
$\mathcal{U}$. These conditions are,

\begin{itemize}
\item[$\bullet$] The only singularities of $p(P)$ are simple poles
at $\{(+1)^{\pm}, (-1)^{\pm}\}$ and the singular behaviour near
these points is given in \eqref{weak dp asymptotics at pm 1}. The
residues are absorbed in the definition of the local coordinates
$w_{\alpha}$ at these points. The absence of simple poles for $dp$
implies that $R^E_{\alpha}=0$ for all $\alpha$. This yields seven
level set conditions.

\item[$\bullet$] The $a$- and $b$-periods of $dp$ are given in
\eqref{AnB periods of dp}. These implies that $\tau_{a_{i}, E}=0$
and $\tau_{b_i,E} = 2\pi n_i$ for $i = 1,\ldots,g$: a total of
$2g$ level set conditions.
\end{itemize}

We will also introduce additional level-set conditions by choosing
appropriate properties for the second Abelian integral $z(P)$.
These conditions will be sufficiently restrictive to allow us to
reconstruct $z(P)$ explicitly. In particular,

\begin{itemize}
\item[$\bullet$] The only singularities of $z(P)$ are simple poles
at the points $\{\infty^{\pm}, 0^{\pm}\}$. The absence of simple
poles in $dz$ implies $R_{\alpha}^{Q}=0$ for all $\alpha$. Thus we
have a further seven level set conditions.

\item[$\bullet$] We will choose the $a$- and $b$-periods of $dz$
to vanish implying that $\tau_{a_i,Q} = \tau_{b_i,Q} = 0$. This
yields $2g$ level set conditions.

\item[$\bullet$] The behaviour of the  differential $QdE = zdp$
near each of the eight singular points yields sixteen constraints
on the remaining coordinates which we choose to be as follows,
\begin{center}
\begin{tabular}{c|cccc}
& $\infty^{\pm}$ & \qquad{}$0^{\pm}$ &  \qquad{}$+1^{\pm}$ &
\qquad{}$-1^{\pm}$ \\
\hline $T_{\alpha, 0}$ & $\frac{4\pi R}{\sqrt{\lambda}}$ &
$-\frac{4\pi L}{\sqrt{\lambda}}$ & $0$ & $0$ \\
$T_{\alpha, 1}$ & $0$& $0$ & $-2$ & $+2$
\end{tabular}
\end{center}

\end{itemize}

At this point we have imposed a total of $4g+29=4K+25$ level set
conditions on the $5g+29=5K+24$ coordinates, thus defining a
subspace of the required dimension $g=K-1$. It remains to show
that these conditions uniquely fix the second Abelian integral
$z(P)$ and that there are no other constraints on the data. In
particular, the only additional constraint placed on the curve in
the text was that it take the hyperelliptic form
\eqref{desingularised curve}. For consistency, it is therefore
necessary that this condition is not independent but is actually a
consequence of the other level-set constraints described above.

The hyperelliptic condition is equivalent to demanding the
existence of a well-defined function on $\Sigma$ with exactly two
simple poles and no other singularities. To demonstrate the
existence of such a function we use the following facts,

\begin{itemize}
\item[{\bf 1}] The condition $\tau_{a_i, Q} = \tau_{b_i, Q} = 0$,
together with $R^Q_{\alpha} = 0$ implies that $dz$ has vanishing
periods and residues on $\Sigma$. Therefore the Abelian integral
$z(P)$ is actually a well-defined function on $\Sigma$.

\item[{\bf 2}] The values of the integers $m_{\alpha}$ at the
punctures $P_{\alpha}$ imply that $z(P)$ has simple poles at the
four points $\{\infty^{\pm},0^{\pm}\}$ and no other singularities
on $\Sigma$. Thus $z(P)$ is a function of degree four, $\text{deg
}z = 4$, which takes each complex value exactly four times on
$\Sigma$ counting multiplicities.

\item[{\bf 3}] The values of the integers $n_{\alpha}$ at the
punctures imply that the meromorphic differential $dp$ has double
poles at each of the four points $\{(+1)^{\pm},(-1)^{\pm}\}$. By
definition of the local coordinates $w_{\alpha}$ we have $dp\sim
-w_{\alpha}^{-2}$ near each of these points.

\item[{\bf 4}] The vanishing of the coordinates $T_{\alpha, 0}$ at
the four points $\{(+1)^{\pm},(-1)^{\pm}\}$ implies that the
differential $zdp$ has zero residues at these points. Given {\bf
3}, this means that the functions $z(P)-z(P_{\alpha})$ must have
at least a double zero at each of these points. Equivalently $dz$
vanishes at each of these points.

\item[{\bf 5}] The values $T_{\alpha, 1}=-2$ at the points
$(+1)^{\pm}$ and $T_{\alpha, 1}=+2$ at the points $(-1)^{\pm}$
imply that $z(P)$ attains the value $+2$ at the first two points
and attains $-2$ at the second two points.

\end{itemize}

We are now ready to define the required function as,
\begin{equation}\label{hyper}
x(P)=\frac{1}{2}\left(z(P)+\sqrt{z(P)^{2}-4}\right)
\end{equation}
This is not obviously single-valued function on $\Sigma$ because
of the apparent branch points at the zeros of the function
$f(P)=z^{2}(P)-4$. More precisely, simple zeros of $f(P)$ (or more
generally zeros of odd order) give rise to branch-points of
$x(P)$. In contrast, double zeros of $f(P)$ do not give rise to
branch points. To show that $x(P)$ is a well-defined function on
$\Sigma$, we will now demonstrate that $f(P)$ has only double
zeros. From {\bf 2} we know that $z(P)$ has degree four and hence
that $f(P)$ has degree eight. Thus $f(P)$ has eight zeros on
$\Sigma$ counting multiplicities. From {\bf 5}, $f(P)$ vanishes at
each of the four points $\{(+1)^{\pm},(-1)^{\pm}\}$ and from ${\bf
4}$ these are at least double zeros. As $f$ has order eight these
can be at most double zeros. Thus $f(P)$ has exactly four double
zeros at the points $\{(+1)^{\pm},(-1)^{\pm}\}$. It follows that
$x(P)$ is well defined on $\Sigma$. Inverting \eqref{hyper} we
obtain,
\begin{equation}\label{hyper2}
z(P)=x(P)+\frac{1}{x(P)}
\end{equation}
As $x(P)$ is a well-defined function it must have a definite
degrees $\text{deg }x$. Hence $x(P)$ attains the values $0$ and
$\infty$ exactly $\text{deg }x$ times. The relation \eqref{hyper2}
implies that $4 = \text{deg }z = 2\text{deg }x$. It follows that
$x(P)$ is a well-defined function on $\Sigma$ of degree two with
exactly two simple poles at the points $\{\infty^{\pm}\}$. Thus we
deduce that $\Sigma$ is hyperelliptic and can be put in the form
\eqref{desingularised curve}. As a byproduct the second Abelian
integral is then uniquely determined by the formula
\eqref{hyper2}. For completeness, one may then check that this
formula for $z(P)$ automatically reproduces all the properties of
$z(P)$ listed above which are used to define the leaf
$\mathcal{L}$.

As discussed in the text, the full moduli space we consider,
denoted $\mathcal{M}_{\mathbb{C}}^{(2g)}$, is the Jacobian
fibration over the leaf $\mathcal{L}$
\begin{equation}
J(\Sigma) \rightarrow \mathcal{M}_{\mathbb{C}}^{(2g)} \rightarrow
\mathcal{L}.
\end{equation}
We can express the coordinates on $\mathcal{L}$ as periods of the
differential,
\begin{equation*}
\alpha = \frac{\sqrt{\lambda}}{4\pi} zdp.
\end{equation*}
In particular, it is convenient to use rescaled coordinates,
\begin{equation}\label{si}
S_i = \frac{1}{2\pi i} \frac{\sqrt{\lambda}}{4\pi} s_i =
\frac{1}{2\pi i} \int_{a_i} \alpha.
\end{equation}
The coordinates on the fibre $\bm{\theta} = (\theta_1, \ldots,
\theta_g)$ are defined as the image of the divisor $\gamma$ under
the Abel map,
\begin{equation}\label{thetai}
\bm{\theta} = 2\pi \sum_{j=1}^g \int_{\infty^+}^{\gamma_j}
\bm{\omega} + \bm{\theta}_0
\end{equation}
where $\bm{\theta}_0$ is the constant vector defined in
\eqref{reality of theta}, and $\bm{\omega}$ is the vector of the
basis holomorphic differentials $\{ \omega_i \}_{i=1}^g$ on
$\Sigma$ with normalisation $\int_{a_i} \omega_j =\delta_{ij}$.

Following Krichever and Phong we can define a symplectic form
$\omega_{2g}$ on the moduli space
$\mathcal{M}_{\mathbb{C}}^{(2g)}$. For this it is useful to first
consider the universal curve bundle $\mathcal{N}$ over the leaf
$\mathcal{L}$
\begin{equation*}
\Sigma \rightarrow \mathcal{N} \rightarrow \mathcal{L},
\end{equation*}
whose fibre above every point of the base $\mathcal{L}$ is the
corresponding curve $\Sigma$. As already mentioned, the
$\{s_i\}_{i=1}^g$ form a set of coordinates on the base
$\mathcal{L}$, and $z$ may be taken as a coordinate on the fibre,
so that $\delta z$ and $\{\delta s_i\}_{i=1}^g$, where $\delta$
denotes the exterior derivative on the total space $\mathcal{N}$,
form a basis of differentials at every point of $\mathcal{N}$. In this basis, the
total exterior derivative of any function $f$ on $\mathcal{N}$ can
be separated as
\begin{equation*}
\delta f = \frac{\partial f}{\partial z} \delta z + \sum_{i=1}^g
(\partial_{s_i} f) \delta s_i \equiv df + \delta^{\mathcal{L}} f,
\end{equation*}
where $\delta^{\mathcal{L}}$ denotes the exterior derivative along
the leaf $\mathcal{L}$. Note that $\delta z = dz$ and $\delta s_i
= \delta^{\mathcal{L}} s_i$. The differential $p\,dz$ on $\Sigma$,
as in fact any differential on $\Sigma$, can be extended to a
differential on $\mathcal{N}$ by setting it to zero along $\delta
s_i$. Consider now the differential $\delta (p\,dz)$ on
$\mathcal{N}$
\begin{equation*}
\delta (p\,dz) = \sum_{i = 1}^g \delta s_i \wedge
\partial_{s_i}(p\,dz).
\end{equation*}
The key observation is that, while the differential $p\,dz$ is
neither single-valued nor holomorphic on $\Sigma$, the ambiguities
in its definition as well as its pole parts are constant along the leaf
$\mathcal{L}$. Therefore $\partial_{s_i}(p\,dz)$ is holomorphic on
$\Sigma$ and can be expanded in the basis of holomorphic
differentials $\{ \omega_i \}_{i=1}^g$ of $\Sigma$
\begin{equation*}
\partial_{s_i}(p\,dz) = \sum_{j = 1}^g \alpha_{ij}
\omega_j, \quad \alpha_{ij} \in \mathbb{C}.
\end{equation*}
The constants $\alpha_{ij}$ can be computed easily
\begin{equation*}
\alpha_{ik} = \sum_{j = 1}^g \alpha_{ij} \int_{a_k} \omega_j =
\int_{a_k} \partial_{s_i}(p\,dz) = \partial_{s_i} \int_{a_k} p\,dz
= - \partial_{s_i} \int_{a_k} \alpha = - \partial_{s_i} s_k = - \delta_{ik},
\end{equation*}
where in the fourth equality we use the fact that $p$ is the Abelian
integral of the normalised differential $dp$. So in fact we have
$\partial_{s_i}(p\,dz) = - \omega_i$. As a result, the differential
$\delta (p\,dz)$ takes the simple form
\begin{equation} \label{differential of pdz}
\delta (p\,dz) = - \sum_{i = 1}^g \delta s_i \wedge \omega_i.
\end{equation}
Now since the Jacobian $J(\Sigma)$ can be identified with the
symmetric product $\Sigma^g/S_g$ of $g$ copies of the curve
$\Sigma$ via the Abel map, the bundle $\mathcal{M}^{(2g)}_{\mathbb{C}}$ can be
naturally viewed as the symmetric product bundle $\mathcal{N}^g$
\begin{equation*}
\Sigma^g/S_g \rightarrow \mathcal{N}^g \rightarrow \mathcal{L},
\end{equation*}
whose fibre above every point of $\mathcal{L}$ is the $g$-th symmetric
power of the curve $\Sigma$. The differential $\delta (p\,dz) = \delta
p \wedge dz$ on $\mathcal{N}$ can be used to define a symplecitc form $\omega_{2g}$ on
$\mathcal{N}^g$ by the following expression, symmetric in the points
$\gamma_i \in \Sigma, i=1,\ldots,g$,
\begin{equation*}
\omega_{2g} = - \frac{\sqrt{\lambda}}{4 \pi i} \sum_{j=1}^g \delta
p(\gamma_j) \wedge dz(\gamma_j).
\end{equation*}
To see how this also defines a symplectic form on the Jacobian bundle
$\mathcal{M}^{(2g)}_{\mathbb{C}}$ however requires a change of
variable. Using \eqref{differential of pdz} it can first of all be
rewritten as
\begin{equation*}
\omega_{2g} = \frac{\sqrt{\lambda}}{4 \pi i} \sum_{i = 1}^g \delta
s_i \wedge \left( \sum_{i=1}^g \omega_i(\gamma_j) \right),
\end{equation*}
or using $\delta$ to now denote the exterior derivative on
$\mathcal{M}^{(2g)}_{\mathbb{C}}$ we can rewrite $\omega_{2g}$
explicitly as a symplectic form on $\mathcal{M}^{(2g)}_{\mathbb{C}}$ with
the help of the Abel map \eqref{thetai}, namely
\begin{equation*}
\omega_{2g} = \sum_{i = 1}^g \delta S_i \wedge \delta \theta_j.
\end{equation*}

\section{Variations of $\mathcal{E}$ and $\mathcal{P}$ on the moduli
space} \label{section: variations of moduli}

In this section we will prove the relations \eqref{variation2b}
which describe the variations of the worldsheet momentum,
$\mathcal{P}$ and energy $\mathcal{E}$ along the leaf ${\cal L}$
defined in the text.

On a smooth Riemann surface of genus $g$, we choose a canonical
basis of one-cycles $\{a_i, b_i \}_{i=1}^g$ and define a pairing
of meromorphic differentials $\Omega_1$ and $\Omega_2$ via the
formula,
\begin{equation}\label{pairing}
\left(\Omega_1 \bullet \Omega_2\right) = \sum_{l=1}^g \left(
\int_{a_l} \Omega_1 \int_{b_l} \Omega_2 - \int_{b_l} \Omega_1
\int_{a_l} \Omega_2\right).
\end{equation}
Let $g_1$ be an Abelian integral of $\Omega_{1}$, i.e. $\Omega_1 =
dg_1$. The Riemann bilinear identity states,
\begin{equation}\label{Riemann}
\left(\Omega_1 \bullet \Omega_2\right) = 2\pi i \sum_{\rm poles}
{\rm Res}\left[g_1 \Omega_2\right].
\end{equation}

We will apply this identity on the spectral curve $\Sigma$ with
the identifications $\Omega_1 = dp$ (thus $g_1 = p$) and $\Omega_2
= \delta^{\mathcal{L}} (zdp)$. As in Appendix \ref{section: The
moduli space} and in the text, $\delta^{\mathcal{L}}$ denotes a
variation along the leaf $\mathcal{L}$. Using \eqref{AnB periods
of dp} for the $a$- and $b$-periods of $dp$ and the definitions
\eqref{independent moduli} of the moduli $S_i$, \eqref{Riemann}
becomes,
\begin{equation*}
2\pi i \frac{8 \pi^2}{\sqrt{\lambda}} \sum_{i=1}^g n_i \delta^{\mathcal{L}} S_i
= 2\pi i \sum_{\rm poles} {\rm Res}\left[p\,\delta^{\mathcal{L}} (zdp)\right].
\end{equation*}

The non-zero residues of $p\,\delta^{\mathcal{L}} (zdp)$ at the
punctures are tabulated below,
\begin{center}
\begin{tabular}{c|cc}&
\quad $(+1)^{\pm}$ & \quad $(-1)^{\pm}$ \\ \hline
${\rm Residue}$ &
$4\pi^2 \kappa_+ \delta^{\mathcal{L}} \kappa_+$ &
$-4\pi^2 \kappa_- \delta^{\mathcal{L}} \kappa_-$
\end{tabular}
\end{center}

Thus we obtain,
\begin{equation}
\sum_{i=1}^g n_i \delta^{\mathcal{L}} S_i = - \frac{\sqrt{\lambda}}{2}
(\kappa_+ \delta^{\mathcal{L}} \kappa_+ - \kappa_-
\delta^{\mathcal{L}} \kappa_-) = -\delta^{\mathcal{L}} \mathcal{P},
\end{equation}
as required.

The second equality in \eqref{variation2b} is obtained by a
similar application of the Riemann bilinear identity
\eqref{Riemann} with $\Omega_1 = dq$ (thus $g_1 = q$) and
$\Omega_2 = \delta^{\mathcal{L}} (z dp)$ as before.

\section{Singular parts at $x = \pm 1$} \label{section: Singular parts}

For any matrix $\lambda = n^A \sigma^A \in \mathfrak{su}(2)$ one has
$\text{det}\, \lambda = - \left( n^A n^A \right) =
-\frac{1}{2}\text{tr}\, \lambda^2$. So the Virasoro constraint
\eqref{Virasoro} may be rewritten as
\begin{equation} \label{Virasoro2}
\text{det}\, j_{\pm} = \kappa_{\pm}^2.
\end{equation}
The eigenvalues of $\frac{1}{2} j_{\pm}$ are determined by
$\text{det}(\mu \mathbf{1} - \frac{1}{2} j_{\pm})=0$ so since
$\text{tr}\, j_{\pm} = 0$ we have $\mu^2 = -\text{det}(\frac{1}{2}
j_{\pm}) = - \frac{1}{4} \text{det}\, j_{\pm} = -\frac{1}{4}
\kappa_{\pm}^2$, using the Virasoro constraints \eqref{Virasoro2}, and
thus $\mu = \frac{i}{2} \kappa_{\pm}$ or $\mu = -\frac{i}{2} \kappa_{\pm}$.

In order to determine the singular parts at $x = \pm 1$ consider
the auxiliary linear problem
\begin{equation} \label{auxiliary linear problem components}
\left[\partial_{\sigma} + \frac{1}{2} \left(\frac{j_+}{1-x} -
\frac{j_-}{1+x}\right)\right] \bm{\psi}(x,\sigma,\tau) = 0, \quad
\left[\partial_{\tau} + \frac{1}{2} \left(\frac{j_+}{1-x} +
\frac{j_-}{1+x}\right)\right] \bm{\psi}(x,\sigma,\tau) = 0.
\end{equation}
We know from \eqref{psi factored scalar} that $\bm{\psi}$ takes
the following form near $x=\pm 1$
\begin{equation*}
\bm{\psi}(x,\sigma,\tau) = e^{g^{\pm}(x,\sigma,\tau)} \left(
\bm{\psi}_0^{\pm}(\sigma,\tau) + \sum_{s=1}^{\infty}
\bm{\psi}_s^{\pm}(\sigma,\tau) (1 \mp x)^s \right) \quad \text{as
} x \rightarrow \pm 1,
\end{equation*}
where $g^{\pm}(x,\sigma,\tau) = \frac{g_1^{\pm}(\sigma,\tau)}{1
\mp x}$ is a singular part at $x=\pm 1$. Consider the point $x=1$
first. Plugging this expression for $\bm{\psi}$ back into
\eqref{auxiliary linear problem components} and solving it to
lowest order in $(1-x)$ we obtain
\begin{equation*}
\left[(\partial_{\sigma}g_1^+) \mathbf{1} + \frac{1}{2} j_+
\right]\bm{\psi}_0^+(\sigma,\tau) = 0, \quad
\left[(\partial_{\tau}g_1^+) \mathbf{1} + \frac{1}{2} j_+
\right]\bm{\psi}_0^+(\sigma,\tau) = 0.
\end{equation*}
If $\bm{\psi}_0^+$ is to be non-zero, this means that
$-(\partial_{\sigma}g_1^+) = -(\partial_{\tau}g_1^+)$ is an
eigenvalue of $\frac{1}{2} j_+$ (with corresponding eigenvector
$\bm{\psi}_0^+$). Hence $(\partial_{\sigma}g_1^+) =
(\partial_{\tau}g_1^+) = \pm \frac{i}{2} \kappa_+$, where the sign
is different on both sheets, and in accordance with the implicit
choice taken in deriving equation \eqref{p asymptotics at pm 1}
one should take the positive eigenvalue on the physical sheet. Doing
the same thing for the point $x=-1$ and working to lowest order in
$(1+x)$ yields
\begin{equation*}
\left[(\partial_{\sigma}g_1^-) \mathbf{1} - \frac{1}{2} j_-
\right]\bm{\psi}_0^-(\sigma,\tau) = 0, \quad
\left[(\partial_{\tau}g_1^-) \mathbf{1} + \frac{1}{2} j_-
\right]\bm{\psi}_0^-(\sigma,\tau) = 0,
\end{equation*}
so that $(\partial_{\sigma}g_1^-) = -(\partial_{\tau}g_1^-) = \pm
\frac{i}{2} \kappa_-$, where again the sign is different on both
sheets and should be positive on the physical sheet.

We make the choice of normalising the singular part of the eigenvector
$\bm{\psi}(x,\sigma,\tau)$ by the following initial condition on
$g^{\pm}(x,\sigma,\tau)$
\begin{equation*}
g^{\pm}(x,0,0) = 0.
\end{equation*}
This can be achieved simply by normalising the solution
$\bm{\psi}(x,\sigma,\tau)$ to the auxiliary linear problem by a
function of $P \in \Sigma$ independent of $\sigma$ and $\tau$.
Integrating up the equations for $g_1^{\pm}$ with these initial
conditions yields the following singular parts at $x = +1$ and $x
= -1$ respectively
\begin{equation} \label{Singular parts}
\left\{
\begin{split}
&S_+(x^{\pm},\sigma,\tau) = \mp \frac{i\kappa_+}{2} \frac{\sigma +
\tau}{x - 1}, \\
&S_-(x^{\pm},\sigma,\tau) = \mp \frac{i\kappa_-}{2} \frac{\sigma -
\tau}{x + 1}.
\end{split}
\right.
\end{equation}

\section{Motion of the dynamical divisor} \label{section: motion of
divisor}

The dynamical divisor $\gamma(\sigma,\tau) = \prod_{i=1}^g
\gamma_i(\sigma,\tau)$ is the divisor of zeroes of the
Baker-Akhiezer function $\varphi(P,\sigma,\tau)$ introduced in
section \ref{section: identifying data}. To determine its
equations of motion on $\Sigma$ we follow \cite{Babelon} and
consider the functions $\frac{\partial_{\sigma} \varphi}{\varphi}$
and $\frac{\partial_{\tau} \varphi}{\varphi}$ on $\Sigma$. These
functions are meromorphic with $g$ poles at $\{
\gamma_i(\sigma,\tau) \}_{i=1}^g$ and 4 further poles at the
points on $\Sigma$ above $x = \pm 1$ with the residues
\begin{align} \label{residues at x=pm 1}
\frac{\partial_{\sigma} \varphi}{\varphi}(x^{\pm})
&\mathop{\sim}_{x \rightarrow
+1} \mp \frac{i\kappa}{2} \frac{1}{x - 1} + O(1), \notag \\
\frac{\partial_{\sigma} \varphi}{\varphi}(x^{\pm})
&\mathop{\sim}_{x \rightarrow
-1} \mp \frac{i\kappa}{2} \frac{1}{x + 1} + O(1) \\
\text{and}\; \frac{\partial_{\tau} \varphi}{\varphi}
&\mathop{\sim}_{x \rightarrow \pm 1} \pm \frac{\partial_{\sigma}
\varphi}{\varphi} \notag
\end{align}
which follow from the behaviour \eqref{BA singular parts} of
$\varphi$ at the essential singularities $x=\pm 1$. By the
Riemann-Roch theorem the space of functions with $g+4$ prescribed
poles and with prescribed residues at four of these poles, as in
\eqref{residues at x=pm 1}, is 1 dimensional. The functions
$\frac{\partial_{\sigma} \varphi}{\varphi}, \frac{\partial_{\tau}
\varphi}{\varphi}$ being meromorphic on the algebraic curve
$\Sigma : y^2 = \prod_{i = 1}^{2 g + 2} (x - x_i)$ they can be
written as rational functions in the variables $x$ and $y$.

The most general rational function with poles at $\{
\gamma_i(\sigma,\tau) \}_{i=1}^g$ and $x = \pm 1$ is
\begin{equation*}
F(x,\sigma,\tau) = \frac{f(x,\sigma,\tau) y +
h(x,\sigma,\tau)}{(x-1)(x+1)\tilde{B}(x)}
\end{equation*}
where $\text{deg}\, f = 1$ and $\text{deg}\, h = g+2$. First note
that it is regular as $x\rightarrow \infty$ (since $y \sim
x^{g+1}$) so that $x = \infty$ isn't a pole of the function, as
required. Now at $x=\pm 1$ there are poles on both sheets with
opposite residues \eqref{residues at x=pm 1}, which imposes the
two conditions $h(\pm 1) = 0$ on $h$ yielding
\begin{equation*}
h(x,\sigma,\tau) = q(x,\sigma,\tau) -
q(+1,\sigma,\tau)\frac{(x+1)\tilde{B}(x)}{2\tilde{B}(1)} -
q(-1,\sigma,\tau)\frac{(x-1)\tilde{B}(x)}{-2 \tilde{B}(-1)},
\end{equation*}
where $\text{deg}\, q = g$. The requirement that the function has
poles at the points $\gamma_i = (x_{\gamma_i}, y_{\gamma_i})$ but
no pole at $(x_{\gamma_i}, -y_{\gamma_i})$ imposes the extra $g$
conditions $q(x_{\gamma_i},\sigma,\tau) = f(x_{\gamma_i})
y_{\gamma_i}$ on $q$ so that $h$ now only contains a single
undetermined constant. The asymptotic behaviour of the function as
$x\rightarrow \pm 1$ is
\begin{equation*}
F(x,\sigma,\tau) \mathop{\sim}_{x \rightarrow +1} \frac{f(+1)
y(+1)}{2\tilde{B}(1)(x-1)}, \quad F(x,\sigma,\tau)
\mathop{\sim}_{x \rightarrow -1} \frac{f(-1)
y(-1)}{-2\tilde{B}(-1)(x+1)}
\end{equation*}
where $y(x)$ is to be understood as a multivalued function taking
different values on both sheets. The remaining linear function $f$
is determined by fixing the residues at the poles $x=\pm 1$. Thus
for the function $\frac{\partial_{\sigma} \varphi}{\varphi}$ we
have
\begin{equation*}
f_{\sigma}(x,\sigma,\tau) = -i\kappa \frac{\tilde{B}(1) (x+1)}{2
y_{+1}} + i\kappa \frac{\tilde{B}(-1) (x-1)}{-2 y_{-1}},
\end{equation*}
and for the function $\frac{\partial_{\tau} \varphi}{\varphi}$ we
have
\begin{equation*}
f_{\tau}(x,\sigma,\tau) = -i\kappa \frac{\tilde{B}(1) (x+1)}{2
y_{+1}} - i\kappa \frac{\tilde{B}(-1) (x-1)}{-2 y_{-1}},
\end{equation*}
where $y_{+1}$ and $y_{-1}$ are the values of $y(+1)$ and $y(-1)$
on the sheet corresponding to $p(x)$.

To determine the equation of motion for the zeroes of $\varphi$ we
note that since the zeroes are simple zeroes one has
$\varphi(x,\sigma,\tau) \sim
(x-x_{\gamma_i})\widetilde{\varphi}(x,\sigma,\tau)$, where
$\widetilde{\varphi}(x_{\gamma_i}) \neq 0$, which implies the
following asymptotic behaviour near the points $\gamma_i$
\begin{equation*}
\frac{\partial_{\sigma,\tau} \varphi}{\varphi} \mathop{\sim}_{P
\rightarrow \gamma_i} - \frac{\partial_{\sigma,\tau}
x_{\gamma_i}}{x-x_{\gamma_i}} + O(1)
\end{equation*}
where $P \in \Sigma$ and $x = \pi(P)$. But from above we also have
\begin{equation*}
\frac{\partial_{\sigma,\tau} \varphi}{\varphi} \mathop{\sim}_{P
\rightarrow \gamma_i} \frac{2 f_{\sigma,\tau}(x_{\gamma_i})
y_{\gamma_i}}{(x_{\gamma_i}+1)(x_{\gamma_i}-1)\prod_{j \neq i}
(x_{\gamma_i} - x_{\gamma_j})} \frac{1}{x - x_{\gamma_i}} + O(1),
\end{equation*}
and comparing the residue of the pole at $\gamma_i$ in the above
two expressions we obtain
\begin{equation} \label{motion of dynamical divisor}
\left\{
\begin{split}
\partial_{\sigma} x_{\gamma_i} &= \frac{-2 f_{\sigma}(x_{\gamma_i})
y_{\gamma_i}}{(x_{\gamma_i}+1)(x_{\gamma_i}-1)\prod_{j \neq i}
(x_{\gamma_i} - x_{\gamma_j})} \\
\partial_{\tau} x_{\gamma_i} &= \frac{-2 f_{\tau}(x_{\gamma_i})
y_{\gamma_i}}{(x_{\gamma_i}+1)(x_{\gamma_i}-1)\prod_{j \neq i}
(x_{\gamma_i} - x_{\gamma_j})}.
\end{split}
\right.
\end{equation}

Suppose we now continuously deform the solution by bringing the
branch points $x_{2g+1}$ and $x_{2g+2}$ together into a single
point. Then by the end of the process the curve $\Sigma$ acquires
a singular point at $x_{2g+1} = x_{2g+2}$ and has genus one less,
namely $y^2 = (x - x_{2g + 1})^2 \prod_{i=1}^{2g} (x - x_i)$. We
know that as a result of shrinking the cut $[x_{2g+1}, x_{2g+2}]$
to a singular point, one of the points of the dynamical divisor
$\gamma(\sigma,\tau)$, say $\gamma_g(\sigma,\tau)$, should cease
to be dynamical. The equations of motion of $x_{\gamma_g}$ then
imply that once the cut is shrunk, $\gamma_g(\sigma,\tau)$ should
end up either at one of the remaining $2g$ branch points $x_i$ of
the singular curve or at the singular point $x_{2g+1}$ in order
that $\partial_{\sigma,\tau} x_{\gamma_g} = 0$. But unless
$\gamma_g(\sigma,\tau)$ ends up at $x_{2g+1}$, the equations of
motion of the remaining $x_{\gamma_i}$ do not take the form
\eqref{motion of dynamical divisor} corresponding to a solution
with a curve of genus $g-1$. The conclusion is that one should
view singular points $x_i, i = -\infty, \ldots, 0$ as `trapped
points' of the dynamical divisor.


\begin{thebibliography}{5}

\bibitem{Bena:2003wd}
  I.~Bena, J.~Polchinski and R.~Roiban,
  ``Hidden symmetries of the AdS(5) x S**5 superstring,''
  Phys.\ Rev.\ D {\bf 69} (2004) 046002,
  arXiv:hep-th/0305116.

\bibitem{Kazakov:2004qf}
  V.~A.~Kazakov, A.~Marshakov, J.~A.~Minahan and K.~Zarembo,
  ``Classical / quantum integrability in AdS/CFT,''
  JHEP {\bf 0405} (2004) 024,
  arXiv:hep-th/0402207.

\bibitem{Beisert:2004ag}
  N.~Beisert, V.~A.~Kazakov and K.~Sakai,
  ``Algebraic curve for the SO(6) sector of AdS/CFT,''
  arXiv:hep-th/0410253.

\bibitem{Beisert:2005bm}
  N.~Beisert, V.~A.~Kazakov, K.~Sakai and K.~Zarembo,
  ``The algebraic curve of classical superstrings on AdS(5) x S**5,''
  arXiv:hep-th/0502226.

\bibitem{Kazakov:2004nh}
  V.~A.~Kazakov and K.~Zarembo,
  ``Classical / quantum integrability in non-compact sector of AdS/CFT,''
  JHEP {\bf 0410} (2004) 060,
  arXiv:hep-th/0410105.

\bibitem{Marsh}  A.~Marshakov, 
``Quasiclassical geometry and integrability of AdS/CFT correspondence,''
  Theor.\ Math.\ Phys.\  {\bf 142} (2005) 222
  [Teor.\ Mat.\ Fiz.\  {\bf 142} (2005) 265]
  [arXiv:hep-th/0406056].

\bibitem{Schafer-Nameki:2004ik}
  S.~Schafer-Nameki,
  ``The algebraic curve of 1-loop planar N = 4 SYM,''
  Nucl.\ Phys.\ B {\bf 714} (2005) 3,
  arXiv:hep-th/0412254.

\bibitem{Alday:2005gi}
  L.~F.~Alday, G.~Arutyunov and A.~A.~Tseytlin,
  ``On integrability of classical superstrings in AdS(5) x S**5,''
  JHEP {\bf 0507} (2005) 002,
  arXiv:hep-th/0502240.

\bibitem{Tseytlin} A.~A.~Tseytlin,
``Semiclassical strings and AdS/CFT,''
arXiv:hep-th/0409296.

\bibitem{Metsaev:1998it}
  R.~R.~Metsaev and A.~A.~Tseytlin,
  ``Type IIB superstring action in AdS(5) x S(5) background,''
  Nucl.\ Phys.\ B {\bf 533} (1998) 109,
  arXiv:hep-th/9805028.

\bibitem{Krichever AdS} I. M. Krichever, ``Two-dimensional
  algebraic-geometrical operators with self-consistent potentials'',
Func. An \& Apps. {\bf 28} (1994) No 1, 26.

\bibitem{KT} M.~Kruczenski and A.~A.~Tseytlin,
  ``Semiclassical relativistic strings in S**5 and long coherent operators  in
  N = 4 SYM theory,'' JHEP {\bf 0409} (2004) 038 [arXiv:hep-th/0406189].

\bibitem{Krichever+Phong}
  I.~M.~Krichever and D.~H.~Phong,
  ``On the integrable geometry of soliton equations and N = 2  supersymmetric
  gauge theories,''
  J.\ Diff.\ Geom.\  {\bf 45} (1997) 349,
  arXiv:hep-th/9604199.

\bibitem{Gleb}
G.~Arutyunov, S.~Frolov, J.~Russo and A.~A.~Tseytlin,
``Spinning strings in AdS(5) x S**5 and integrable systems,''
Nucl.\ Phys.\ B {\bf 671} (2003) 3
[arXiv:hep-th/0307191].

\bibitem{WIP} N. Dorey and B. Vicedo, work in progress.

\bibitem{Maillet} J.~M.~Maillet,
  ``Kac-Moody algebra and extended Yang-Baxter relations in the O(N)
  non-linear $\sigma$-model'',
  Phys. Lett. {\bf 162}B, (1985) 137-142. \\
  J.~M.~Maillet,
  ``Hamiltonian Structures For Integrable Classical Theories From
  Graded Kac-Moody Algebras,''
  Phys. Lett. B {\bf 167} (1986) 401. \\
  J.~M.~Maillet,
  ``New integrable canonical structures in two-dimensional models'',
  Nucl. Phys. B {\bf 269}, (1986) 54-76.

\bibitem{BVII} N. Dorey and B. Vicedo, ``A symplectic structure on the
  space of finite-gap solutions'', in preparation.

\bibitem{DHN} R.~F.~Dashen, B.~Hasslacher and A.~Neveu,
``The Particle Spectrum In Model Field Theories From Semiclassical Functional
Integral Techniques,''
Phys.\ Rev.\ D {\bf 11} (1975) 3424.

\bibitem{Babelon}
  O.~Babelon, D.~Bernard, M.~Talon,
  \textit{Introduction to Classical Integrable Systems},
  Cambridge University Press (2003)

\bibitem{Krichever1}
  I.~M.~Krichever,
  ``Integration of non-linear equations by methods of algebraic geometry,''
  Funct. Anal. Appl. {\bf 11} (1) (1977), 12-26

\bibitem{Krichever2}
  I.~M.~Krichever,
  ``Methods of algebraic geometry in the theory of non-linear equations,''
  Russian Math. Surveys {\bf 32} (6) (1977), 185-213


\bibitem{Krichever:1997sq}
  I.~M.~Krichever and D.~H.~Phong,
  ``Symplectic forms in the theory of solitons,''
  arXiv:hep-th/9708170.

\bibitem{Belokolos}
  E.~D.~Belokolos, A.~I.~Bobenko, V.~Z.~Enol'skii, A.~R.~Its, V.~B.~Matveev,
  \textit{Algebro-Geometric Approach to Nonlinear Integrable Equations},
  Springer-Verlag Telos (1994)

\bibitem{Sklyanin:1995bm}
  E.~K.~Sklyanin,
  ``Separation of variables - new trends,''
  Prog.\ Theor.\ Phys.\ Suppl.\  {\bf 118} (1995) 35.


\end{thebibliography}
\end{document}